\documentclass[onecolumn,aps,superscriptaddress,nofootinbib]{revtex4}

\usepackage{amssymb}
\usepackage{amsmath}
\usepackage{graphicx}
\usepackage[normalem]{ulem}
\usepackage[dvips]{color}
\usepackage{verbatim}
\usepackage{rotating}

\usepackage[normalem]{ulem}
\UseRawInputEncoding

\begin{document}

\title{Comparing pion production in transport simulations of heavy-ion collisions at $270A$ MeV under controlled conditions}

\author{Jun Xu}
\email{junxu@tongji.edu.cn}
\affiliation{School of Physics Science and Engineering, Tongji University, Shanghai 200092, China}

\author{Hermann Wolter}
\email{hermann.wolter@physik.uni-muenchen.de}
\affiliation{Faculty of Physics, University of Munich, D-85748 Garching, Germany}

\author{Maria Colonna}
\email{colonna@lns.infn.it}
\affiliation{INFN-LNS, Laboratori Nazionali del Sud, 95123 Catania, Italy}

\author{Mircea Dan Cozma}
\email{dan.cozma@theory.nipne.ro}
\affiliation{IFIN-HH, Reactorului 30, 077125 M\v{a}gurele-Bucharest,
Romania}

\author{Pawel Danielewicz}
\email{danielewicz@frib.msu.edu}
\affiliation{Facility for Rare Isotope Beams, Michigan State University, East Lansing, Michigan 48824, USA}
\affiliation{Department of Physics and Astronomy, Michigan State University, East
Lansing, Michigan 48824, USA}

\author{Che Ming Ko}
\email{ko@comp.tamu.edu}
\affiliation{Cyclotron Institute and Department of Physics and
Astronomy, Texas A$\&$M University, College Station, Texas 77843,
USA}

\author{Akira Ono}
\email{ono@nucl.phys.tohoku.ac.jp}
\affiliation{Department of Physics, Tohoku University, Sendai
980-8578, Japan}

\author{ManYee Betty Tsang}
\email{tsang@frib.msu.edu}
\affiliation{Facility for Rare Isotope Beams, Michigan State University, East Lansing, Michigan 48824, USA}
\affiliation{Department of Physics and Astronomy, Michigan State University, East
Lansing, Michigan 48824, USA}


\author{Ying-Xun Zhang}
\email{zhyx@ciae.ac.cn}
\affiliation{China Institute of Atomic Energy, Beijing 102413, China}
\affiliation{Guangxi Key Laboratory Breeding Base of Nuclear Physics and Technology, Guilin 541004, China}

\author{Hui-Gan Cheng}
\affiliation{School of Physics and Optoelectronic Technology,
South China University of Technology, Guangzhou 510641, China}

\author{Natsumi Ikeno}
\affiliation{Department of Life and Environmental Agricultural Sciences, Tottori University, Tottori 680-8551, Japan}
\affiliation{RIKEN Nishina Center, 2-1 Hirosawa, Wako, Saitama 351-0198, Japan}

\author{Rohit Kumar}
\affiliation{Facility for Rare Isotope Beams, Michigan State University, East Lansing, Michigan 48824, USA}

\author{Jun Su}
\affiliation{Sino-French Institute of Nuclear Engineering $\&$
Technology, Sun Yat-sen University, Zhuhai 519082, China}

\author{Hua Zheng}
\affiliation{School of Physics and Information Technology, Shaanxi Normal University, Xi¡¯an 710119, China}

\author{Zhen Zhang}
\affiliation{Sino-French Institute of Nuclear Engineering $\&$ Technology, Sun Yat-sen University, Zhuhai 519082, China}

\author{Lie-Wen Chen}
\affiliation{School of Physics and Astronomy, Shanghai Key Laboratory for Particle Physics and Cosmology,
and Key Laboratory for Particle Astrophysics and Cosmology (MOE),
Shanghai Jiao Tong University, Shanghai 200240, China}

\author{Zhao-Qing Feng}
\affiliation{School of Physics and Optoelectronic Technology,
South China University of Technology, Guangzhou 510641, China}

\author{Christoph Hartnack}
\affiliation{SUBATECH, UMR 6457, IMT Atlantique, IN2P3/CNRS Universit\'e de Nantes, 4 rue Alfred Kastler, 44307 Nantes, France}

\author{Arnaud Le F\`evre}
\affiliation{GSI Helmholtzzentrum f\"{u}r Schwerionenforschung, Planckstr. 1, 64291 Darmstadt, Germany}

\author{Bao-An Li}
\affiliation{Department of Physics and Astronomy, Texas A$\&$M
University-Commerce, Commerce, TX 75429-3011, USA}

\author{Yasushi Nara}
\affiliation{Akita International University, Akita 010-1292, Japan}

\author{Akira Ohnishi\footnote{Deceased.}}
\affiliation{Yukawa Institute for Theoretical Physics, Kyoto University, Kyoto 606-8502, Japan}

\author{Feng-Shou Zhang}
\affiliation{Key Laboratory of Beam Technology of Ministry of Education, College of Nuclear Science and Technology, Beijing Normal University, Beijing 100875, China}
\affiliation{Institute of Radiation Technology, Beijing Academy of Science and Technology, Beijing 100875, China}

\collaboration{TMEP Collaboration}

\begin{abstract}
Within the Transport Model Evaluation Project (TMEP), we present a detailed study of the performance of different transport models in Sn+Sn collisions at $270A$ MeV, which are representative reactions used to study the equation of state at suprasaturation densities. We put particular emphasis on the production of pions and $\Delta$ resonances, which have been used as probes of the nuclear symmetry energy. In this paper, we aim to understand the differences in the results of different codes for a given physics model to estimate the uncertainties of transport model studies in the intermediate energy range. Thus, we prescribe a common and rather simple physics model, and follow in detail the results of 4 Boltzmann-Uehling-Uhlenbeck (BUU) models and 6 quantum molecular dynamics (QMD) models. The nucleonic evolution of the collision and the nucleonic observables in these codes do not completely converge, but the differences among the codes can be understood as being due to several reasons: the basic differences between BUU and QMD models in the representation of the phase-space distributions, computational differences in the mean-field evaluation, and differences in the adopted strategies for the Pauli blocking in the collision integrals. For pionic observables, we find that a higher maximum density leads to an enhanced pion yield and a reduced $\pi^-/\pi^+$ yield ratio, while a more effective Pauli blocking generally leads to a slightly suppressed pion yield and an enhanced $\pi^-/\pi^+$ yield ratio. We specifically investigate the effect of the Coulomb force, and find that it increases the total $\pi^-/\pi^+$ yield ratio but reduces the ratio at high pion energies, although differences in its implementations do not have a dominating role in the differences among the codes. Taking into account only the results of codes that strictly follow the homework specifications, we find a convergence of the codes in the final charged pion yield ratio to a $1\sigma$ deviation of about $5\%$. However, the uncertainty is expected to be reduced to about $1.6\%$ if the same or similar strategies and ingredients, i.e., an improved Pauli blocking and calculation of the non-linear term in the mean-field potential, are similarly used in all codes. As a result of this work, we identify the sensitive aspects of a simulation with respect to pion observables, and suggest optimal procedures in some cases. This work provides benchmark calculations of heavy-ion collisions to be complemented in the future by simulations with more realistic physics models, which include the momentum-dependence of isoscalar and isovector mean-field potentials and pion in-medium effects.
\end{abstract}

\maketitle

\section{Introduction}
\label{introduction}

Transport models are indispensable tools for extracting information on the nuclear matter equation of state (EOS) from experimentally measured observables in heavy-ion collisions~\cite{Ono19r,Xu19r,Mar20r,Her21r}. The EOS of isospin symmetric nuclear matter has already been rather well constrained by giant monopole resonances in finite nuclei~\cite{GMR} as well as by the kaon yield~\cite{kaon,Fuchs06} and nucleon collective flows~\cite{Dan02,Fev16} in heavy-ion collisions up to beam energies of the order of $1A$ GeV. On the other hand, there are still uncertainties in the isospin-dependent part of the EOS, i.e., the nuclear symmetry energy, particularly at suprasaturation densities, as compared to the better constrained EOS at subsaturation densities (see, e.g., Ref.~\cite{Lyn22}).  Information on the density dependence of nuclear symmetry energy is essential for understanding problems in nuclear structure, nuclear reactions, and nuclear astrophysics~\cite{Bar05,Ste05,Lat07,BCK08}. As first proposed in Refs.~\cite{Li02a,Li02b}, the nuclear symmetry energy at suprasaturation densities could be determined from the $\pi^-/\pi^+$ yield ratio in intermediate-energy heavy-ion collisions. This is because the $\pi^-/\pi^+$ yield ratio is essentially determined by the neutron/proton ratio in the high-density region of the collisions, which is affected by the stiffness
of the nuclear symmetry energy. Studying the $\pi^-/\pi^+$ yield ratio in intermediate-energy heavy-ion collisions has since attracted considerable interest both theoretically and experimentally. However, based on the same FOPI data on the $\pi^-/\pi^+$ yield ratio~\cite{FOPI}, the extracted density dependence of the nuclear symmetry energy was very different from studies based on different transport models (see, e.g., Refs.~\cite{Xia09,Fen10,Xie13}). It is not easy to understand whether these different conclusions were due to the inclusion of different physics effects, or due to different treatments of same physics inputs in the simulation of heavy-on collisions in transport models. For transport models to be useful for providing information on the EOS and, in particular, on the nuclear symmetry energy, it is thus imperative to require that similar physics observables should be obtained from different transport models with same physics inputs and under the same controlled conditions, or, if this is not the case, to understand the differences due to different approximations made in the simulation of heavy-ion collisions. To achieve this goal has led to the international collaborative studies on the Transport Model Evaluation Project (TMEP).

The activities of the TMEP started in 2014 at Shanghai Jiao Tong University (Transport2014). With investigations examining more closely the details in participant transport models, further workshops were held in 2017 at Michigan State University (Transport2017)~\cite{transport2017,icnt2017} and in 2019 at ECT* (Transport2019)~\cite{transport2019}. In Transport2014, the focus was mainly on the stability of initial nuclei, the collision rate and the Pauli blocking, and on the comparison of observables such as the stopping and the transverse flow in Au+Au collisions at $100A$ and $400A$ MeV.  Benchmark calculations were carried out to quantify the theoretical uncertainties of the slope parameter of the transverse flow in transport models~\cite{Xu16}. These uncertainties, which are larger at low collision energies, could be largely attributed to different initializations and implementations of Pauli blocking, as well as competition between the nuclear mean-field potential and nucleon-nucleon (NN) collisions. To better understand these differences, calculations of nuclear matter in a box with the periodic boundary condition were thus called for, because common initializations are easy to implement in this case. Also, box calculations provide the possibility of evaluating the accuracy of a transport model by comparing its results from simulations to known theoretical limits or exact numerical calculations. This is different from heavy-ion simulations where results may only be compared among different transport codes. Reference \cite{Zha18} presents the first study of the TMEP for a box system, where the NN collision rate as well as the Pauli blocking rate are compared with the theoretical limits. The study in Ref.~\cite{Ono19} is a direct extension of that in Ref.~\cite{Zha18}, by incorporating also inelastic collisions. There, the production of pion-like particles, i.e., pions and $\Delta$ resonances, is compared with that from the rate equations and to a thermal model with ideal gas mixture. Starting from a sinusoidally distorted density distribution, Ref.~\cite{Mar21} investigated the dissipation during the density evolution and compared the main component of the oscillation frequency with that predicted by the Landau parameter of the mean-field interaction. For recent overviews of the TMEP activities, we refer the reader to Refs.~\cite{Xu19r,Her21r}. The previous studies of nuclear matter in a box~\cite{Zha18,Ono19,Mar21} have helped to optimize the implementation of NN elastic and inelastic collisions as well as the momentum-independent mean-field potentials in transport simulations. The aim of this paper is to incorporate this knowledge into transport simulations of heavy-ion collisions to see whether one obtains more robust conclusions.

Recently, the S$\pi$RIT Collaboration has performed a detailed experimental study of Sn+Sn collisions with different isospin contents at $270A$ MeV focusing mainly on pion observables to constrain the symmetry energy at suprasaturation densities. The experiment measured not only pion multiplicities, spectra, and charged pion ratios, but also bulk observables like light cluster yields and spectra, as well as nucleon momentum distributions, which will be published shortly. This data set provides an excellent opportunity to study in detail the potential of pion observables to constrain the nuclear symmetry energy at densities above saturation. To obtain an impression on the status of such efforts, the TMEP collaboration performed a comparison by many codes to predict pion yields and ratios. In the previous study~\cite{Spirit20}, the codes used their ``best'' models to calculate the pion production without prior knowledge of the data. The results confirmed the divergence of conclusions from different studies. The prediction of the codes differed in such a way from each other and from the data, revealed afterwards, that a constraint on the symmetry energy was hardly possible.

Pion production in transport approaches involves many issues on the physics model, which are still being debated and were also discussed in Ref.~\cite{Spirit20}: Different mechanisms contribute to pion production, e.g., a non-resonant mechanism via processes of the type $NN\rightarrow NN\pi$, and a resonant mechanism proceeding via the excitation of a $\Delta$ resonance. In this study, as in the previous box calculation~\cite{Ono19} and realistic heavy-ion calculations~\cite{Spirit20,Spirit21}, we exclusively use the resonant mechanism. For cross sections of non-resonant channels and their relative importance to resonance ($\Delta$) channels, we refer the reader to Refs.~\cite{Eng96,Shy98,Eff97}. The strong potential of the $\Delta$ resonance is not directly accessible and is parametrized in different ways. Also, the pion potential has s- and p-wave components, which are employed in different ways~\cite{Hon14,Zha17,Xu10,Xu13}. The pion and $\Delta$ potentials should be momentum- and isospin-dependent, just as the nucleon potentials. The $\Delta$ potential is either calculated or extracted from comparison to experimental data (see, e.g., Refs.~\cite{Con90,Bod20,Hir77,Hor80,Ose87,Gar91,Jon92,Bal94}). So far, only the isoscalar momentum-independent $\Delta$ potential has been studied, and tensions between theoretical and empirical strengths of the potential are observed. In inelastic collisions, there is a discontinuous change in the particle potentials, resulting in the change of the potential energies. Energy conservation has to be taken into account in a 2-body collision, and this in particular leads to a shift of thresholds in inelastic collisions, which can be crucial in collisions at energies close to production thresholds~\cite{Son15,Coz16}. The finite width of $\Delta$ resonances (as, in principle, also of all other particles in a non-equilibrium situation) in transport models is discussed in Refs.~\cite{Dan96,Dav99,Lar02}. It is usually approximated by sampling a mass distribution, and has to be considered in the detailed balance condition~\cite{pBUU1}. Clustering and light cluster formation may have considerable effects in intermediate-energy heavy-ion collisions, and it has been shown that they may also affect pion production~\cite{Ike16}.

We emphasize that the present study is not aimed at studying these effects. We aim to understand the differences in the results of different codes for a given physics model to estimate the uncertainties of transport model studies in the intermediate energy range. It still allows us to probe the sensitivity of pion production to important physics ingredients, such as the question of the influence of fluctuations, the Pauli blocking including surface corrections, aspects of using effective many-body terms, and inelastic cross sections including a strength function for the $\Delta$ resonance. We note that the pion multiplicities and charged pion yield ratios are in the range of the experimental values and the study is thus qualitatively realistic. It does not include other important issues mentioned above, particularly the momentum dependence of the potentials and corresponding threshold effects. These are studied in a parallel ongoing box calculation in a simpler environment~\cite{Dan22}. In this study, we keep the nuclear symmetry energy fixed and thus do not study the sensitivity of nucleon and pion observables to the symmetry energy as in Ref.~\cite{Spirit20}. We felt that this should be done in a more realistic study of pion production, including, in particular, the momentum-dependent mean-field potential. We also note, however, that in the meantime, Ref.~\cite{Spirit21} with just one transport model (TuQMD/dcQMD) has shown that a more realistic treatment of the physics ingredients, like momentum-dependent mean-field potential, threshold shifts, and pion and $\Delta$ potentials, can yield constraints on the symmetry energy.


\begin{table}[h]\small
  \centering
  \caption{Code names, authors and correspondents, and representative references of 4 BUU-type and 6 QMD-type codes participating in the present study. SMF and ImQMD-L only participate in the comparison of nucleon observables and are therefore listed separately in the last row.
}
    \begin{tabular}{|c|c|c|c|c|c|c|}
    \hline
    BUU-type & code correspondents  & reference  &&  QMD-type & code correspondents & reference\\
    \hline
     IBUU\footnote{A new version using the lattice Hamiltonian framework, recently developed in Ref.~\cite{IBUU-L}, is mainly used in the present study. The original version
(called IBUU-O here) is described in Refs.~\cite{BCK08,IBUU}.}  &  J. Xu, L.W. Chen, B.A. Li & \cite{BCK08,IBUU,IBUU-L} && IQMD-BNU  & J. Su, F.S. Zhang & \cite{IQMD-BNU} \\
     pBUU   &  P. Danielewicz & \cite{pBUU1,pBUU2} && IQMD-IMP\footnote{Also known as LQMD in literature.} & H.G. Cheng, Z.Q. Feng & \cite{IQMD-IMP}\\
     RVUU   & Z. Zhang, C.M. Ko &  \cite{RVUU} && IQMD & R. Kumar, Ch. Hartnack, A. Le F\`evre & \cite{Har98}\\
      & &   && JAM & N. Ikeno, A. Ono, Y. Nara, A. Ohnishi & \cite{Ike16,JAM}\\
      & &   && TuQMD\footnote{This code provides both traditional and accurate calculations of the non-linear density-dependent term in the mean-field potential, with the latter dubbed ``TuQMD-L'' in the present study. The dcQMD model, used in Ref.~\cite{Spirit21} to describe the pion spectra from the S$\pi$RIT experiment, is a recent offspring of the TuQMD transport code.} & M. D. Cozma & \cite{TuQMD}\\
      \hline
     SMF    & M. Colonna, H. Zheng & \cite{SMF} && ImQMD-L\footnote{A lattice version of ImQMD with a more accurate calculation of the non-linear density-dependent term in the mean-field potential recently developed in Ref.~\cite{ImQMD-L} is used in the present study. The original version is described in Ref.~\cite{ImQMD}.} & Y.X. Zhang & \cite{ImQMD,ImQMD-L} \\
    \hline
    \end{tabular}
  \label{T1}
\end{table}

To better understand the model dependence of transport simulations for Sn+Sn collisions at $270A$ MeV seen in Ref.~\cite{Spirit20}, we compare in the present study results from 10 transport models with better controlled setups. Thus one can consider this work a follow-up investigation of the physics issues on pion production in the uncontrolled comparison of Ref.~\cite{Spirit20}. For the 4 BUU models and 6 QMD models participating in the present study, the code names, authors and/or correspondents, and representative references are all summarized in Table~\ref{T1}.

The list of previous works attempting to understand the robustness of transport model predictions is extensive. Below, we discuss briefly some relevant examples. In early studies, the main emphasis was on the comparison of BUU- and QMD-type approaches. A first comparison of bulk observables, including a comparison with the time-dependent Hartree-Fock approach at a relatively low energy of about $80A$ MeV, was performed in Ref.~\cite{Aic89plb}. A good agreement of all methods was found. However, at this low energy the evolution is dominated by the mean field, and collisions are largely suppressed by the Pauli principle. A conference report in Ref.~\cite{Har89p} showed a similar study. Here the ability to reproduce bulk observables in low-energy heavy-ion collisions by a BUU and a QMD code was demonstrated. An early example of a comparison at intermediate energies is that at $800A$ MeV with different versions of BUU codes, one QMD code, and one cascade code~\cite{Aic89prl}, where agreement in the bulk evolution was found. The in-medium cross sections, which dominate at these energies, were the main object of the study, while pion production was not considered. A comparison of pion and kaon production at energies above $1A$ GeV was performed in Ref.~\cite{Kol05} (for a review, see Ref.~\cite{Her21r}), where results among different codes were only qualitatively consistent. However, the physics models were not strictly prescribed as in the present study.

Common to these works, and in contrast to the comparisons performed by TMEP, very few codes were compared for a limited number of observables, and usually not for a well-prescribed physics model. The comparisons resulted in qualitative statements on agreement among the codes. The comparisons by TMEP and this paper follow a more extensive aim. Since the comparisons are performed with controlled conditions, the differences among codes, which do exist, must be due to the simulation strategies. We aim to understand these differences in detail, with the support of box calculations and by following non-observables, e.g., time evolution of densities, asymmetries, and collisions. In this way, we hope to trace back the differences to different strategies in the simulation, and, hopefully, to identify sensitive issues of simulations and to recommend successful strategies. The TMEP comparisons and the present paper would then help to improve transport code interpretations of heavy-ion collisions.

On the other hand, we do not intend to present a converged result of all the codes or to identify the ¡°best¡± code in this work. Rather, we aim to quantify the systematic uncertainties of transport analyses that originate from various sources, ranging from numerical issues to fundamental model assumptions for complicated many-body problems. While trying to quantify differences, determine their origins among the codes, and discuss optimized strategies for some simulation issues, we must leave it to the code owners to implement lessons from these comparisons. Different simulation strategies often include effects beyond what the underlying equations prescribe. Thus, the optimal strategy or the best model assumptions are not always clear. Some indications are given by comparing the results from box calculations with exact results. Still, a box calculation of infinite matter and dynamic heavy-ion collisions are very different systems, and strategies that work well in box calculations may not necessarily be optimal in heavy-ion collisions. There exist open questions in transport approaches to heavy-ion collisions that we cannot solve by simple comparisons, even in box calculations. The most important one is the question of realistically including fluctuations, which affect the results of collisions but go beyond the canonical BUU equations and are treated empirically in QMD-like codes, thus constituting a basic difference between these two families of codes. Code comparisons, like those performed by the TMEP collaboration, can highlight these various aspects and are valuable for understanding open questions such as the above, but cannot completely solve the problem of achieving converged predictions. A possible way to achieve this would be to construct a common code within the collaboration, which could be done in the future but appears unrealistic at the present stage. In the conclusion, we will discuss our ideas to obtain at least uncertainty quantifications for transport analyses.

The rest of this paper is organized as follows. The theoretical framework of BUU and QMD transport models, including the implementation of the Coulomb potential and the revisiting of the Pauli blocking, is briefly reviewed in Sec.~\ref{theory}. Section~\ref{homework} gives the details of the homework description that was assigned to code correspondents, especially those related to pion production. We first compare nucleon observables in Sec.~\ref{nucleon} to check the agreement of the bulk evolution of the collision, and then pion observables in Sec.~\ref{results}. In these discussions, we attempt to follow in great detail the differences among the codes and to understand underlying reasons for the differences. For a first reading of the paper, these detailed explanations may not be necessarily needed to obtain the main message of the paper. In this case we recommend to skip Sec.~\ref{results} except to take a look at Figs.~\ref{pimul} and \ref{ratio}, which contain the main results of the paper with respect to pion observables, and to continue with Sec.~\ref{discussion}, where the results of the comparison are discussed from a more global point of view, and finish with the conclusion and outlook in Sec.~\ref{summary}.

\section{Theoretical framework of transport models}
\label{theory}

In this section, we briefly review the main features of BUU and QMD transport models for the simulations of intermediate-energy heavy-ion collisions. The major focus will be on the Coulomb potential and the Pauli blocking and their effects on the $\pi^-/\pi^+$ yield ratio. One may consult Ref.~\cite{Her21r} and the corresponding references for more details and default setups in the participated codes listed in Table.~\ref{T1}.

\subsection{The Boltzmann-Uehling-Uhlenbeck approach}
\label{buu}

The BUU approach solves numerically the BUU equation for the one-body phase-space distribution function $f_a(r,p,t)$ which is typically of the form
\begin{equation}
  \label{eq:boltz}
  \frac{\partial f_a(\vec{r},\vec{p},t)}{\partial t}
  +\frac{\vec{p}}{\sqrt{m_a^2+\vec{p}^2}}\cdot
  \frac{\partial f_a(\vec{r},\vec{p},t)}{\partial \vec{r}}-\frac{\partial U_a}{\partial \vec{r}}\cdot\frac{\partial f_a(\vec{r},\vec{p},t)}{\partial \vec{p}}
  =I_a(\vec{r},\vec{p},t).
\end{equation}
The above equation is written for relativistic kinematics, where the index $a$ labels the different particle species, and $m_a$ are their rest masses. For fully covariant forms of this equation used in pBUU, see Refs.~\cite{pBUU1,pBUU2}. In the present study, we include only nucleons, pions, and $\Delta$ resonances of different charge states or isospins, i.e., $a\in N\cup\pi\cup\Delta$ with the sets
\begin{align}
  N&=\{n,\ p\},\\ \pi&=\{\pi^-,\ \pi^0,\ \pi^+\},\\
  \Delta &= \{\Delta^-,\ \Delta^0,\ \Delta^+,\ \Delta^{++}\}.
\end{align}
While nucleons and pions have fixed masses, we take masses of $\Delta$ resonances, $m_{\Delta}$, smoothly distributed according to a spectral function of the Breit-Wigner form $A(m_\Delta)$ with a finite width [see Eq.~(\ref{BW}) for details]. $U_a$ denotes the mean-field potential for species $a$, which is non-zero only for nucleons and taken to be momentum-independent in this study. The collision term $I_a(\vec{r},\vec{p},t)$ in Eq.~(\ref{eq:boltz}) includes elastic binary collisions between nucleons and $\Delta$ resonances, inelastic collisions with the production and absorption of $\Delta$ resonances, and the production and absorption of pions via respectively the decay and formation of $\Delta$ resonances. Suppressing the isospin indices for different particle species, the collision term can be expressed for each species as
\begin{eqnarray}
I_N &=& I_N^{NN \leftrightarrow NN} + I_N^{N\Delta \leftrightarrow N\Delta} + I_N^{NN \leftrightarrow N\Delta} + I_N^{N\pi \leftrightarrow \Delta}, \\
I_\Delta &=& I_\Delta^{\Delta \Delta \leftrightarrow \Delta \Delta} + I_\Delta^{N\Delta \leftrightarrow N\Delta} + I_\Delta^{NN \leftrightarrow N\Delta} + I_\Delta^{N\pi \leftrightarrow \Delta}, \\
I_\pi &=& I_\pi^{N\pi \leftrightarrow \Delta}.
\end{eqnarray}
In the above, $I_N^{NN \leftrightarrow NN}$, $I_N^{N\Delta \leftrightarrow N\Delta}$, $I_\Delta^{\Delta \Delta \leftrightarrow \Delta \Delta}$, and $I_\Delta^{N\Delta \leftrightarrow N\Delta}$ represent elastic scatterings between baryons, with Pauli blocking effects only present for nucleons. The terms related to pion production can be expressed in terms of the $\Delta$ production cross sections ($\sigma$) and/or decay widths ($\Gamma$) as
\begin{eqnarray}
I_\Delta^{NN \leftrightarrow N\Delta} &=& -I_N^{NN \leftrightarrow N\Delta}\nonumber \\ \
&=& \frac{g_N^2}{(1+\delta_{N_1N_2})g_\Delta^{}}\int\frac{d^3p_2}{(2\pi\hbar)^3}\int d\Omega
\ f_{N_1}f_{N_2}
v_{mol}\frac{d\sigma^{}_{N_1N_2\to N_3\Delta}}{d\Omega}
(1-n_{N_3}^{}) \nonumber \\
&-&g_N^{}\int\frac{d^3p_2}{(2\pi\hbar)^3}\int d\Omega
\ f_{N_1}f_\Delta
v_{mol}\frac{d\sigma^{}_{N_1\Delta\to N_3N_4}}{d\Omega}
(1-n_{N_3}^{})(1-n_{N_4}^{}), \label{eq:IN-NNND}\\
I_\Delta^{N\pi \leftrightarrow \Delta}&=&
\frac{g_N^{}g_\pi^{}}{g_\Delta^{}}\int \frac{dm_\Delta}{2\pi} A(m_\Delta)\int\frac{d^3p_2}{(2\pi\hbar)^3}f_Nf_\pi v_{mol}\sigma_{N\pi\to\Delta}^{}
-\int\frac{d\Omega}{4\pi}f_\Delta\Gamma'_{\Delta\to N\pi}(1-n_N^{})
,\label{eq:IDelta-DNpi}\\
I_\pi^{N\pi \leftrightarrow \Delta} &=&
\frac{g_\Delta^{}}{g_\pi^{}}\int\frac{d\Omega}{4\pi}f_\Delta\Gamma'_{\Delta\to N\pi}(1-n_N^{})
-g_N^{}\int \frac{dm_\Delta}{2\pi} A(m_\Delta)\int\frac{d^3p_2}{(2\pi\hbar)^3}f_Nf_\pi v_{mol} \sigma_{N\pi\to\Delta}^{}.\label{eq:Ipi-NpiD}
\end{eqnarray}
In above equations, the integrals are generally taken over the momentum $\vec{p}_2$ of the second scattering particle in $1+2 \rightarrow 3+4$ or $1+2 \rightarrow 3$ channels, $g_N^{}=2$, $g_\Delta^{}=4$, and $g_\pi^{}=1$ are the spin degeneracy factors, $v_{mol}$ is the M{\o}ller velocity between two scattering particles given by
\begin{equation}
v_{mol}=\frac{\sqrt{[s-(m_1+m_2)^2][s-(m_1-m_2)^2]}}{2E_1E_2},
\end{equation}
and the angular integrals are taken in the center-of-mass (C.M.) frame of the particles. Moreover, $s$ is the square of the invariant mass of the scattering particle pair, and $E_{1(2)}=\sqrt{m_{1(2)}^2+\vec{p}_{1(2)}^{\ 2}}$ are their energies. The width $\Gamma'_{\Delta\to N\pi}=\frac{m_\Delta}{\sqrt{m_\Delta^2+\vec{p}_\Delta^{\ 2}}}\Gamma_{\Delta\to N\pi}$ is that of the $\Delta$ resonance in the computational frame, which is the rest frame of the nuclear matter in the box calculations or the C.M. frame of colliding nuclei in heavy-ion collisions, with $\vec{p}_\Delta$ being the $\Delta$ momentum in the computational frame and $\Gamma_{\Delta\to N\pi}$ being the $\Delta$ width in its rest frame. The Pauli blocking factors $(1-n_N^{})$ are introduced only for nucleons, with $n_N^{}$ being their occupation probability to be given in Sec.~\ref{pb}. We do not consider Pauli blocking for $\Delta$ resonances or Bose enhancement for pions, since their occupation factors are very small in the reactions considered in the present study.

In the BUU approach, Eq. (\ref{eq:boltz}) is solved using the TP method by expressing the particle phase-space distribution functions according to~\cite{Won82,Ber88}
\begin{equation}
f_a (\vec{r},\vec{p},t) = \frac{(2\pi\hbar)^3}{g_a N_{TP}} \sum_{i \in a}^{A_a N_{TP}} G(\vec{r}-\vec{r}_i) G'(\vec{p}-\vec{p}_i),
\end{equation}
where $A_a$ is the particle number for species $a$, $N_{TP}$ is the number of test particles (TPs) per particle, $\vec{r}_i$ and $\vec{p}_i$ are the centroid coordinate and momentum of the $i$th TP, and $G$ and $G'$ are the shape functions in coordinate and momentum space, respectively. While $G'$ is typically chosen as a $\delta$ function, the use of a Gaussian function or a triangular function has often been adopted for $G(\vec{r}-\vec{r}_i)$. For point TPs, this leads to Hamiltonian equations of motion as shown in Ref.~\cite{Ber88}. For finite-size TPs, typically in the lattice Hamiltonian framework~\cite{Len89}, the coordinate space is divided into cubic cells with the volume $l^3$, and the density at the site $\vec{r}_\alpha$ of the lattice is then given by
\begin{equation}\label{rhol}
\rho_L (\vec{r}_\alpha) = \frac{1}{N_{TP}}\sum_{i \in a}^{A_a N_{TP}} G(\vec{r}_\alpha-\vec{r}_i).
\end{equation}
For example, in IBUU and SMF the shape function is defined as
\begin{equation}\label{shape}
G(\vec{r}_\alpha-\vec{r}_i) = \frac{1}{(nl)^6}g(x)g(y)g(z)
\end{equation}
with $x=x_\alpha-x_i$, $y=y_\alpha-y_i$, $z=z_\alpha-z_i$, and
\begin{equation}\label{gfunc}
g(q)=(nl-|q|)\Theta(nl-|q|),
\end{equation}
where $l$ is the lattice spacing, $n$ determines the range of $G$, and $\Theta$ is the Heaviside function. pBUU uses a modified lattice Hamlitonian framework with a different $g$ function in calculating the density and mean-field potential~\cite{pBUU2} as follows: in the interior of the computational
volume, pBUU uses $g(q)=0.5$ for $|q|/l<0.5$, $g(q)=0.75-0.5|q|/l$ for $0.5<|q|/l<1.5$, and $g(q)=0$ for $1.5<|q|/l$. At a forward edge, pBUU uses $g(q)=0.75+0.5q/l$ for $-1.5<q/l<0.5$, and $g(q)=0$ outside of that interval. The corresponding coefficient in Eq.~(\ref{shape}) is modified accordingly to satisfy $l^3\sum_\alpha G(\vec{r}_\alpha-\vec{r}_i) =1$. This code, however, modifies its default mean-field potential with a scalar nucleon mass~\cite{pBUU2} to fit the nuclear matter properties described by the homework setup to be given later, so it uses effectively a momentum-dependent mean-field potential in the non-relativistic reduction.

The total potential energy of the system is the sum of the potential energy in each cubic cell, i.e., $H^{pot}=l^3 \sum_\alpha \epsilon^{pot}_\alpha$, with $\epsilon^{pot}_\alpha$ being the potential energy density at site $\alpha$. For a momentum-independent mean-field potential $U$, the drift part of the BUU equation, i.e., the left-hand-side of Eq.~(\ref{eq:boltz}), is solved by the canonical equations of motion for TPs, i.e.,
\begin{eqnarray}
\frac{d\vec{r}_i}{dt} &=& \frac{\vec{p}_i}{\sqrt{m_i^2+\vec{p}_i^2}},\\
\frac{d\vec{p}_i}{dt} &=& -l^3 \sum_\alpha \frac{\partial \epsilon^{pot}_\alpha}{\partial \rho_L} \frac{\partial \rho_L}{\partial \vec{r}_i} = -\frac{l^3}{N_{TP}} \sum_\alpha U[\rho_L(\vec{r}_\alpha)] \frac{\partial G(\vec{r}_\alpha-\vec{r}_i)}{\partial \vec{r}_i},\label{eom}
\end{eqnarray}
where $U[\rho_L(\vec{r}_\alpha)]$ is the mean-field potential. In the case of using point TPs, i.e., $G(\vec{r}-\vec{r}_i)=\delta(\vec{r}-\vec{r}_i)$, Eq.~(\ref{eom}) reduces to
\begin{eqnarray}
\frac{d\vec{p}_i}{dt}=-\frac{\partial U}{\partial \vec{r}_i}.
\end{eqnarray}

The collision term in Eqs. (\ref{eq:IN-NNND})-(\ref{eq:Ipi-NpiD}) is simulated stochastically. This is discussed in great detail for elastic collisions in Ref.~\cite{Zha18} and for inelastic collisions and $\Delta$ resonance decay in Ref.~\cite{Ono19} (as an overview, see Ref.~\cite{Her21r}). Briefly, it is simulated by stochastic particle (or test-particle for BUU) collisions. This involves a determination of a collision probability, the evaluation of the Pauli blocking factors, and the consideration of energy-momentum conservation. The Pauli blocking is discussed in Sec.~\ref{pb} below. The collision probability is treated in different ways in the codes represented here. In most codes, a geometric criterion is used based on the total cross section, while some codes choose collision partners randomly in a cell based on the mean free path depending on the density (see Sec.~\ref{specifics} below). The total momentum is conserved in each collision. The total energy is conserved in each elastic collision, and in the mean-field evolution within the numerical accuracy as checked in Ref.~\cite{Mar21}. Still, such constraint could be violated in inelastic collisions or decays (see the homework description in Sec.~\ref{homework}), and possible effects due to momentum-dependent potentials are not considered here (see remarks at the end of Sec.~\ref{homework} below). The effect of angular momentum conservation in transport simulations of heavy-ion collisions was first discussed in Ref.~\cite{Gal90} and recently revisited in Ref.~\cite{Liu23a}. To put the constraint of the angular momentum conservation more consistently, one should, in principle, also incorporate a spin degree of freedom (see, e.g., Ref.~\cite{Liu24b}). The way to incorporate properly the total angular momentum conservation in elastic and decay channels needs further investigation, and the effects on pion production are largely unknown. For this study, our main concern is that the treatment of the codes is comparable based on a canonical setup that all code practitioners can practically incorporate.

\subsection{The quantum molecular dynamics approach}

In the QMD approach~\cite{Har89,Aic91}, instead of considering the time evolution of the one-body phase-space distribution function as in the BUU approach, the phase-space distribution function of each particle is followed in time. Approximating the wave function of the $i$th particle by a Gaussian wave packet of spatial width $\Delta x$, i.e.,
\begin{equation}
\phi_i(\vec{r};t) = \frac{1}{(2\pi \Delta x)^{4/3}} \exp\left[ -\left(\frac{\vec{r}-\vec{r}_i(t)}{2\Delta x}\right)^2+ \frac{i\vec{p}_i(t) \cdot \vec{r}}{\hbar}\right],
\end{equation}
where $\vec{r}_i(t)$ and $\vec{p}_i(t)$ are its centroid coordinate and momentum at time $t$, its phase-space distribution function is calculated from the Wigner transformation of the wave function and has the form
\begin{eqnarray}\label{wigner}
f_i(\vec{r},\vec{p}) &=& \int \phi^*_i(\vec{r}-\vec{s}/2) \phi_i(\vec{r}+\vec{s}/2) \exp(-i\vec{p}\cdot\vec{s}\,) d^3s \notag\\
&=& 8\exp \left[ -\frac{(\vec{r}-\vec{r}_i)^2}{2(\Delta x)^2} - \frac{2(\Delta x)^2(\vec{p}-\vec{p}_i)^2 }{\hbar^2}\right].
\end{eqnarray}

To obtain the mean-field potential exerted on a particle, the QMD approach starts from a many-body Hamiltonian with the potential part expressed as $H^{pot}=\frac{1}{2} \sum_{i \ne j} \langle V \rangle_{ij} $, where $V$ is an effective two-body interaction. Since the total wave function of a system in most QMD models\footnote{This is not the case for the antisymmetrized molecular dynamics model~\cite{AMD} and the Fermionic molecular dynamics model~\cite{FMD}, where the total wave function is antisymmetrized.} is taken to be the direct product of single-particle wave functions, i.e., $\Phi(\vec{r};t)=\Pi_i \phi(\vec{r},\vec{r}_i,\vec{p}_i;t)$, the expectation value of the effective interaction can be calculated using the single-particle phase-space distribution function given by Eq.~(\ref{wigner}), i.e.,
\begin{eqnarray}
\langle V \rangle_{ij} \ &=& \frac{1}{(2\pi\hbar)^6} \int f_i(\vec{r},\vec{p}) f_j(\vec{r}',\vec{p}') V(|\vec{r}-\vec{r}'|) d^3r d^3r' d^3p d^3p'. \label{U2}
\end{eqnarray}

The centroid coordinate and momentum of the $i$th-particle phase-space distribution function in the QMD approach evolve in time according to the canonical equations based on the many-body Hamiltonian. For a momentum-independent effective two-body interaction as in the present study, the corresponding equations of motion are
\begin{eqnarray}
\frac{d\vec{r}_i}{dt} &=& \frac{\vec{p}_i}{\sqrt{m_i^2+\vec{p}_i^{\ 2}}},  \\
\frac{d\vec{p}_i}{dt} &=& -\frac{1}{2}\sum_{j(\ne i)}\frac{\partial \langle V \rangle_{ij} }{\partial \vec{r}_i} = -\frac{\partial H^{pot} }{\partial \vec{r}_i}. \label{qmddp}
\end{eqnarray}
Generally, a density-dependent potential is often derived from an energy-density functional and used in Eq.~(23), which often contains a non-linear term due to density-dependent forces. In the present study, to be specified later, it is proportional to $\rho^\gamma$ with $\gamma>1$. One then has to calculate the spatial average $\overline{\rho^\gamma}$. In the traditional QMD calculation, the approximation $\overline{\rho^\gamma} \approx \overline{\rho}^\gamma$ is generally used, but such approximation may not be valid in the case of large density fluctuations and large $\gamma$. It was seen in the box study of Ref.~\cite{Mar21}, that this approximation effectively leads to weaker gradients of this repulsive mean-field potential. In ImQMD-L~\cite{ImQMD-L} and also in the lattice version of TuQMD (TuQMD-L), the $\overline{\rho^\gamma}$ term is calculated exactly by numerical methods. In BUU models, where the event average through the TP method is used, there is no corresponding approximation employed in calculating the repulsive non-linear density-dependent term in the mean-field potential, in contrast to the event-by-event mean-field calculation in QMD models. We will discuss later the impact of these different treatments of this non-linear term on the density evolution and the pion yield.

\subsection{Coulomb potential}

For the electromagnetic field acting on a particle in transport simulations of heavy-ion collisions, one should ideally include its retardation effect. Since the implementation of this effect is technically involved and computationally time-consuming, most transport models for low- and intermediate-energy heavy-ion collisions only include the static Coulomb potential acting on a charged particle. The static Coulomb potential is calculated by different methods in different codes, which will be seen later to influence the final results to some extent, especially the $\pi^-/\pi^+$ yield ratio. In the following, we briefly discuss these different methods.

In IBUU and RVUU models, the Coulomb force acting on the $i$th particle is calculated from the Coulomb potential for point particles as
\begin{equation}\label{coubuu}
\vec{F}_i^{cou} = \frac{Z_i e^2}{N_m} \sum_{j (\ne i), r_{ij}>r_c} Z_j \frac{\vec{r}_{ij}}{r_{ij}^3},
\end{equation}
where $Z_{i(j)}$ is the charge number of the $i(j)$th particle, and $\vec{r}_{ij}=\vec{r}_i-\vec{r}_j$ is the relative coordinate vector. To avoid a divergence or a very large value of the Coulomb force, for the case that two TPs are close together\footnote{This may happen in a decay process, e.g., $\Delta^{++} \rightarrow p + \pi^+$, where the two charged particles in the final state are created at the same point. In such cases, the Coulomb force is often calculated in the next time step after the propagation of the charged particles.}, a cut-off radius $r_c$ is introduced, below which the Coulomb force is taken to be that at $r_{ij}=r_c$. An additional procedure is sometimes applied by taking the summation not only over the particles in one event, but over all particles in $N_m$ parallel events, i.e., mixing $N_m$ events, and in this way the Coulomb force is further smoothed. Thus, the Coulomb potential may depend on the two smoothing parameters $r_c$ and $N_m$.

To be more consistent with the TP method in BUU models, the Coulomb potential can be calculated in the lattice Hamiltonian framework, as has been done for IBUU in this study. Here the Coulomb force acting on the $i$th particle is calculated from the Coulomb potential energy density $\epsilon^{cou}_\alpha$ according to
\begin{equation}\label{coulombf}
\vec{F}_i^{cou} = -l^3 Z_i \sum_\alpha \frac{\partial \epsilon^{cou}_\alpha}{\partial \vec{r}_i},
\end{equation}
with the summation over the lattice sites $\vec{r}_\alpha$. Including the direct contribution as well as the exchange contribution based on the local density approximation from the Coulomb interaction, the Coulomb potential energy density $\epsilon^{cou}_\alpha$ can be expressed as
\begin{equation}\label{coulombv}
\epsilon^{cou}_\alpha = \frac{l^3}{2} e^2 \sum_{\alpha' (\neq \alpha)} \frac{\rho_L^c(\vec{r}_\alpha) \rho_L^c(\vec{r}_{\alpha'}) }{|\vec{r}_\alpha - \vec{r}_{\alpha'}|}
- \frac{3}{4} e^2 \left[ \frac{3 \rho_L^c (\vec{r}_\alpha)}{\pi} \right]^{4/3}.
\end{equation}
In the above, $\rho_L^c(\vec{r}_\alpha)$ is the net-charge number density at the lattice site $\vec{r}_\alpha$ and is calculated in a way similar to Eq.~(\ref{rhol}) for the particle density.  Substituting Eq.~(\ref{coulombv}) into Eq.~(\ref{coulombf}) leads to the following Coulomb force acting on the $i$th charged particle
\begin{equation}\label{coulat}
\vec{F}_i^{cou} = - \frac{l^3 Z_i e^2}{N_{TP}} \sum_\alpha \left\{ l^3\sum_{\alpha' (\neq \alpha)} \frac{\rho_L^c(\vec{r}_\alpha)}{|\vec{r}_\alpha - \vec{r}_{\alpha'}|} \frac{\partial G(\vec{r}_{\alpha'} - \vec{r}_i)}{\partial \vec{r}_i} - \frac{3}{\pi} \left[ \frac{3 \rho_L^c (\vec{r}_\alpha)}{\pi}\right]^{1/3} \frac{\partial G(\vec{r}_\alpha - \vec{r}_i)}{\partial \vec{r}_i}
\right\}.
\end{equation}
We note that the exchange contribution represented by the second term in Eqs.~(\ref{coulombv}) and (\ref{coulat}) is usually neglected in transport model calculations, since it is generally much smaller than the direct contribution.

In the pBUU model, the static electric field $\phi$ due to the Coulomb interaction is obtained from the Poisson equation
\begin{equation}
\nabla^2 \phi(\vec{r}) = -4 \pi e \rho^c(\vec{r}),
\end{equation}
where $\rho^c$ is the charge density. The above equation is solved numerically on the three-dimensional spatial grid, with the boundary condition determined by comparing to the integral representation of the potential field
\begin{equation}
\phi(\vec{r}) = e \int d^3 r' \frac{\rho^c(\vec{r}')}{|\vec{r}-\vec{r}'|},
\end{equation}
with its values on the boundaries evaluated using the multipole expansion of $1/|\vec{r}-\vec{r}'|$. An iterative relaxation method is employed in Ref.~\cite{PD} by updating $\phi$ at each time step of the evolution of the system. For SMF, while the calculation with the Coulomb interaction is not compared in the present study, this code also solves the Poisson equation but with simpler boundary conditions, corresponding to the potential obtained at the boundaries of the box assuming that the total charge of the system is concentrated at the center of the box.

In the QMD approach, the Coulomb interaction between two particles at respective positions $\vec{r}$ and $\vec{r}'$ has the standard form
\begin{equation}
V^{cou} = \frac{Z_i Z_j e^2}{|\vec{r}-\vec{r}'|}.
\end{equation}
Substituting it into Eq.~(\ref{U2}) leads to the following Coulomb force acting on the $i$th particle
\begin{eqnarray}\label{couqmd}
\vec{F}_i^{cou} = - \frac{\partial }{\partial \vec{r}_i} \left[\frac{e^2}{2} Z_i \sum_{j (\ne i)} \frac{Z_j}{r_{ij}} {\rm erf}\left( \frac{r_{ij}}{2\Delta x}\right)\right]
= -\frac{e^2}{2} Z_i \sum_{j (\ne i)} Z_j \left[ \frac{\exp[(-r_{ij}/2\Delta x)^2] }{\sqrt{\pi} r_{ij}\Delta x} - \frac{1}{r^2_{ij}}{\rm erf}\left( \frac{r_{ij}}{2\Delta x}\right) \right] \cdot \frac{\vec{r}_{ij}}{r_{ij}},
\end{eqnarray}
where $\vec{r}_{ij}=\vec{r}_i-\vec{r}_j$ is the relative position vector between the centroid coordinates of particle $i$ and $j$, and ${\rm erf}(x) = \frac{2}{\sqrt{\pi}} \int^x_0 e^{-u^2} du$ is the error function. The above standard method is used for most QMD models in the present work. It is similar to the lattice Hamiltonian method described above, and the summation is over all particles in the same event. Since the Gaussian shape of the wave packets is explicitly used, the strength of the Coulomb force depends on the width parameter $\Delta x$. In the JAM calculation for the present work, both the electric and magnetic force acting on the $i$th particle are taken into account, and their effects on its motion are calculated, as in Ref.~\cite{Ike23}, according to
\begin{equation}
\left(\frac{dp_i^\mu}{dt}\right)_{EB} = eZ_iF_i^{\mu\nu}\frac{dx_{i,\nu}}{dt}
,\qquad
F_i^{\mu\nu}=\sum_{j (\ne i)}F_{ij}^{\mu\nu}
\end{equation}
by using the electromagnetic field tensor $F_i^{\mu\nu}$ at the 4-coordinate of the particle $x_{i,\nu}=(t,-x_i,-y_i,-z_i)$.  The field tensor $F_{ij}^{\mu\nu}$ created by the $j$th particle is obtained by first using the electrostatic Coulomb and magnetic fields for point charges, i.e., $\vec{E}'=eZ_j\vec{r}_{ij}'/(r_{ij}')^3$ and $\vec{B}'=0$, in the rest frame of the $j$th particle and then Lorentz transforming them to the computational frame. The electric field $\vec{E}'$ is assumed to be zero for $r_{ij}'\le$ 2 fm, instead of assuming a constant Coulomb force for $r_{ij}'=|\vec{r}_{ij}'|$ less than a cutoff distance in the rest frame of the $j$th particle as in some of the BUU models.

\subsection{Pauli blocking}
\label{pb}

The Pauli blocking of nucleons in $N+N\leftrightarrow N+\Delta$ and $\Delta \rightarrow N+\pi$ channels has a large effect on pion production, and affects the $\pi^-/\pi^+$ yield ratio in collisions of nuclei with unequal proton and neutron numbers~\cite{Ike20}. In most transport models, the Pauli blocking probability in a scattering or decay process is evaluated according to $1-(1-n_i)(1-n_j)$, where $n_{i(j)}$ is the occupation probability of the $i(j)$th particle in the final state of the scattering or decay. Because of fluctuations in the calculation of $n_{i(j)}$, its value can sometimes be greater than one and is taken to be 1 when this happens. The occupation probability is further calculated separately for neutrons and protons in all transport models participating in the present study. We briefly discuss in the following the methods for the Pauli blocking used in different transport models as mentioned in Ref.~\cite{Zha18}, which are applied for nucleons in not only elastic scatterings but also inelastic scatterings and $\Delta$ decay processes.

In IBUU, RVUU, and SMF, the occupation probability for the $i$th nucleon is calculated from
\begin{equation}\label{pbbuu}
n_i = \frac{(2\pi\hbar)^3}{g_N V_r V_p} \int_{i \in V_r,V_p} f(\vec{r}, \vec{p}\,) d^3r d^3p,
\end{equation}
where the integral is taken over the volume $V_{r(p)}$ of the phase-space cell in coordinate (momentum) space around that of the $i$th nucleon. While the phase-space cell is taken to be cubic in the spatial and momentum spaces with $V_r = (\Delta r)^3$ and $V_p = (\Delta p)^3$ in the IBUU model, it is taken to be spherical with $V_r = 4\pi(\Delta r)^3/3$ and $V_p = 4\pi(\Delta p)^3/3$ in the RVUU and SMF model. In both IBUU and RVUU models, an interpolation method by smoothing the distribution in neighbouring cells is used in evaluating the occupation probability, while a Gaussian weight of TPs in momentum space is applied in SMF. In the pBUU model, the Pauli blocking probability is calculated by parameterizing the local momentum distribution in a cell in terms of two deformed Fermi distributions of finite temperature, associated, respectively, with the target and projectile distributions. Specifically, one first establishes the local frames for the two groups of nucleons and then calculates their momentum tensors $\langle p^i p^j \rangle$ in their respective frames. After diagonalizing $\langle p^i p^j \rangle$, the local momenta are rescaled to make the respective tensors isotropic, and the Fermi distribution parameters $\mu'$ ($\mu''$) and $T$ ($T''$) are determined for projectile (target) nucleons. Finally, the parameterized occupation probability at $\bar{p}$ is calculated as~\cite{pBUU1}
\begin{equation}
n_i(\bar{p}) = \frac{A'}{\exp[(\sqrt{m^2+p'^2}-\mu')/T'] + 1} + \frac{A''}{\exp[(\sqrt{m^2+p''^2}-\mu'')/T''] + 1}.
\end{equation}
In the above, $p'$ ($p''$) represents $\bar{p}$ subjected to the same transformation as projectile (target) nucleon momenta, and $A'$ ($A''$) is a scaling factor to yield the correct density of projectile (target) nucleons. The above method is referred to as the ``local statistic method" in later discussions. Although the lattice Hamiltonian framework is used to calculate the density distribution and the gradient of the mean-field potential in IBUU and pBUU, point TPs are used to evaluate the occupation probability for the Pauli blocking of nucleons.

In IQMD-BNU, IQMD-IMP\footnote{IQMD-IMP uses this method only for the purpose of the present study, but usually the one given in Ref.~\cite{Zha18}.}, and JAM models, the occupation probability is calculated from the overlap of the wave packets, i.e.,
\begin{equation}\label{pb1qmd}
n_i = \frac{(2\pi\hbar)^3}{g_N^{}} \frac{1}{(\pi \hbar)^3} \sum_{j (\ne i)} \exp \left[ -\frac{(\vec{r}_j-\vec{r}_i)^2}{2(\Delta x)^2} - \frac{2(\Delta x)^2(\vec{p}_j-\vec{p}_i)^2 }{\hbar^2}\right],
\end{equation}
where the summation calculates the contributions from all particles to the local phase space but with the self-contribution subtracted. In IQMD and TuQMD, the occupation probability is calculated using the hard sphere overlap
\begin{equation}\label{pb2qmd}
n_i = \sum_{j (\ne i)} \frac{O_{ij}^{(r)}}{\frac{4}{3}\pi (\Delta r)^3} \frac{O_{ij}^{(p)}}{\frac{4}{3}\pi (\Delta p)^3},
\end{equation}
where $O_{ij}^{(r(p))}$ is the volume of the overlap region of spheres with the radius $\Delta r$ ($\Delta p$) for nucleons $i$ and $j$ in coordinate (momentum) space. For the case of a nucleon close to the surface, only part of the classically available phase-space is allowed. To correct for this effect, the occupation probabilities of these nucleons are calculated in a different way in TuQMD and IQMD (see, e.g., Ref.~\cite{Dan18} for more details). The criterion of determining surface nucleons can be based on, e.g., a single-particle energy lower than 5 MeV, as is done in the present comparison, possibly together with certain cuts for the local density and momentum of the nucleon.

\subsection{Summary of model specifics}
\label{specifics}

\begin{table}[h]\small
  \centering
  \caption{Summary of model specifics in the 4 BUU models and 6 QMD models used in the present study: TP shape/size, implementation of the Coulomb potential and Pauli blocking, and the time step. SMF and ImQMD-L participated in the calculation without Coulomb and for nucleon observables only. See text for more details.}
    \begin{tabular}{|c|c|c|c|c|c|c|}
    \hline
     & TP shape/size  & Coulomb potential &  Pauli blocking & $\Delta t$ (fm/c) \\
    \hline
     IBUU & $n=2$, $l=1$ fm & $N_m=1$, $r_c=1$ fm& cubic cell, $\Delta r = 2$ fm, $\Delta p = 100$ MeV/c, 100 TPs & 0.2 \\
     pBUU & modified $g$, $l=0.92$ fm & Poisson equation & fit with a superposition of two FD distributions & 0.2 \\
     RVUU & point & $N_m=10$, $r_c=0.1$ fm & spherical cell, $\Delta r = 1$ fm, $\Delta p = 150$ MeV/c, 1000 TPs & 0.2 \\
    \hline
     SMF & $n=2$, $l=1$ fm & Poisson equation & spherical cell, $\Delta r = 2.53$ fm, $\sigma_p = 29$ MeV/c, 100 TPs & 0.5 \\
    \hline
    IQMD-BNU & $\Delta x = 1.41$ fm & standard & overlap of wave packets & 1 \\
    IQMD-IMP & $\Delta x = 1.41$ fm & standard & overlap of wave packets & 0.2 \\
    IQMD & $\Delta x = 1.41$ fm & standard & \small $\Delta r = 3$ fm, $\Delta p = 100$ MeV/c, surface correction & 0.2\\
    JAM & $\Delta x = 1.41$ fm & EB fields with $r_c=2$ fm & overlap of wave packets & 0\\
    TuQMD & $\Delta x = 1.41$ fm & standard & \small $\Delta r = 3$ fm, $\Delta p = 100$ MeV/c, surface correction & 0.2 \\
    \hline
    ImQMD-L & $\Delta x = 1.41$ fm & standard & overlap of wave packets & 0.2 \\
    \hline
    \end{tabular}
  \label{T2}
\end{table}

Table~\ref{T2} summarizes the model specifics on the TP shape/size, the Coulomb potential, the Pauli-blocking factor, and the time step in the 4 BUU models and 6 QMD models used in the present study. This information is given here for completeness, and it does not imply that all the specific settings are important in the comparison. This has partly been checked in the box comparisons, and we summarize in the following some of these findings. Most codes have checked that the results are sufficiently stable under a shortening of the time steps given in the table. JAM is a time-step free code (see Ref.~\cite{Ono19}). A more realistic but time-consuming comparison could be done in the future by extrapolating the results to $\Delta t \rightarrow 0$ as in Ref.~\cite{Ono19}. For the TP size, which enters in the mean-field calculation, while it is taken to be zero, i.e., point TPs, in RVUU, the parameters $n=2$ and $l=1$ fm in the shape function [Eq.~(\ref{shape})] are used in the lattice Hamiltonian formulations of IBUU and SMF, and $l=0.92$ fm is used in pBUU with a modified shape function in the lattice Hamiltonian framework (see text below Eq.~(\ref{gfunc})). In Ref.~\cite{Mar21} it was seen that for BUU calculations by a proper choice of TP size and number the results in the mean-field propagation are stable. In the QMD models participating in the present study, the same default widths of $\Delta x = 1.41$ fm are used in the Gaussian wave packets. This is a reasonable choice in QMD codes, which was found in the past to initialize well the surface of the initial nuclei and give a reasonable amount of fluctuation to reproduce fragmentation. Results from simulations using different TP sizes will be compared. For the calculation of the Coulomb potential, the number of mixed events is 1 and the default cut-off distance in Eq.~(\ref{coubuu}) is 1 fm in IBUU, and the corresponding values in RVUU are 10 and 0.1 fm. Results from IBUU using an improved calculation based on the lattice Hamiltonian framework as in Eq.~(\ref{coulat}) (denoted as ``cou-LH'') will also be compared. In pBUU, the Poisson equation is solved to obtain the electrostatic field with a consistent boundary condition from an iterative relaxation method. For the QMD models in the present study, the standard treatment is used in calculating the Coulomb potential as in Eq.~(\ref{couqmd}) except for JAM, which calculates the electromagnetic field with a cut-off distance. We will find, as discussed in Sec.~\ref{discussion}, that the particular method of incorporating the Coulomb potential influences the pion observables to some extent, but does not contribute significantly to the differences among the codes. For the Pauli blocking probability, IBUU uses cubic phase-space cells with $\Delta r = 2$ fm and $\Delta p = 100$ MeV/c, RVUU uses spherical phase-space cells with $\Delta r = 1$ fm and $\Delta p = 150$ MeV/c, SMF uses spherical phase-space cells with $\Delta r = 2.53$ fm in coordinate space, and pBUU fits the phase-space distributions around the final state of the scattered particle by summing two Fermi-Dirac distributions. Typically, IBUU and RVUU use, respectively, 100 and 1000 TPs for evaluating the occupation probability with an interpretation method to smooth the phase-space distributions among neighboring cells, SMF uses 100 TPs for evaluating the occupation probability with a width of $\sigma_p = 29$ MeV/c for the Gaussian weight in the momentum space, and pBUU fits the phase-space distribution with 1000 TPs. For QMD models, IQMD-BNU, IQMD-IMP, and JAM calculate the occupation probability from the overlap of wave packets [Eq.~(\ref{pb1qmd})], while IQMD and TuQMD calculate the occupation probability from the overlap of the hard spheres [Eq.~(\ref{pb2qmd})] with $\Delta r = 3$ fm and $\Delta p = 100$ MeV/c. It was seen in Ref.~\cite{Zha18} that the particular parameters used for the Pauli blocking for BUU do not lead to significant differences. However, the method used in pBUU increases the effectiveness of Pauli blocking. For QMD the width of the wave packet influences the fluctuations and thus also the Pauli blocking~\cite{Zha18}. As seen in Ref.~\cite{Mar21} it also affects the forces in the mean-field propagation. In IQMD and TuQMD, surface corrections for Pauli blocking are introduced for bound nucleons and thus important in the presence of the mean-field potential. Such a correction was already introduced early in the development of transport models~\cite{Aic91}, but the way to determine surface nucleons and the calculation method for the occupation probabilities of these nucleons can be sources of model dependence. Most codes employ the geometric method for the attempted collisions using a maximum distance determined from the total cross section, except SMF and pBUU which employ a stochastic method using the collision rate at the local density (see corresponding discussions in Ref.~\cite{Zha18} for details). SMF is mainly developed for low-energy nuclear reactions where non-relativistic kinematics is used, and in the present study it has been modified to relativistic kinematics for the propagation part, though the NN elastic collision part is still treated non-relativistically. Moreover, the parallel ensemble treatment of NN collisions in SMF may lead to additional differences compared with the full ensemble treatment in pBUU, to be shown later. Settings of the Coulomb interaction and the Pauli blocking in SMF and ImQMD-L are also listed in the table, though they may not be actively turned on in the present comparison, and these two codes have provided baseline calculations for nucleon dynamics only. 

\section{Homework description}
\label{homework}

\begin{figure}[ht]
\includegraphics[scale=0.3]{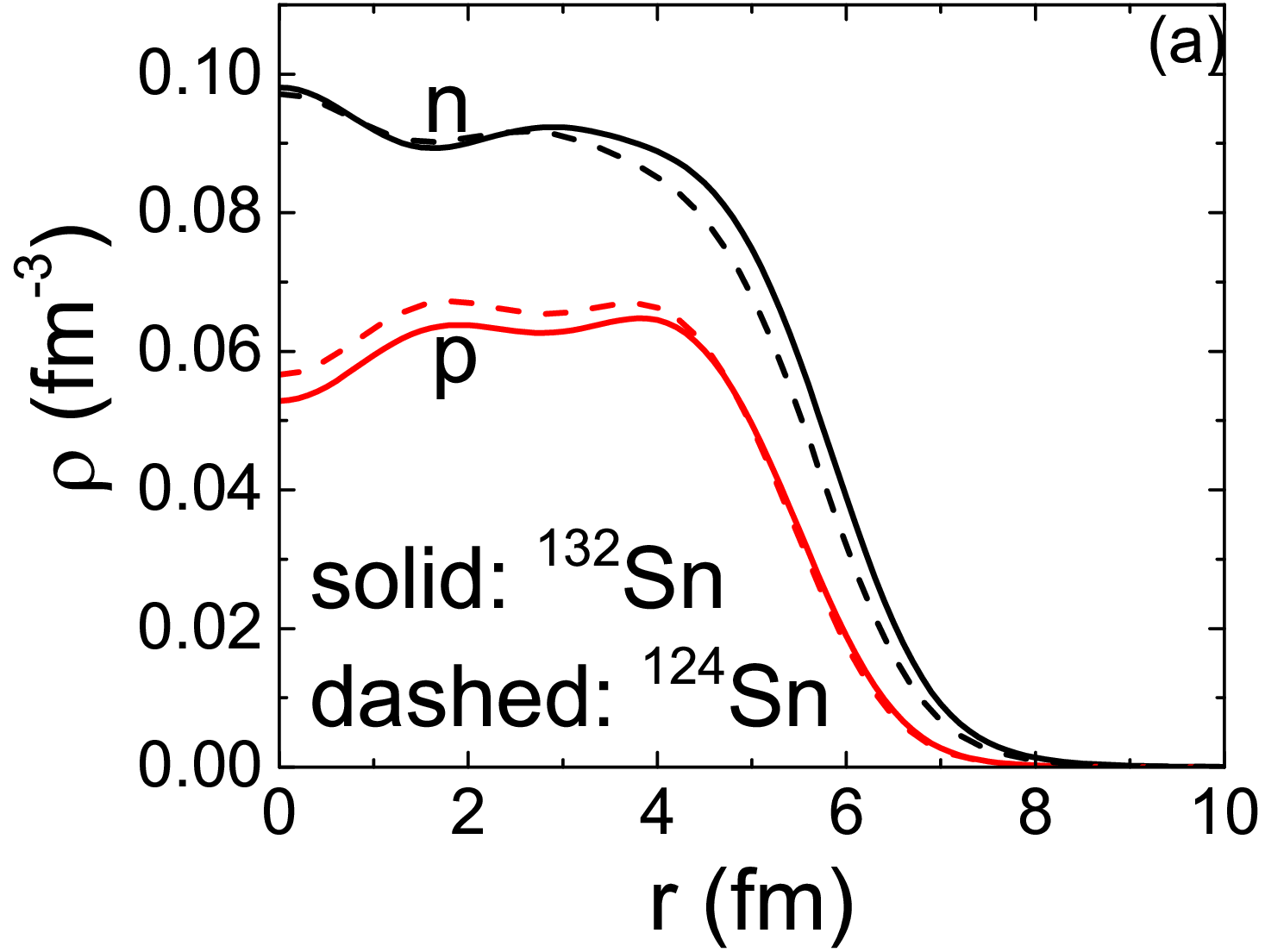}
\includegraphics[scale=0.3]{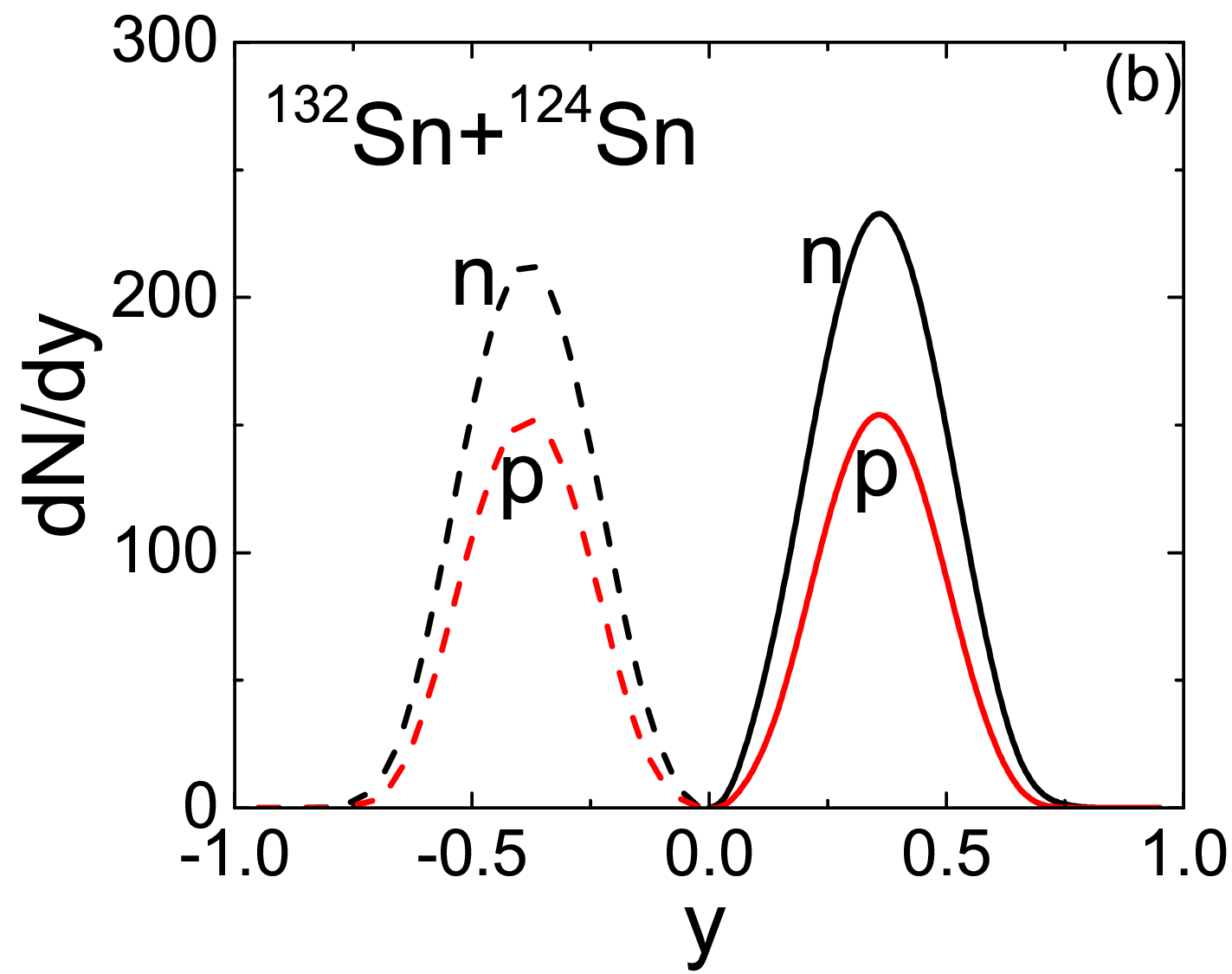}
\caption{(Color online) Left: Distributions of centroid coordinates for neutrons and protons in the radial direction of $^{132}$Sn and $^{124}$Sn; Right: Distributions of centroid rapidities for neutrons and protons in the initial stage of a $^{132}$Sn+$^{124}$Sn collision.
} \label{ini}
\end{figure}

In the present homework calculation, we incorporate inelastic collisions related to the production of pion-like particles, as in Ref.~\cite{Ono19}, into heavy-ion simulations with a similar setup of nucleon dynamics as in Ref.~\cite{Xu16} but for the reactions in the S$\pi$RIT experiment. In particular, we consider the reaction of $^{132}$Sn+$^{124}$Sn at the incident energy $270A$ MeV and an impact parameter $\text{b}=4$ fm. Results for a less neutron-rich reaction of $^{112}$Sn+$^{108}$Sn at the same incident energy and impact parameter are also considered to evaluate the double ratio of observables of interest. To avoid the difference in final results due to the use of different initial conditions adopted by individual codes as found in Ref.~\cite{Xu16}, results shown in the present study are obtained from a common initialization for all participant codes. Specifically, the coordinates of initial nucleons in the colliding nuclei are sampled according to the density profile generated by the Skyrme-Hartree-Fock (SHF) calculation~\cite{Vau72} using the MSL0 force~\cite{MSL0}. The distance between the centers of the projectile and the target in the beam direction is taken to be 16 fm. The momenta of initial nucleons are sampled isotropically in the local Fermi sphere, with the Fermi momentum depending on the local density. The distributions are then boosted according to the incident energy. The distributions of the centroid coordinates in the nucleus rest frame and the centroid rapidities after Lorentz boost for neutrons and protons are displayed in Fig.~\ref{ini} for reference. Note that the physical density distribution depends further on the size of the TPs in BUU models or on the width of the Gaussian wave packet in QMD models. Collision information and particle phase-space information are provided from simulations by each individual code in the different calculation modes, in order to compare separately effects of different ingredients on the final results. These calculation modes are simulations with particle collisions only (Cascade) and full calculations with both particle collisions and mean-field potentials (Full), with and without the Coulomb potential and/or Pauli blocking. Table~\ref{T3} summarizes the different simulation modes considered in this study. We note that using a common initialization eliminates uncertainties from initializations in the Cascade modes. This is not strictly true in the Full-mode calculations, because the density distributions of initial nuclei are calculated with a different force compared to the one used in the transport calculation, and in different models the nuclei may evolve differently before they touch each other, due to different implementations of the mean-field potential, as we will see later. Detailed comparison will be based on the collision information and the phase-space information of final pions from 100000 events, and the phase-space information of nucleons from 1000 events from BUU models using the TP method and QMD models using 1000 separate events.

\begin{table}[h]\small
  \centering
  \caption{Different simulation modes in the present study.
}
    \begin{tabular}{|c|c|c|c|c|c|c|}
    \hline
        & Coulomb potential &  Pauli blocking & Mean-field potential\\
    \hline
     Cascade-nopb & off & off & off \\
     Cascade-nopb-cou & on & off & off \\
     Cascade & off & on & off \\
     Cascade-cou & on & on & off \\
    \hline
     Full-nopb & off & off & on \\
     Full-nopb-cou & on & off & on \\
     Full & off & on & on \\
     Full-cou & on & on & on \\
    \hline
    \end{tabular}
  \label{T3}
\end{table}

The simulation of the Sn+Sn reaction is carried out until $t=70$ fm/$c$. The masses of nucleons and pions are fixed as $m_N^{}=938$ MeV and $m_\pi^{}=139$ MeV, respectively. Compared to the study in Ref.~\cite{Xu16}, additional details are specified in this homework calculation, e.g., allowing the collisions between nucleons within the same nucleus, and turning off all artificial thresholds or cuts on the C.M. energy and distance in the treatment of scatterings or decays. Also, the code correspondents in the present study have tried to incorporate what we have learnt in Refs.~\cite{Zha18} and \cite{Ono19} by removing the spurious collisions and randomizing the NN collision order list in each time step, which was seen to be a source of differences in not only the collision rate but also the pion yield. A constant and isotropic elastic scattering cross section of $\sigma=40$ mb for collisions between baryons has been specified. Production channels for particles other than nucleons, $\Delta$ resonances, and pions, as well as processes that involve direct production of pions, i.e., $N+N \leftrightarrow N+N+\pi$, have been turned off. The energy-dependent isospin-averaged isotropic cross section for $N+N \rightarrow N+\Delta$ is taken from Ref.~\cite{Ber88}, which has the form
\begin{equation}
 \sigma_{NN\rightarrow N\Delta}
  =\begin{cases}
    \frac{20(\sqrt{s}-2.015)^2}{0.015+(\sqrt{s}-2.015)^2}, & \sqrt{s} \geq 2.015;\\
    0, & \sqrt{s}<2.015,
  \end{cases}
\end{equation}
with $\sigma$ in mb and the C.M. energy $\sqrt{s}$ in GeV. The mass distribution of $\Delta$ resonances is taken to have a Breit-Wigner form
\begin{equation}\label{BW}
A(m_\Delta^{}) = \frac{1}{\tilde{A}} \frac{4 {m_\Delta^0}^2 \Gamma(m_\Delta^{})}{(m_\Delta^2-{m_\Delta^0}^2)^2+{m_\Delta^0}^2 \Gamma^2(m_\Delta^{})},
\end{equation}
where $m_\Delta^0=1.232$ GeV is the pole mass, $\tilde{A} \approx 0.95$ is the normalization factor chosen to satisfy $\int_{m_N+m_\pi}^\infty A(m_\Delta^{}) \frac{dm_\Delta}{2\pi} = 1$. The mass-dependent width function $\Gamma(m_\Delta^{})$ has the form
\begin{equation}\label{gwidth}
\Gamma(m_\Delta^{}) = \frac{0.47 q^3}{m_\pi^2 + 0.6q^2},
\end{equation}
with $q = \frac{\sqrt{(m_\Delta^2-m_N^2 - m_\pi^2)^2-4m_N^2m_\pi^2}}{2m_\Delta}$, and it vanishes at $m_\Delta^{}=m_N^{}+m_\pi^{}$. When a $\Delta$ resonance is produced based on the above cross section, the mass of the resonance is sampled within the range of $m_N^{}+m_\pi^{} < m_\Delta^{} < \sqrt{s} - m_N^{}$ according to the probability
\begin{equation}\label{pd}
P(m_\Delta^{}) = \frac{p_\Delta^{} m_\Delta^{} A(m_\Delta^{})}{\int_{m_N^{}+m_\pi^{}}^{\sqrt{s}-m_N^{}} p_\Delta^{} m_\Delta^{} A(m_\Delta^{}) \frac{dm_\Delta^{}}{2\pi}},
\end{equation}
where $p_\Delta^{} = \sqrt{\frac{(s+m_\Delta^2-m_N^2)^2}{4s}-m_\Delta^2}$ is the $\Delta$ momentum in the C.M. frame of the collision. For the cross section of the inverse reaction, it can be obtained from the detailed balance condition as~\cite{pBUU1}
\begin{equation}
\sigma_{N\Delta \rightarrow NN}^{} = \frac{1}{g} \frac{m_\Delta^{} p_f^2}{p_i(m_\Delta^{})}\sigma_{NN\rightarrow N\Delta}^{} / \int_{m_N^{}+m_\pi^{}}^{\sqrt{s}-m_N^{}} m A(m) p_i(m) \frac{dm}{2\pi},
\end{equation}
with $p_f=\sqrt{\frac{s}{4} -m_N^2}$ and $p_i(m)=\sqrt{ \frac{(s+m^2-m_N^2)^2}{4s} -m^2 }$ being respectively the final and initial momentum in the C.M. frame of the collision, and $g$ being the spin-isospin factor.

In the $\Delta$ decay process $\Delta \rightarrow N+\pi$, the same mass-dependent width given in Eq.~(\ref{gwidth}) is used, and the momenta of the nucleon and the pion are sampled isotropically in the C.M. frame of the $\Delta$ resonance. The isospin-averaged cross section for the inverse reaction, i.e., $\pi+N\rightarrow \Delta$, is obtained from the detailed balance condition, i.e.,
\begin{equation}
\sigma_{N\pi \rightarrow \Delta} = \frac{4\pi\hbar^2}{3 p_{cm}^2} \Gamma(\sqrt{s}) A(\sqrt{s}),
\end{equation}
with $p_{cm}$ being the pion or nucleon momentum in their C.M. frame, $\sqrt{s}$ being the nucleon-pion C.M. energy, and $A(\sqrt{s})$ and $\Gamma(\sqrt{s})$ being, respectively, the Breit-Wigner distribution [Eq.~(\ref{BW})] and the width of $\Delta$ [Eq.~(\ref{gwidth})] with mass equal to the C.M. energy. All channels mentioned above are isospin-dependent and satisfy the detailed balance condition. For the isospin-dependent $\Delta$ resonance production cross sections, they are:
\begin{eqnarray}
&&\sigma_{p+p \leftrightarrow \Delta^{++} + n}=\sigma_{n + n \leftrightarrow \Delta^- + p}=\sigma_{NN\leftrightarrow N\Delta};\notag\\
&&\sigma_{p+p \leftrightarrow \Delta^{+} + p}=\sigma_{n + n \leftrightarrow \Delta^0 + n}=\sigma_{NN\leftrightarrow N\Delta}/3;\notag\\
&&\sigma_{n+p \leftrightarrow \Delta^{+} + n}=\sigma_{n + p \leftrightarrow \Delta^0 + p}=\sigma_{NN\leftrightarrow N\Delta}/3. \nonumber
\end{eqnarray}
Similarly, the isospin-dependent $\Delta$ decay widths and the $\Delta$ production cross sections from pion-nucleon scatterings are:
\begin{eqnarray}
&&\Gamma_{\Delta^{++} \rightarrow p + \pi^+}=\Gamma_{\Delta^{-} \rightarrow n + \pi^-}=\Gamma; ~\sigma_{p+\pi^+\to\Delta^{++}}=\sigma_{n+\pi^-\to\Delta^-}=3\sigma_{N\pi \rightarrow \Delta}/2; \nonumber\\
&&\Gamma_{\Delta^{+} \rightarrow p + \pi^0}=\Gamma_{\Delta^{0} \rightarrow n + \pi^0}=2\Gamma/3; ~\sigma_{p+\pi^0\to\Delta^{+}}=\sigma_{n+\pi^0\to\Delta^0}=\sigma_{N\pi \rightarrow \Delta}; \nonumber\\
&&\Gamma_{\Delta^{+} \rightarrow n + \pi^+}=\Gamma_{\Delta^{0} \rightarrow p + \pi^-}=\Gamma/3; ~\sigma_{n+\pi^+\to\Delta^{+}}=\sigma_{p+\pi^-\to\Delta^0}=\sigma_{N\pi \rightarrow \Delta}/2. \nonumber
\end{eqnarray}
In the present comparison study, the time step $\Delta t$ in the simulation, which may affect the production of pion-like particles, as discussed in detail in Ref.~\cite{Ono19}, is given in Table~\ref{T2} for each code. Except for SMF, IQMD-BNU, and JAM, all codes use $\Delta t=0.2$ fm/c, as recommended in the homework description. SMF checked the convergence already at $\Delta t=0.5$ fm/c.

In the Full-mode calculations, a momentum-independent mean-field potential\footnote{Although RVUU uses by default a covariant mean-field potential, it adopts for the present study the prescribed non-relativistic mean-field potential.} is included for nucleons, similar to that in Ref.~\cite{Xu16}, i.e.,
\begin{equation}\label{mf}
U_{n/p} = \alpha \left(\frac{\rho}{\rho_0}\right) + \beta \left(\frac{\rho}{\rho_0}\right)^\gamma \pm 2 S_{pot} \left(\frac{\rho}{\rho_0}\right)^{1.1}\delta.
\end{equation}
In the above, the `$+(-)$' sign is for neutrons (protons), $\rho=\rho_n+\rho_p$ is the nucleon number density with $\rho_{n(p)}$ being the neutron (proton) number density, and $\delta = (\rho_n -\rho_p)/\rho$ is the isospin asymmetry. The parameters are chosen to be $\alpha = -209.2$ MeV, $\beta = 156.4$ MeV, and $\gamma = 1.35$, and the potential part of the symmetry energy at saturation density is set to be $S_{pot}=18$ MeV. These parameters lead to the saturation density $\rho_0 = 0.16$ fm$^{-3}$, the binding energy at saturation density $E_0=-16$ MeV, the incompressibility of symmetric nuclear matter $K_0=240$ MeV, the symmetry energy at saturation density $E_{sym}^0 = 30.3$ MeV, and its slope parameter at saturation density $L=84$ MeV.

We used some simplifications in the treatment of the $\Delta$ and $\pi$ propagation and collisions to avoid complications in comparing the codes, which, however, will be important in realistic descriptions of experiments. Thus, pions and $\Delta$ resonances propagate as free particles which are not affected by nuclear mean-field potentials but only by the Coulomb force if included. This could not be easily implemented in the pBUU code, which is constructed in such a way that it necessarily uses both isoscalar and isovector potentials for $\Delta$ resonances and the s-wave isovector potential for pions. Therefore, the pBUU results for different pion charge states are not comparable to those from the other codes\footnote{We still keep this code in this comparison since the results of this well-established and extensively used code for the nucleonic observables are of interest, where this difference is unimportant. We also show the results of this code for the pionic observables since we think that it is of interest how these observables are affected by the potentials of pion-like particles. However, in the final evaluation of the differences and convergence of the codes in Sec.~\ref{discussion}, these results are not included.}. In principle, threshold shifts may occur due to different mean-field potentials of initial- and final-state particles in inelastic collisions, especially in the presence of an isospin- and/or momentum-dependent mean-field potential, and here in addition due to the neglect of potentials for $\Delta$ resonances and pions. In the present study, threshold effects are required to be neglected by all participating codes, which results in energy non-conservation in inelastic collisions. We note, however, that the overall dynamics of the studied reaction is not much affected by the neglect of this effect, since very few pion-like particles are produced at the collision energy considered here. A more rigorous investigation on the energy conservation in the production of pion-like particles with momentum-dependent mean-field potentials is in progress~\cite{Dan22}.

\section{Nucleon observables}
\label{nucleon}

We begin by showing results for nucleons to understand the overall heavy-ion collision dynamics, which is expected to affect the production of pion-like particles. Particularly, we revisit the results on density evolution, stopping, and transverse flow, as in Ref.~\cite{Xu16}. The Coulomb interaction is turned off in this section, so that we can have a clear view on the effects of NN collisions and the strong mean-field force.

\subsection{Density evolution}

\begin{figure}[ht]
\includegraphics[scale=0.5]{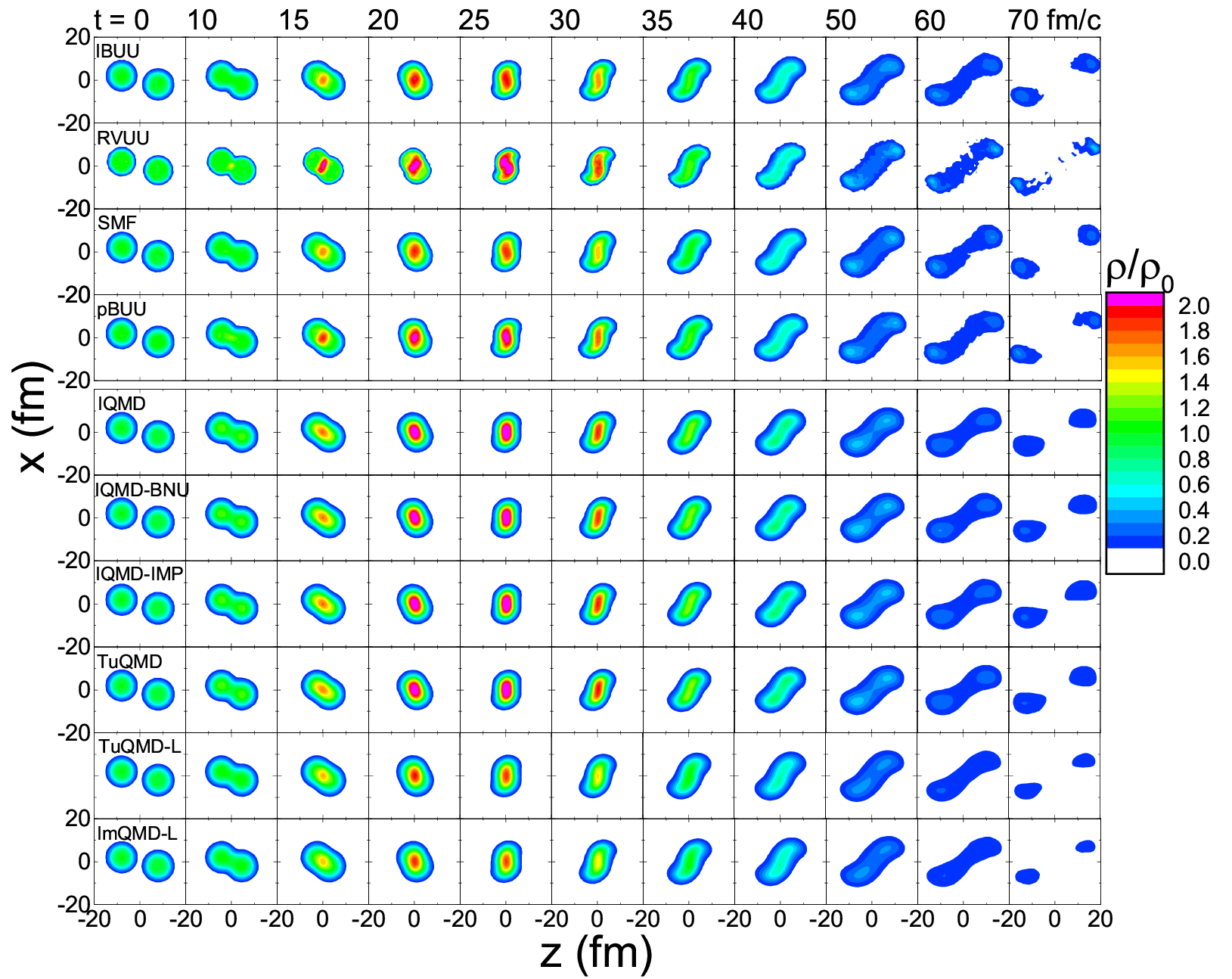}
\caption{(Color online) Contours of reduced densities $\rho/\rho_0$ in the x-0-z plane at different indicated times in the Full-nopb mode.
} \label{dencon}
\end{figure}

To have a global picture of the nucleon dynamics in heavy-ion collisions, we first compare the density contours in the x-0-z plane at different times from BUU and QMD models for the typical Full-nopb mode. Results from averaging the density evolutions of 1000 events are displayed in Fig.~\ref{dencon}, with the method of calculating the physical density as used in each individual code, i.e., using the shape functions of the (test) particles in that code. As mentioned above, at $t=0$ fm/$c$, the same centroid coordinates and momenta for initial nucleons or TPs are used in all codes. For BUU models, it is seen that IBUU, SMF, and pBUU give an initially more diffuse physical density distribution than RVUU, which is due to the larger TP size used in the former cases. 
QMD models using the same width parameter $\Delta x$ for the wave packets give the same initial density distribution and also have similar density evolutions. At around $t=20$ fm/$c$ QMD models overall lead to higher densities than BUU models, except for ImQMD-L and TuQMD-L which are more similar to BUU models due to the more exact evaluation of the non-linear term in the mean-field potential. As mentioned in Ref.~\cite{Xu16}, one should also keep in mind that in BUU models 100 or more parallel events give the same smooth density evolution, while in QMD models the density evolution fluctuates strongly on an event-by-event basis.

\begin{figure}[ht]
\includegraphics[scale=0.3]{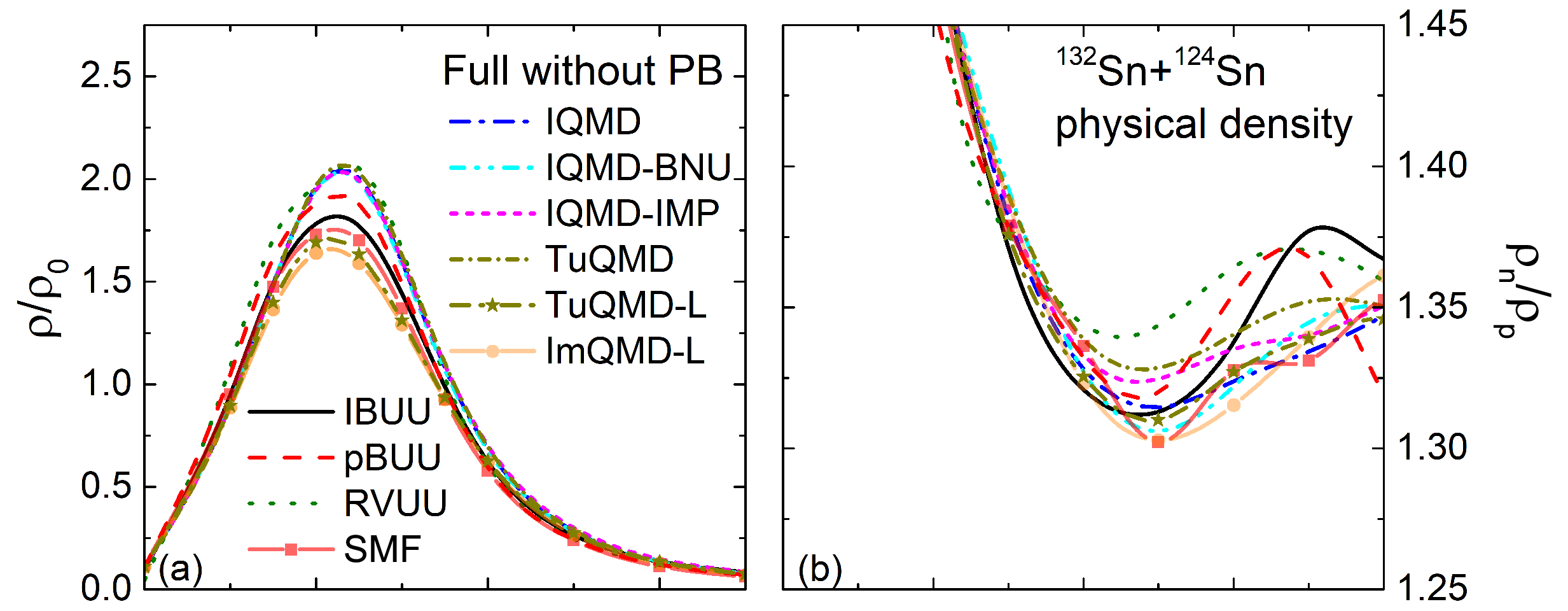}\\
\includegraphics[scale=0.3]{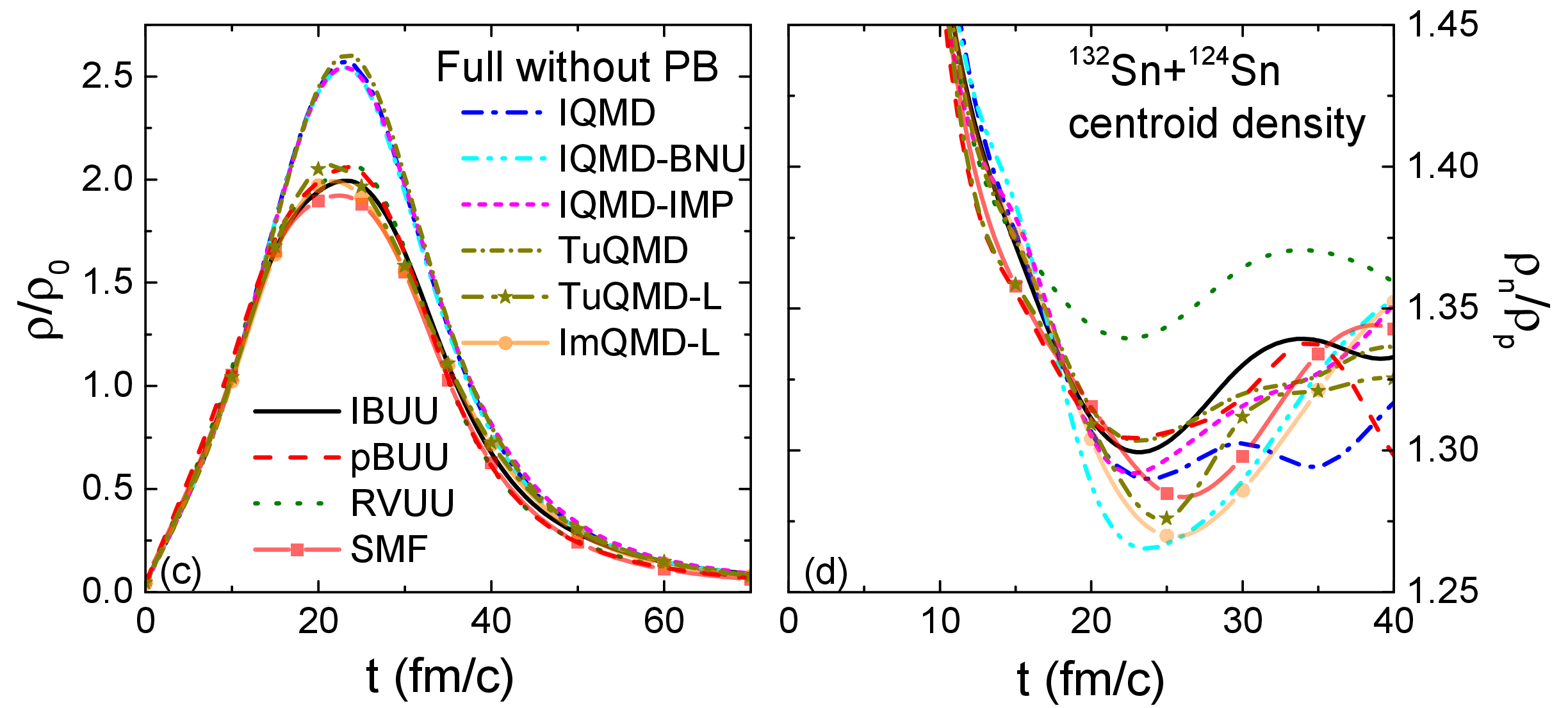}
\caption{(Color online) Time evolution of reduced densities $\rho/\rho_0$ (left) and asymmetries $\rho_n/\rho_p$ (right) in the central region of a volume $5\times5\times5$ fm$^3$ in the Full-nopb mode. Upper panels are for physical densities calculated from finite-size (test) particles, while lower panels are for centroid densities calculated from point (test) particles. Asymmetries are only plotted up to 40 fm/c, since beyond this they fluctuate strongly.
} \label{den}
\end{figure}

To have a more quantitative view of the evolution of the collision, we show in Fig.~\ref{den} the average reduced density $\rho/\rho_0$ (left column) and the neutron-proton density ratio $\rho_n/\rho_p$ (right column) in the central region ($5\times 5 \times 5$ fm$^3$ volume) of $^{132}$Sn+$^{124}$Sn collisions as a function of time from both BUU and QMD models. We show the time evolutions of physical densities in the upper panels of Fig.~\ref{den}, and the time evolutions of centroid densities, i.e., densities of centroid coordinates of (test) particles, in the lower panels of Fig.~\ref{den}. The centroid coordinates are propagated in the simulations and thus determine the evolution of the system, in particular the elastic and inelastic collisions and the production of pions. The physical densities are calculated from folding the centroid densities with the shapes/sizes of the (test) particles, and are used to calculate the mean-field potentials. Thus, starting from the same distribution of centroid coordinates, nucleons may evolve differently, mostly due to different calculations of the mean-field potential. From the time evolution of the centroid coordinates (lower left panel) it is seen that the conventional QMD models with the common initialization and with the same width parameter have a very similar evolution. The BUU models also have a similar evolution among themselves, but lead to a lower maximum density compared to conventional QMD models. This is mainly a consequence of the approximate calculation of the non-linear component in the mean-field potential in conventional QMD codes, which leads to weaker repulsive forces and thus to more compression, as already shown in the box-Vlasov study~\cite{Mar21}. This is supported by the observation that the lattice versions of QMD, i.e., ImQMD-L and TuQMD-L, which calculate the spatial average of the non-linear term in the mean-field potential accurately by numerical integration, lead to a similar centroid density evolution as BUU models, but small differences remain because of the different TP shapes/sizes of nucleons and intrinsic difference between BUU and QMD models. BUU-type models also do not agree exactly due to the use of different TP shapes/sizes and different treatments of the collision term, e.g., in SMF. This is seen more clearly in the physical densities in the upper left panel. Generally they are lower than the centroid densities, because of the smearing with the (test) particle shapes. The physical and centroid densities are identical only for RVUU, which uses point TPs, and this code has a higher physical density and thus evolves differently from other BUU codes. 
The conventional QMD codes all have the same physical densities, as expected. The central physical densities rise to a maximum of about 1.7 to 2 times $\rho_0$ at about $t = 25$ fm/c and become very small at about 70 fm/c, when this calculation stops.

The neutron-proton density ratio, or the asymmetry, is shown in the right column of Fig.~\ref{den}. It is very large initially when the neutron-rich surfaces start to overlap in the center of the collision region. It then reaches a minimum of about 1.3 at the time of maximum compression, which is lower than the overall N/Z ratio of 1.56. This is because the symmetry potential is repulsive for neutrons especially at high densities, and expels pre-equilibrium neutrons from the central region, lowering its asymmetry. The asymmetry then rises slightly, but begins to fluctuate strongly due to statistical fluctuations, since the density in the central region becomes very low. Therefore we display the asymmetry for times only up to 40 fm/c. The minimum of the centroid asymmetries in Fig.~\ref{den} (d) is deeper compared to that of the physical asymmetries in Fig.~\ref{den} (b) because of the smearing by finite-size (test) particles, since the asymmetry is lowest in the most compressed region and rises towards the surface. The asymmetries generally behave rather similarly among all codes up to the maximum compression stage. An exception to this general behavior is RVUU, for which the centroid and physical asymmetries are again identical. It gives a substantially higher centroid asymmetry compared to other codes at the maximum compression stage, but comes closer to the other codes for the physical asymmetries. This must be due to a somewhat different evolution with respect to the other BUU codes when using point TPs, as remarked above. The asymmetry differs more strongly among codes than the densities, and does not show a systematic difference between QMD and BUU models. However, one should keep in mind that the vertical scale is very much enlarged, particularly when comparing it to the overall N/Z ratio 1.56. On this scale there are differences, even within QMD models, e.g., IQMD-BNU and ImQMD-L give slightly smaller asymmetries compared to other QMD models. We also note here that the statistical errors are relatively large for 1000 events.

Although the time evolution of the nucleon densities and asymmetries is not directly observable, the differences seen in Fig.~\ref{den} will have consequences for observables. Their effects on nucleon final momentum distributions, like the stopping and flow, will be discussed in the next subsection. As for pion production, a higher central centroid density is expected to enhance $N+N \rightarrow N+\Delta$ collisions and increase the pion multiplicity, and a larger centroid asymmetry in the high-density phase would favor more $n+n \rightarrow p + \Delta^-$ than $p+p \rightarrow n + \Delta^{++}$ reactions and thus increase the $\pi^-/\pi^+$ yield ratio.

\subsection{Stopping and transverse flow}

In this section we discuss the rapidity distribution (stopping) and the transverse flow of nucleons. These two observables represent the measurable asymptotic momentum distributions of nucleons in the z- and x-directions, respectively. The comparison gives the observational consequences of the differences in the evolution of nucleon distributions in space and momentum, which may have some influence on observables related to pions. To disentangle contributions from different transport ingredients, we discuss these observables first without the mean-field potential (Cascade).

\begin{figure}[ht]
\includegraphics[scale=0.3]{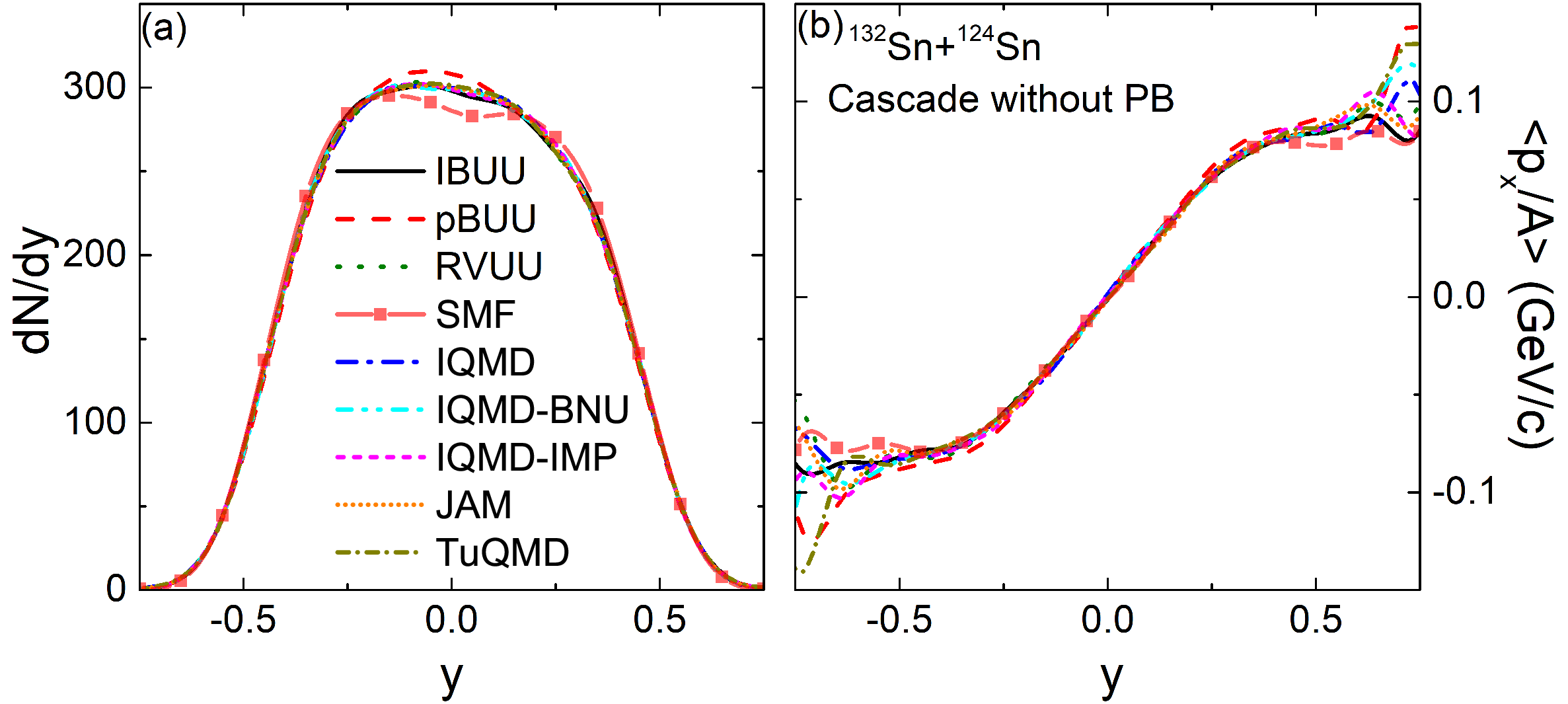}\\
\includegraphics[scale=0.3]{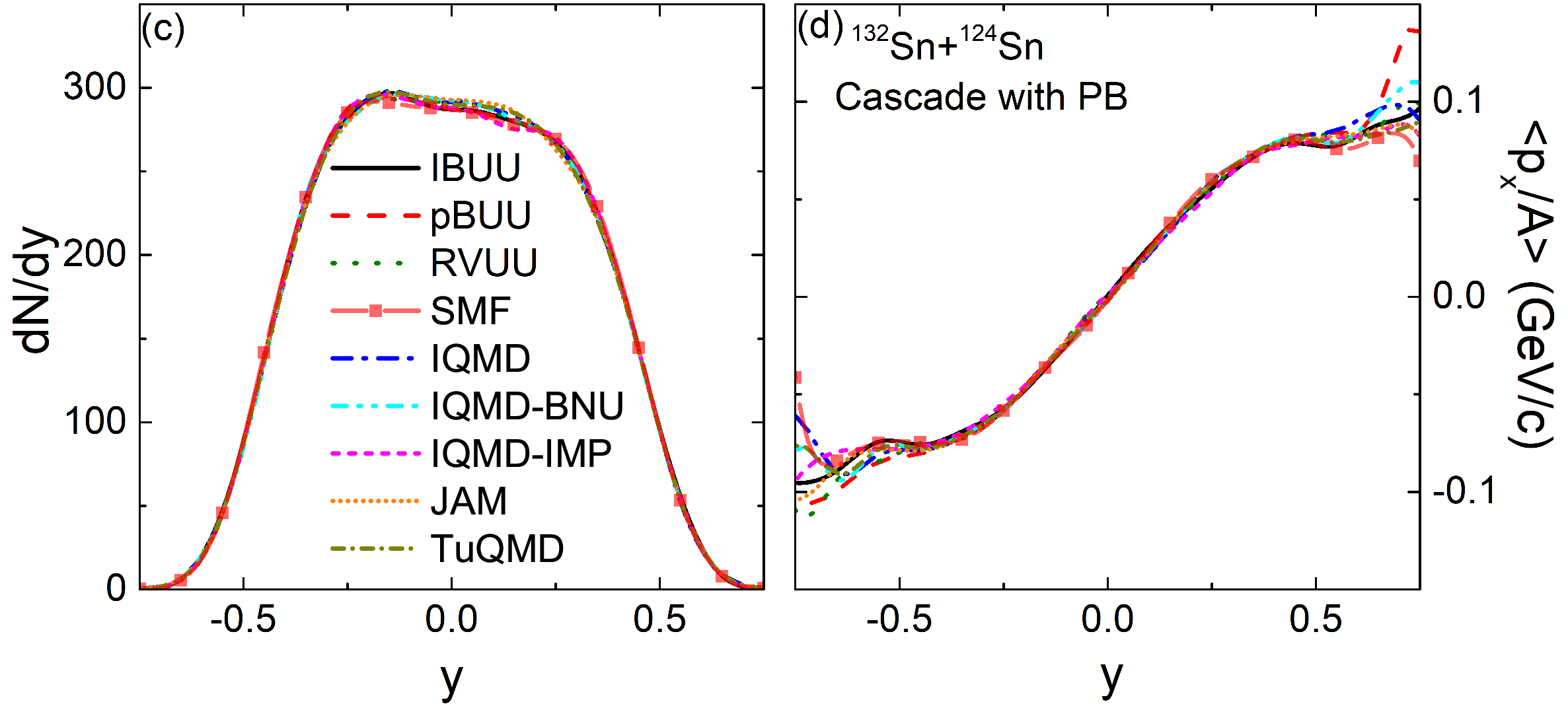}\\
\caption{(Color online) Final nucleon rapidity distribution (left) and transverse flow (right) in the Cascade mode, without Pauli blocking (upper) and with Pauli blocking (lower).
} \label{dNdypx_Cascade}
\end{figure}

Figure~\ref{dNdypx_Cascade} displays the final rapidity distribution of nucleon multiplicities (left) and transverse flows (right) from the Cascade mode without and with Pauli blocking, in the upper and lower panels, respectively. With a common initialization and without the mean-field potential, the stopping and the transverse flow from different codes are very similar. At this collision energy and without the mean-field potential, the Pauli blocking is not very effective, and the results with Pauli blocking only show a slightly weaker stopping and transverse flow, compared to those without Pauli blocking. Although Pauli blocking is implemented differently in different transport models, the resulting differences in nucleon observables are seen to be small in the Cascade mode. pBUU gives a slightly stronger stopping and a larger transverse flow than other codes especially in the case without Pauli blocking. This may be due to the different (stochastic) treatment of the collision probability of the $\Delta$ absorption reaction, for which the parameterized cross section used here is singular at the threshold. SMF shows weaker stopping, as seen in Fig.~\ref{dNdypx_Cascade} (a), since it does not include inelastic channels, while this effect is not very evident in Fig.~\ref{dNdypx_Cascade} (c) since the inelastic processes are partly Pauli blocked.

\begin{figure}[ht]
\includegraphics[scale=0.3]{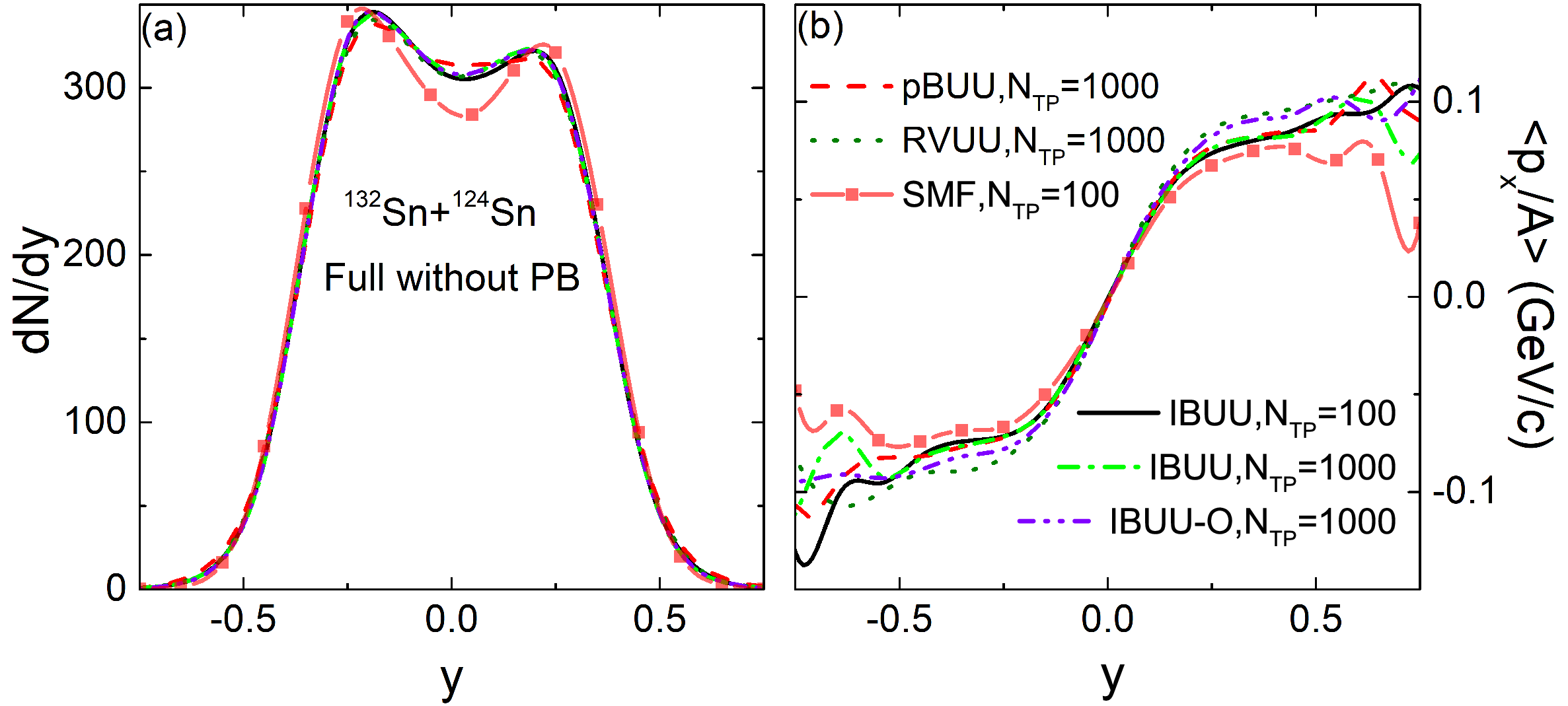}\\
\includegraphics[scale=0.3]{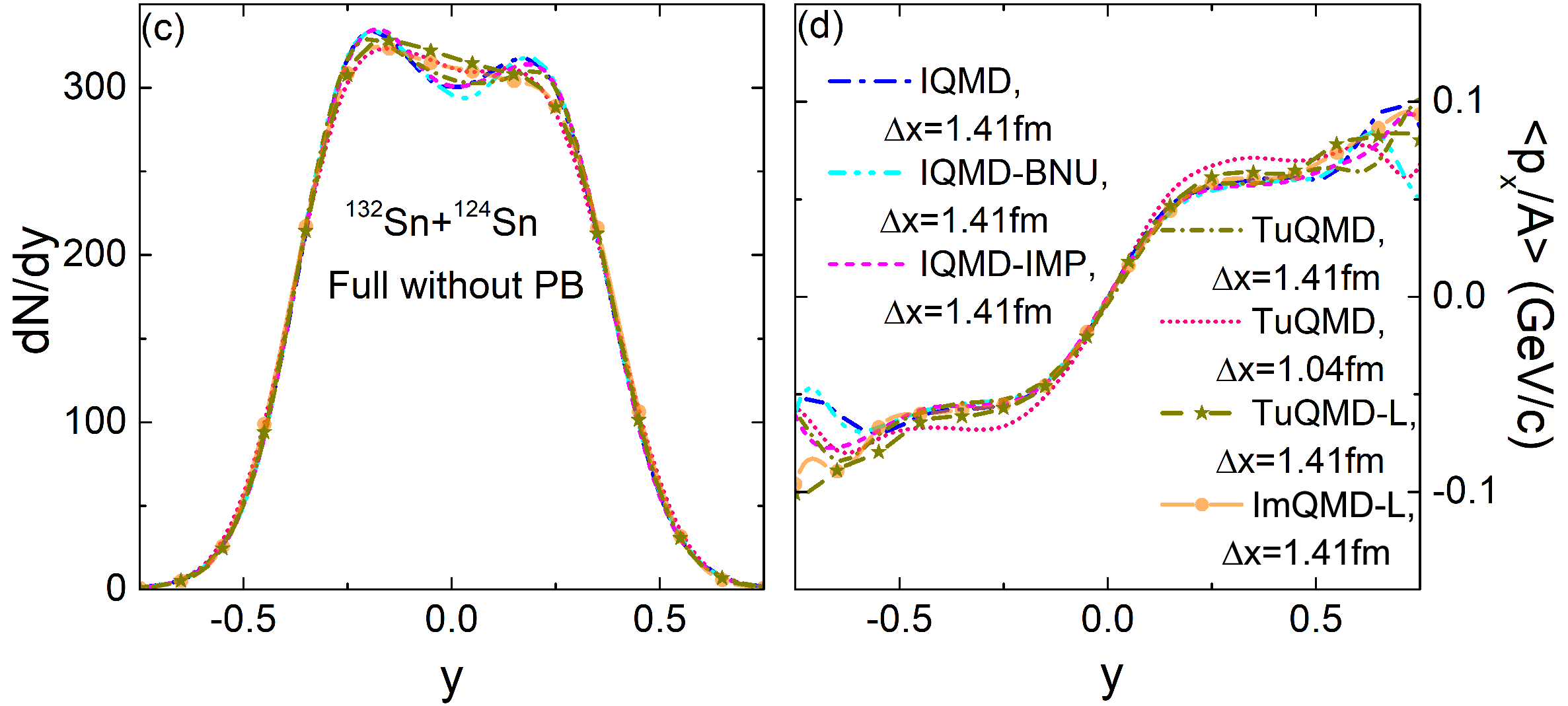}\\
\includegraphics[scale=0.3]{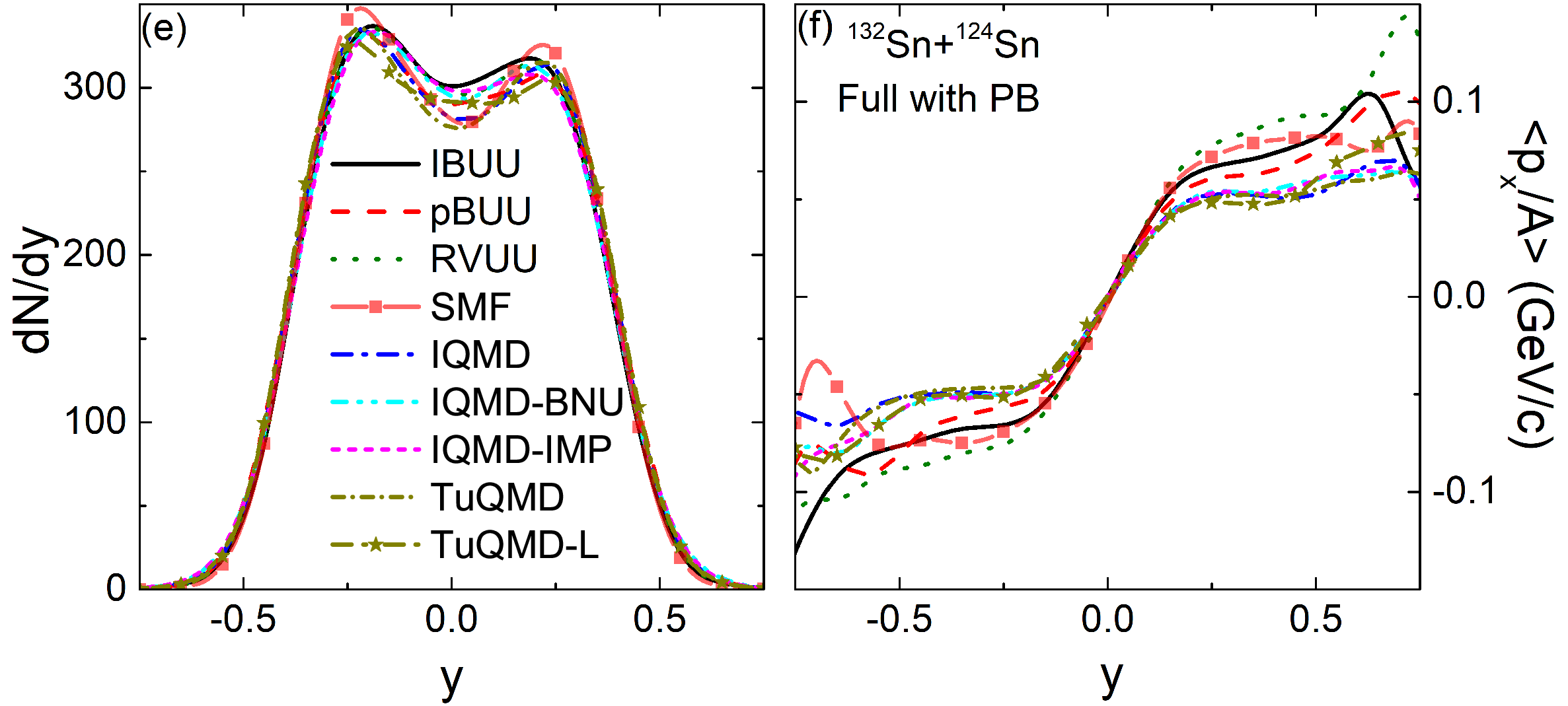}
\caption{(Color online) Final nucleon rapidity and transverse flow distributions as in Fig.~\ref{dNdypx_Cascade} but in the Full mode, in the top and middle rows without Pauli blocking, and in the bottom row with Pauli blocking. The top row shows results for BUU models using different TP shapes and numbers, the middle row shows results for QMD models with different widths of the wave packet, and the bottom row shows results for all codes in their standard implementations and with standard parameters.} \label{dNdypx_Full}
\end{figure}

Similar comparisons as in Fig.~\ref{dNdypx_Cascade} are shown in Fig.~\ref{dNdypx_Full} for different scenarios with the inclusion of nucleon mean-field potentials. The left and right columns again display the rapidity distribution and transverse flow of nucleons, respectively. The upper two rows show results from calculations without Pauli blocking, where the top row gives results for BUU models using different TP shapes and numbers, and the middle row shows results for QMD models using different widths of the wave packet and lattice implementations. The bottom row shows results for all participant transport models in the Full mode with Pauli blocking in their standard implementations and with standard parameters.

In the upper panels of Fig.~\ref{dNdypx_Full}, results from IBUU, pBUU, and SMF using the lattice Hamiltonian framework, and RVUU using point TPs, are compared. Results from IBUU are shown for 3 different calculations, using the lattice version (IBUU) with 100 and 1000 TPs, where the shape of TPs is given by Eq.~(\ref{shape}), and using the original version (IBUU-O) with 1000 TPs, where each nucleon contributes to the density in local and neighboring cells of the volume 1 fm$^3$. IBUU-O thus has effectively a smaller (larger) TP size than IBUU (RVUU). The comparison of the lattice versions shows that 100 TPs in the calculation of the mean-field potential are sufficient for both the stopping and flow. Comparing the top left and right panels, it is seen that the TP shape/size does not affect much the stopping in BUU models, but affects the nucleon transverse flow, making it weaker when the TP size becomes larger (see the decreasing flows from RVUU to IBUU-O and to IBUU with the increasing size of the TPs). This is understandable since a smaller TP size leads to a sharper density distribution near the surface of the compressed matter, as seen in Ref.~\cite{Mar21}, which thus results in a stronger density gradient. For SMF, although the mean-field calculation is expected to be the same as the lattice version of IBUU, it gives a weaker stopping and transverse flow, due to the absence of inelastic channels and the treatment of the collision integral with the mean-free path method in parallel ensemble (see Sec.~\ref{specifics}).

For the results from QMD models in the middle row, the choice of an identical initialization and the same width $\Delta x = 1.41$ fm of the wave packets gives very similar results. A calculation of TuQMD using $\Delta x=1.04$ fm leads to stronger fluctuations and thus more dissipation and consequently a stronger stopping, and to a larger density gradient and thus a stronger transverse flow, similar to the TP size effect in BUU models. The more accurate calculations of the non-linear repulsive term in the lattice versions of QMD, i.e., ImQMD-L and TuQMD-L, lead only to a slight increase in stopping and flow compared to the traditional method, although it has significant effects on the density evolution as shown in Figs.~\ref{dencon} and \ref{den}. Compared to TuQMD-L, a slightly weaker stopping from ImQMD-L may again be due to the absence of inelastic channels.

Results from Full-mode calculations with Pauli blocking for both BUU and QMD models using the standard choices of (test) particle shapes and the calculation method for the mean-field potential (see Table \ref{T2}) are displayed in the bottom panels of Fig.~\ref{dNdypx_Full}. Comparing BUU and QMD models, a substantially stronger flow is seen for BUU. This is consistent with the observation in Fig.~\ref{den} that BUU models lead to a lower density as a result of the stronger repulsion of the accurately calculated non-linear force. This difference is further enhanced by the use of an effectively larger particle size in QMD, which reduces the gradients of the mean-field potential. The differences within BUU models also follow from the same connection that a smaller TP size leads to a stronger force and thus to a stronger stopping and flow. Using the point TP, RVUU has the largest flow and a large stopping. The difference among BUU models is larger than that in Fig.~\ref{dNdypx_Full} (b), indicating the effect from different Pauli blocking prescriptions. Among the BUU models, pBUU has a small flow and weak stopping as a result of its particularly effective Pauli blocking which reduces the number of successful NN collisions, while SMF has a weaker stopping again due to its special treatment of collisions. QMD models in the standard versions agree well among themselves for the transverse flow and a little less well for the stopping. Here, differences due to different strategies in treating the Pauli blocking are seen, in particular, the use of a surface correction in the Pauli blocking by TuQMD and IQMD. Compared to TuQMD, a stronger stopping and a similar transverse flow are seen for TuQMD-L.

Summarizing this section, we compared the properties of nucleonic matter in $^{132}$Sn+$^{124}$Sn collisions at $270A$ MeV to check to what extent the codes describe a common evolution of the reaction as a whole. The finding from Fig.~\ref{den} for non-observables, i.e., the density and asymmetry evolution, and from Fig.~\ref{dNdypx_Full} for observables, i.e., the stopping and flow, is that the evolution is not identical in spite of a common initialization and common physics setups. We are able to trace the origins of essentially all these discrepancies. First, there is a systematic difference between BUU and QMD models, which is due to the approximate evaluation of the non-linear potential term in the conventional QMD strategy. Thus, these QMD models effectively use a different, and actually, a weaker repulsive term of the interaction than BUU models. We see that this can be avoided by using a more elaborate lattice formulation for QMD. The larger fluctuations in QMD models, previously found from the box studies~\cite{Zha18}, are expected to lead to weaker Pauli blockings relative to BUU models. Differences in the Pauli blocking effect are seen within BUU models depending on the strategy used in its implementation (see, e.g., pBUU), and also within QMD models depending on, e.g., whether a surface correction is implemented (see IQMD and TuQMD). Thus, the differences in the evolution of a heavy-on collision can, in principle, be understood. They could be, in principle, better controlled, but, as discussed in the introduction, constructing a unified code is not the aim here. The basic difference between BUU and QMD codes in treating fluctuations does not seem to affect the results in the present reaction by much. Since pion production in heavy-ion collisions is sensitive to the evolution of the nucleonic matter through the reached densities and asymmetries, and consequently to the rate of isospin-dependent NN collisions, we have to expect also differences in pion observables. This will be seen in the next section.

\section{Pion production} \label{results}

With the nucleon dynamics in Sn+Sn collisions at the incident energy $270A$ MeV investigated in the previous section, in this section we compare in detail the results on pion production from different codes and try to understand the differences. Specifically, we compare the rate of inelastic collisions related to pion production, the pion multiplicity, and the $\pi^-/\pi^+$ yield ratio. We focus on the model dependence caused by different implementations of the Pauli blocking and the Coulomb potential. Our discussions are mostly on the heavier $^{132}$Sn+$^{124}$Sn system, and we only use the results from $^{112}$Sn+$^{108}$Sn system to study the relevant double ratios from two different collision systems.

\subsection{Collision rate and Pauli blocking}

\begin{figure}[h]
\includegraphics[scale=0.25]{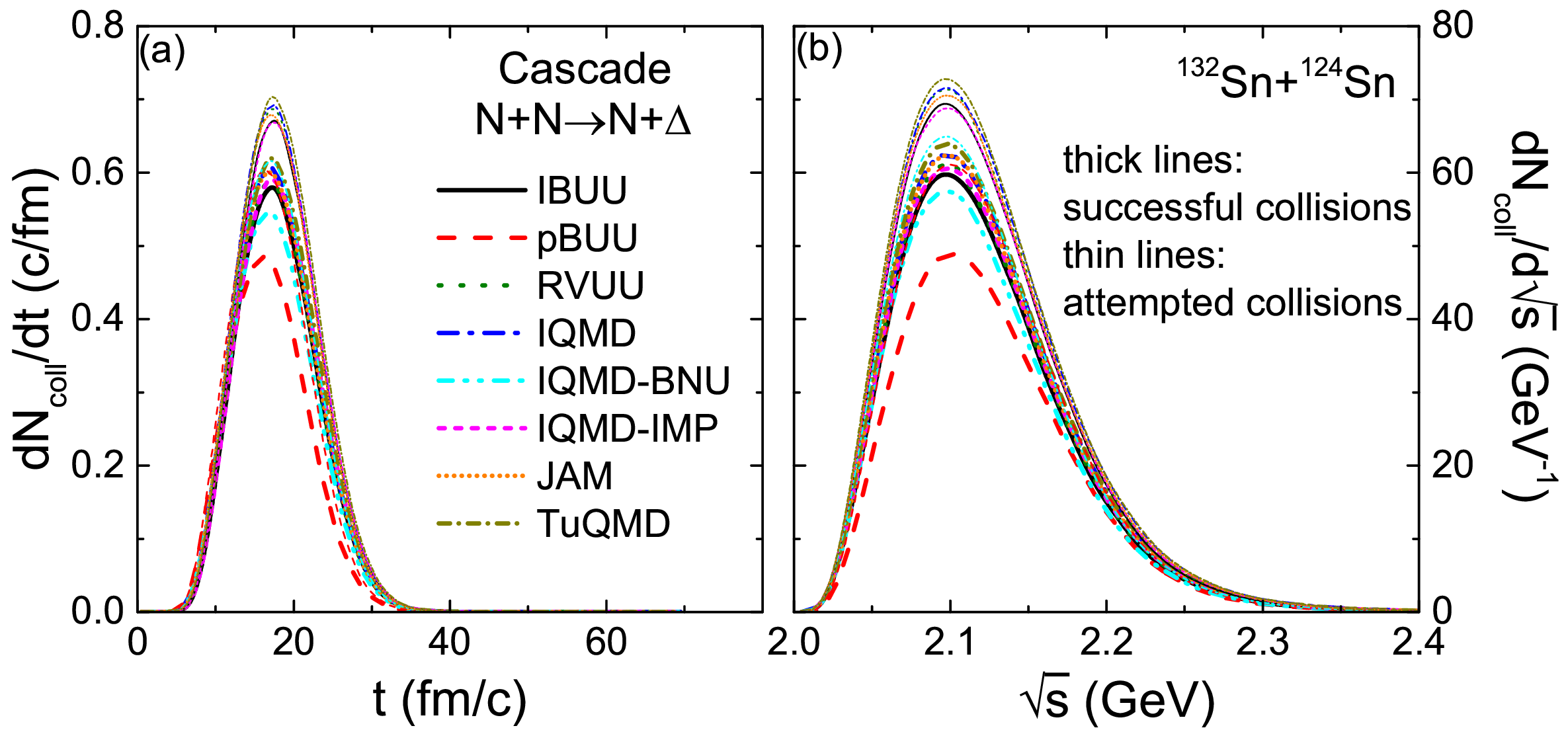}\\
\includegraphics[scale=0.25]{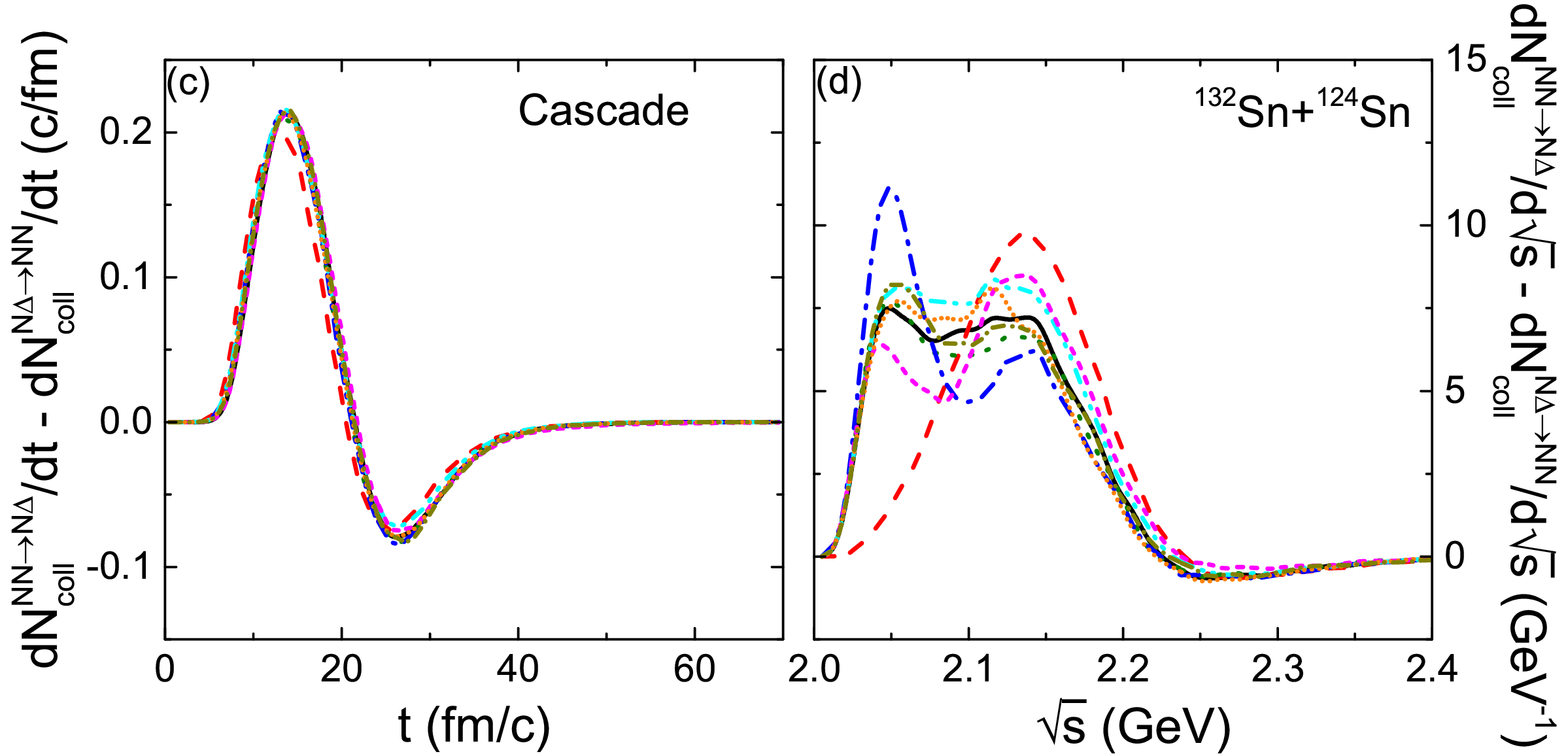}\\
\includegraphics[scale=0.25]{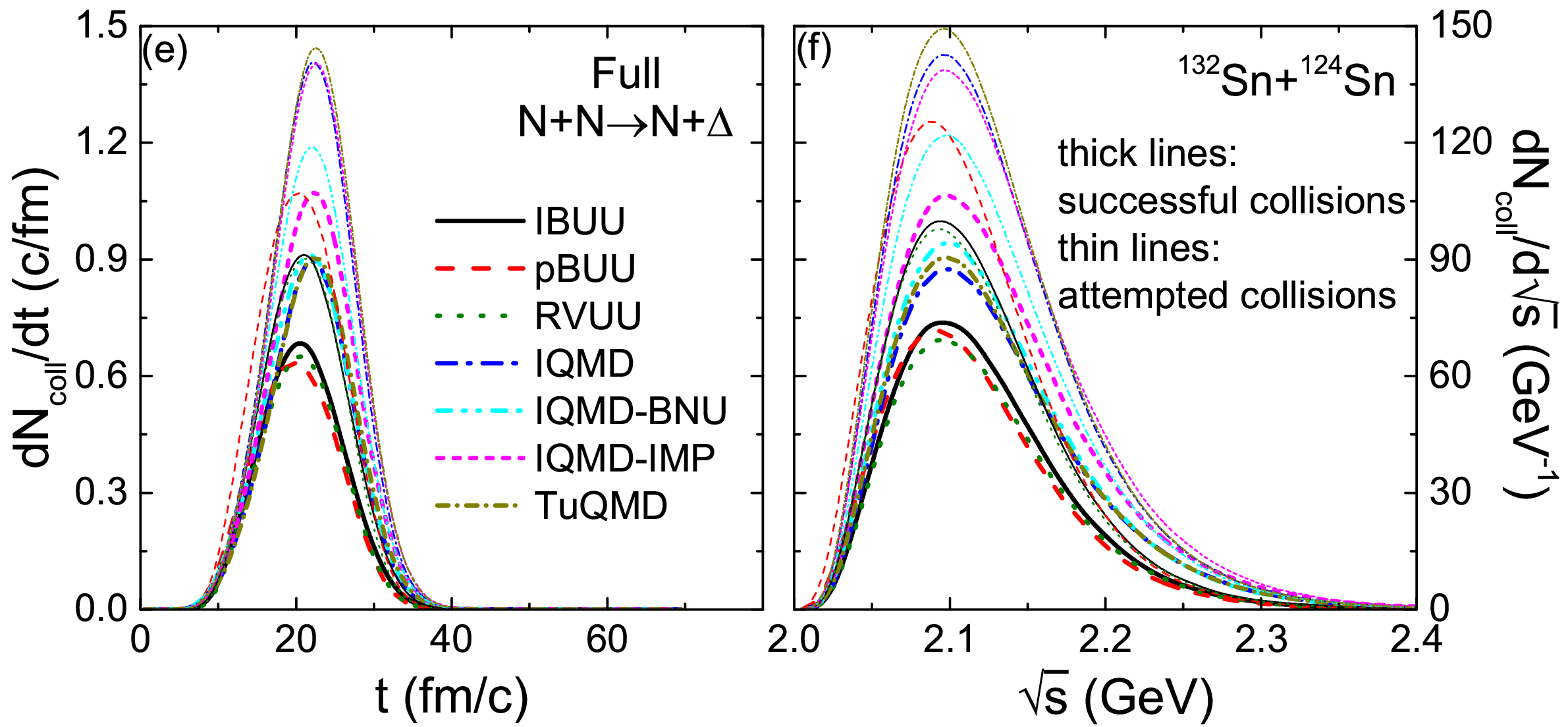} \\
\includegraphics[scale=0.25]{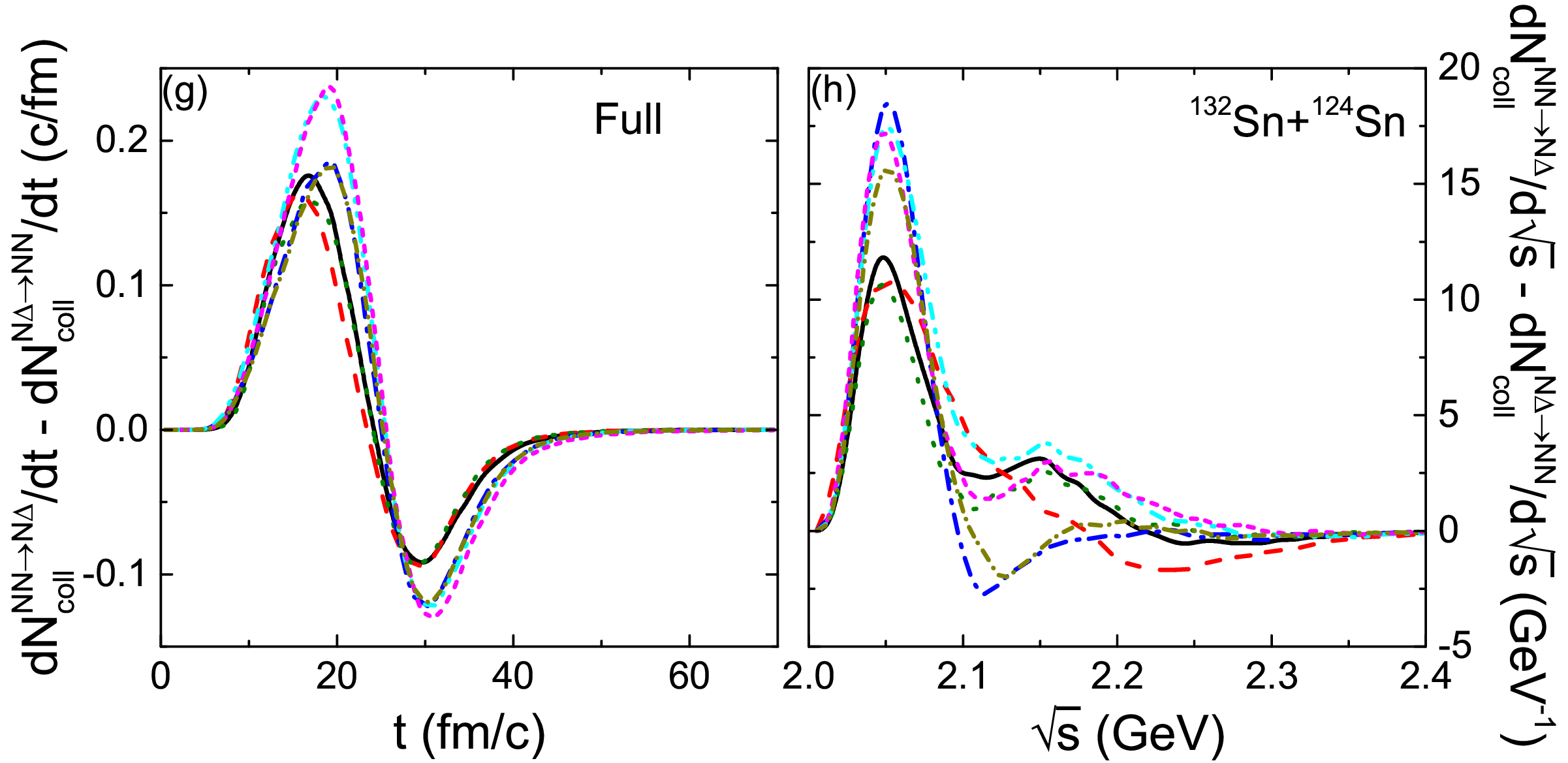}
\caption{(Color online) Successful and attempted rate of $NN \rightarrow N\Delta$ reactions and the net $\Delta$ production rate as a function of time and C.M. energy in the Cascade [(a)-(d)] and Full [(e)-(h)] mode with Pauli blocking.
} \label{NNND}
\end{figure}

Figure~\ref{NNND} shows the inelastic $NN \rightarrow N\Delta $ collision rates for the Cascade mode in the upper part [panels (a)-(d)] and for the Full mode in the lower part [panels (e)-(h)]. We note that the scales for rates are different between those from the Cascade-mode and Full-mode calculations. The rates in the Full-mode calculation are generally higher because of the higher densities reached in the presence of the mean-field potential. They are shown as functions of time or C.M. energy $\sqrt{s}$ in the right and left columns, respectively, for each case. The upper row of panels shows by thin lines the collision rates for the attempted collisions, i.e., without considering the Pauli blocking in the final state, and by thick lines for the successful collisions, with different colors and types for the different codes as given in the legend. The lower row of panels gives the net $\Delta$ production rates in the $NN \leftrightarrow N\Delta $ channel, i.e., the difference between the production of $\Delta$ particles and the inverse absorption reaction, again as functions of time or C.M. energy. We note that the attempted inelastic collision rates (thin lines in the upper panels) should be roughly proportional to the attempted elastic $NN \rightarrow NN$ collision rates with the ratio similar to the total inelastic to elastic cross sections, especially when considered as a function of $\sqrt{s}$, since the multiplicities of $\Delta$ particles and pions are small at this energy and they do not influence significantly the nucleonic evolution. As non-observables, these collision rates also characterize the nucleonic evolution of the heavy-ion simulation, similarly as the density and asymmetry evolution in the last section.

Let us first discuss the results of the Cascade-mode calculation in panels (a) and (b) of Fig.~\ref{NNND}. The attempted $NN \rightarrow N\Delta$ collision rates (thin lines), and thus also indicative of the attempted elastic $NN$ collision rates, generally agree well among the codes, as should be expected with an identical initialization and using the same inelastic cross sections (exceptions are the pBUU and IQMD-BNU codes, see below). As a function of time, the rates peak around the time of maximum overlap. As a function of C.M. energy, the peak is at an energy higher than the C.M. energy of the initial boosted nucleon pair of about 2.01 GeV due to the Fermi motion. The low rate from pBUU could be due to the stochastic method used for calculating the collision probability. The successful collisions (thick lines) show some small differences, which should be due to the treatment of the Pauli blocking on the nucleons in the final state of the $NN \rightarrow N\Delta $ reactions, and the exceptions are still pBUU and IQMD-BNU, which already give lower attempted collision rates.

Panels (c) and (d) of Fig.~\ref{NNND} show the net $\Delta$ production rates for the Cascade mode as functions of time and C.M. energy, i.e., the difference between the $\Delta$ production rates shown in panels (a) and (b) and corresponding rates for the inverse reaction. The time dependence is very similar among the codes, showing net $\Delta$ production at early times, and net $\Delta$ absorption at later times in the decompression phase, as expected. A similar behavior is also seen for pBUU which gives a lower successful inelastic collision rate, indicating that the absolute scale in $\Delta$ production largely cancels in its net production. The energy distribution has a more complex behavior, consisting of two components at energies lower and higher than the peak energy of the total $NN\rightarrow N\Delta$ collision rates in panel (b).

\begin{figure}[ht]
\includegraphics[scale=0.45]{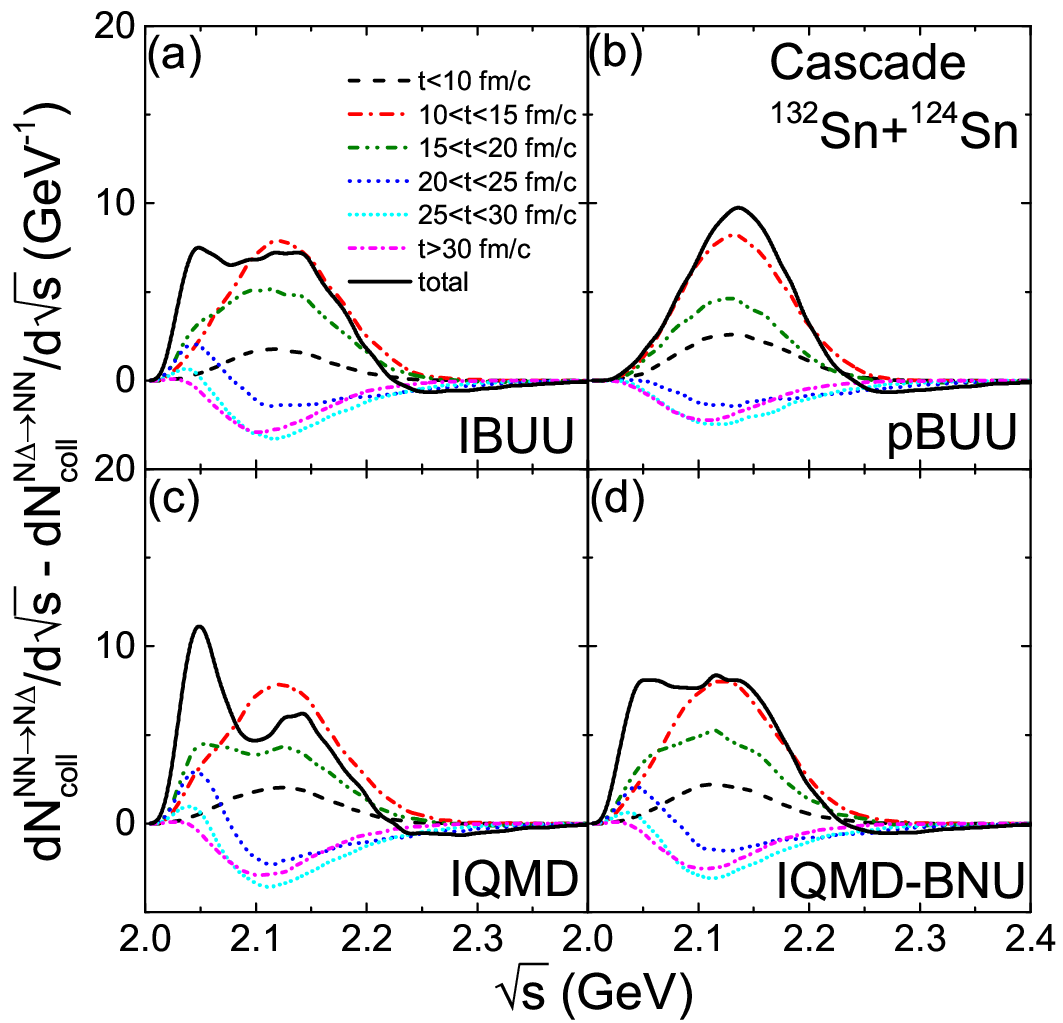}
\includegraphics[scale=0.45]{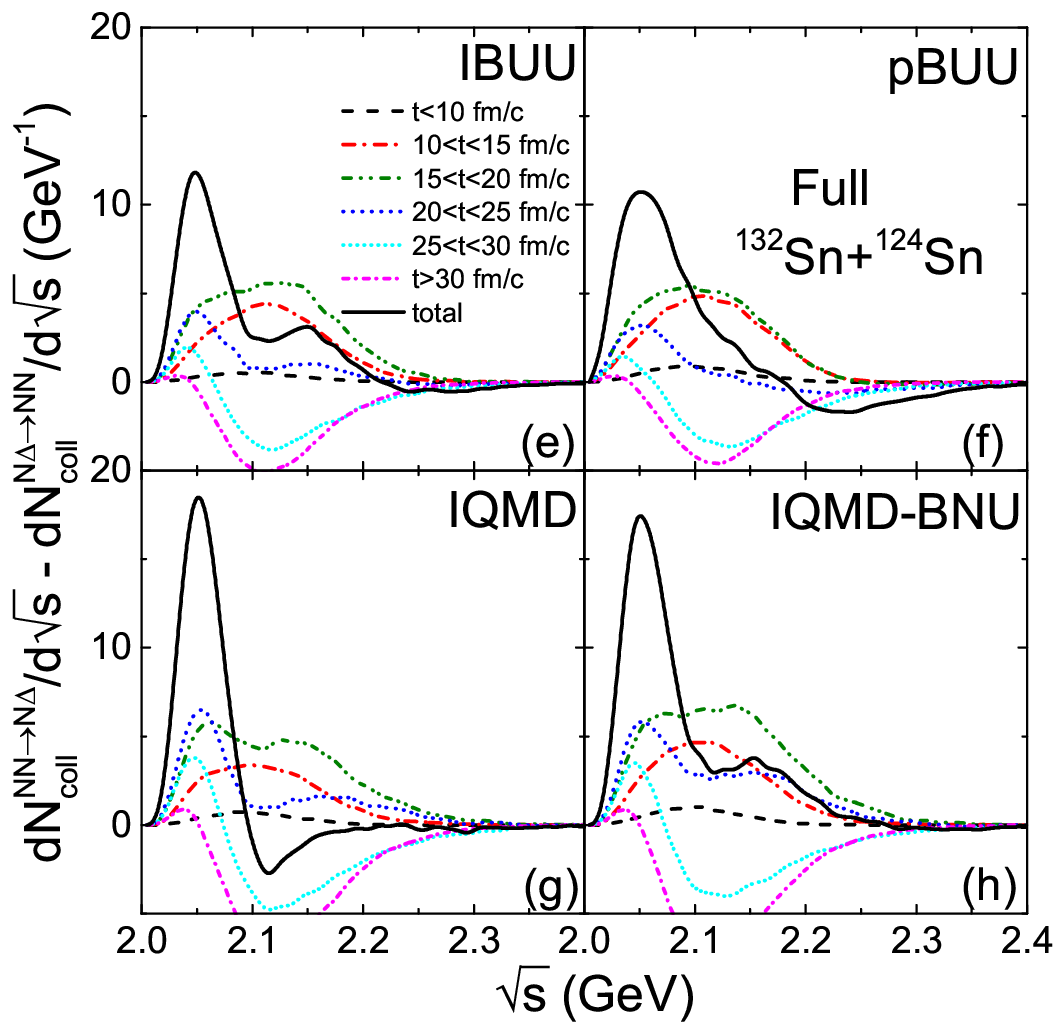}
\caption{(Color online) C.M. energy dependence of net $\Delta$ production at different time slots in the Cascade (left) and Full (right) mode for IBUU, pBUU, IQMD, and IQMD-BNU. The net production summed over time is shown as a dashed line.
} \label{timeslot}
\end{figure}

It may be worthwhile to investigate the above behavior in more detail, since different codes show large differences. This is done in Fig.~\ref{timeslot}, where the energy distributions of the net $\Delta$ production rates are shown in different time slots of length 5 fm/c and for the initial and final times for 4 selected codes of IBUU, pBUU, IQMD, and IQMD-BNU, for the Cascade mode in the left window and the Full mode in the right window. The distributions of the net production rates summed over the time slots are shown for each code as dashed lines, which are the same as the ones shown in panels (d) and (h) of Fig.~\ref{NNND} for the respective codes. We first discuss these time-slot plots for IBUU in the Cascade mode in panel (a) of Fig.~\ref{timeslot}. It is seen that at early times $\Delta$ resonances are produced at higher C.M. collision energies centering at $\sqrt{s}=2.105$ GeV, which reach the highest rate at 10-15 fm/c and then start to decrease. These are first-chance inelastic $NN$ collisions which produce high-energy and massive $\Delta$ resonances. At later times, these $\Delta$ particles are destroyed to a considerable amount by the inverse $N\Delta\rightarrow NN$ reactions. In the time interval around 20 fm/c which is the time of maximum overlap, a second peak starts to appear at a lower collision energy of about $\sqrt{s}=2.05$ GeV. These collisions have to be interpreted as inelastic $NN$ collisions in the dense compression zone among $NN$ pairs that have already lost some of their energy in previous elastic or inelastic collisions. One could call this the second-chance $\Delta$ production. These $\Delta$ resonances are mostly not destroyed by $N\Delta \rightarrow NN$ reactions, since there are almost no negative values for the net $\Delta$ production in this energy region at later times, showing that they decay preferentially via the $\Delta \rightarrow N\pi$ channel. Summing the net $\Delta$ production over time then results in two peaks of about equal height for IBUU. The behaviors of the two QMD codes in panels (c) and (d) show a roughly similar patten, except that the second-chance peak is more pronounced for IQMD. pBUU in the Cascade mode has a rather different behavior, in that the second-chance peak does not appear. This could indicate that the Pauli blocking in pBUU is so strong that it has blocked most secondary inelastic collisions.


We now return to Fig.~\ref{NNND} for the collision rates from the Full-mode calculation including the mean-field potential in panels (e) and (f) of the figure. Now the rates peak at the time of maximum compression (see Fig.~\ref{den}), while the peak in energy is the same as that from the Cascade-mode calculation. It is seen that the attempted collision rates (thin lines) in this case already show larger differences among the codes, which, according to the above argument, also indicates different attempted elastic $NN$ collision rates. In fact, there is a strong correlation between the centroid density evolution in Fig.~\ref{den} (c) and the attempted collision rates. The QMD models, which reach higher densities for the reasons discussed there, also have higher $NN$ collision rates (IQMD-BNU is again somewhat an exception), while the BUU models with lower densities have lower rates. One sees that results from IBUU and RVUU agree rather well, while pBUU has a considerably larger attempted $NN \rightarrow N\Delta$ rate. This is opposite to the behavior in the Cascade mode, and could be again due to the stochastic method used in pBUU, which may have a different influence in the Full-mode calculation, where the density dissolves in a slower pace compared to that in the Cascade-mode calculation. The successful collision rates (thick lines) follow largely the behavior of the attempted collision rates with some variations, indicating differences in the Pauli blocking effect. Compared with the QMD models, the BUU models again have lower successful collision rates, and agree with each other fairly well in the present case.

The net $\Delta$ production rates in the Full-mode calculation are shown in panels (g) and (h) of Fig.~\ref{NNND}. The time behavior in panel (g) is more similar among the QMD and BUU models than the forward reaction rates in panel (e), showing again some cancellation between the forward and backward reactions. However, differences among the codes remain. The QMD models generally show a systematically higher net production and absorption rates, especially in the late absorption phase. On the other hand, the BUU models are close together. The energy distribution of the net $\Delta$ production rate in panel (h) again shows the two components as in the Cascade-mode calculation [panel (d)], but now the low-energy component is more enhanced by all codes. Thus the inclusion of a mean-field potential affects the two components differently.

For a more detailed understanding of these energy distributions in the Full mode, we again go to Fig.~\ref{timeslot} and examine panels (e)-(h). We first discuss again IBUU in panel (e). The first-chance production of $\Delta$ resonances is reduced, since the repulsive mean-field potential at high densities suppresses the production of high-energy $\Delta$ particles. Also, the produced high-energy $\Delta$ particles are destroyed to a greater degree at later times because of the higher density, such that the summed contribution of the high-energy peak is reduced relative to the situation in the Cascade mode. On the other hand, the mean-field potential keeps the system together longer and thus enhances the second-chance $\Delta$ production, compared to the Cascade-mode calculation. Thus in the summed net production rate, the low-energy peak is enhanced and the high-energy peak is reduced relative to the results from the Cascade mode. This is similar or even enhanced for the two QMD codes, where the density is higher and the second-chance production is enhanced, while at the same time the destruction of high-energy $\Delta$ particles is stronger. For IQMD (and similarly for TuQMD not shown here), the surface correction to the Pauli blocking suppresses even more the first-chance production and results in a negative total balance (dashed line) for this component. For pBUU one now also observes a similar second-chance component as in other codes, which is consistent with the more convergent behavior of pBUU for the Full-mode calculation in Fig.~\ref{NNND}. Here also the high-energy $\Delta$ resonances are completely destroyed, and the negative tail in the C.M. energy dependence of net production reaches out to higher energies than that in other codes, indicating that some $\Delta$ particles are accelerated during their evolution in the expanding nuclear matter, due to the $\Delta$ potential incorporated only in pBUU.


\begin{figure}[ht]
\includegraphics[scale=0.22]{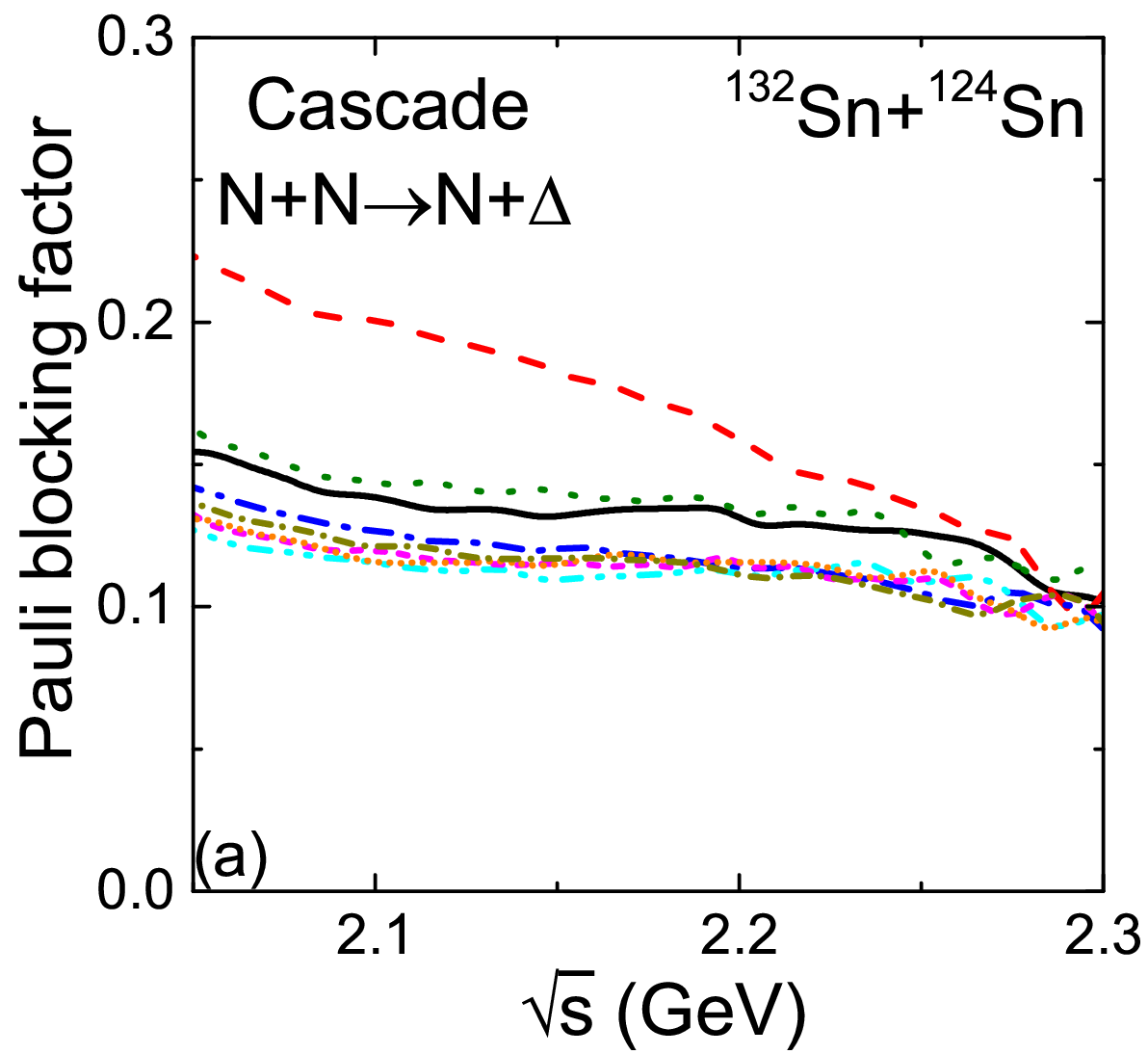}
\includegraphics[scale=0.22]{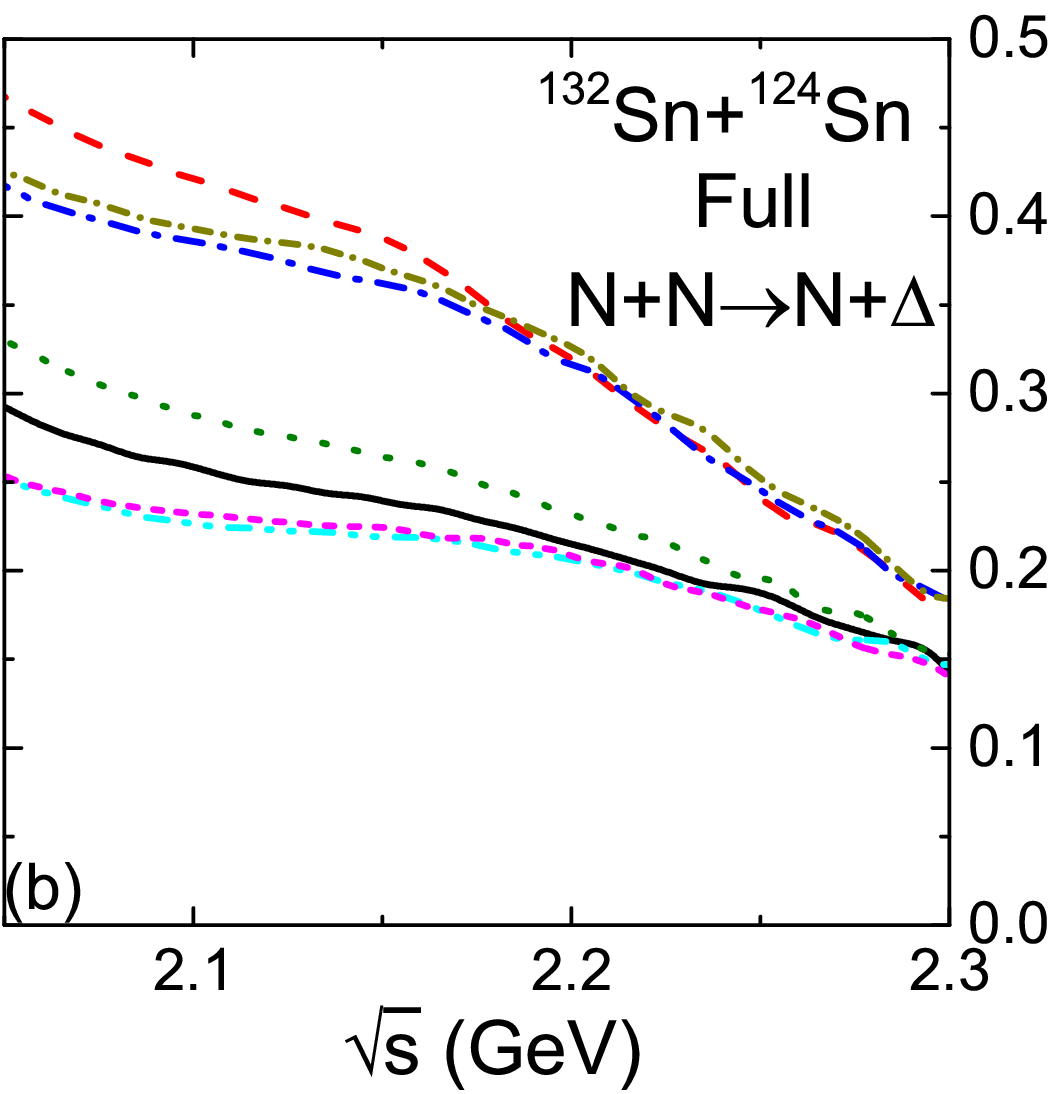}
\includegraphics[scale=0.22]{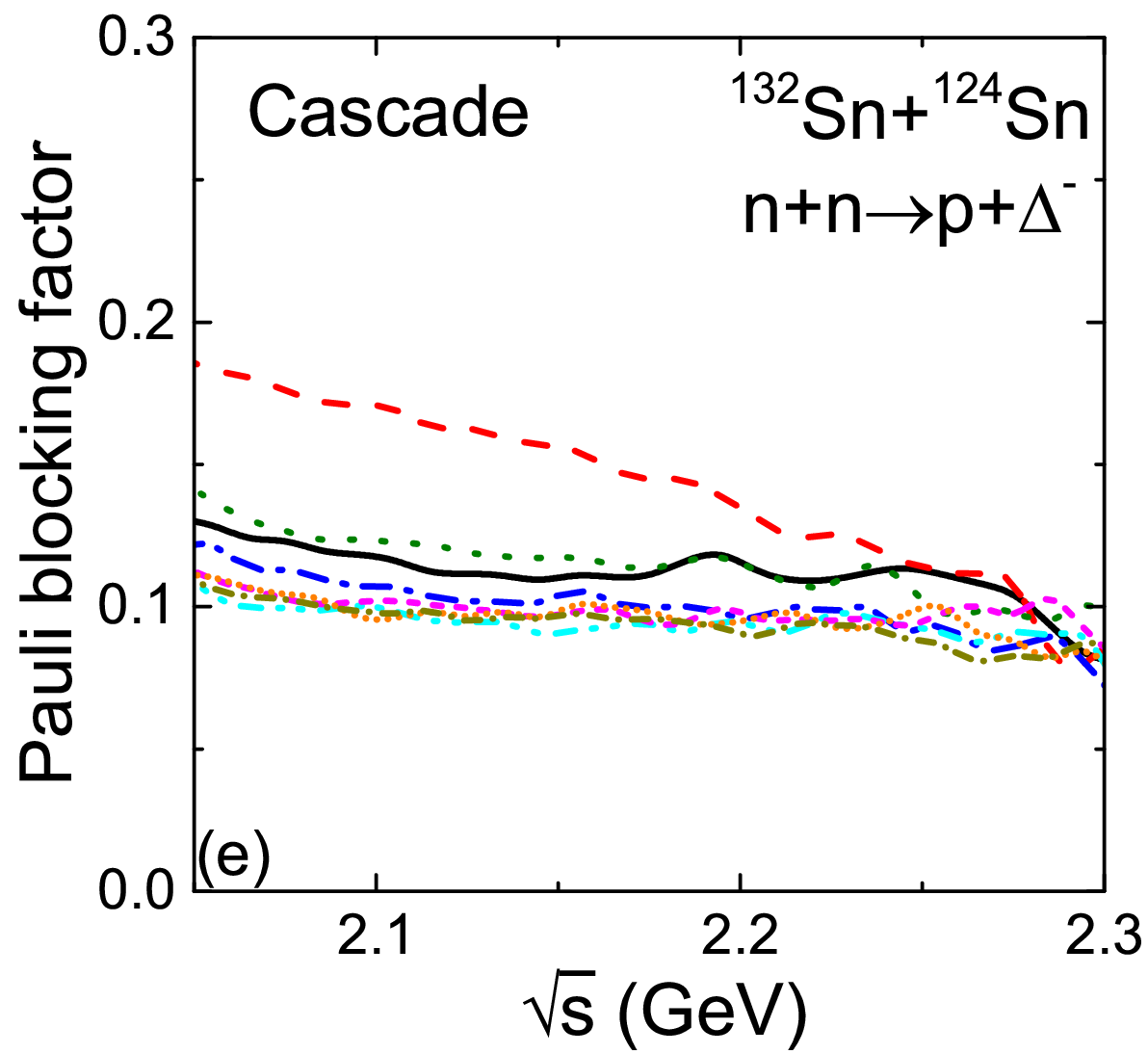}
\includegraphics[scale=0.22]{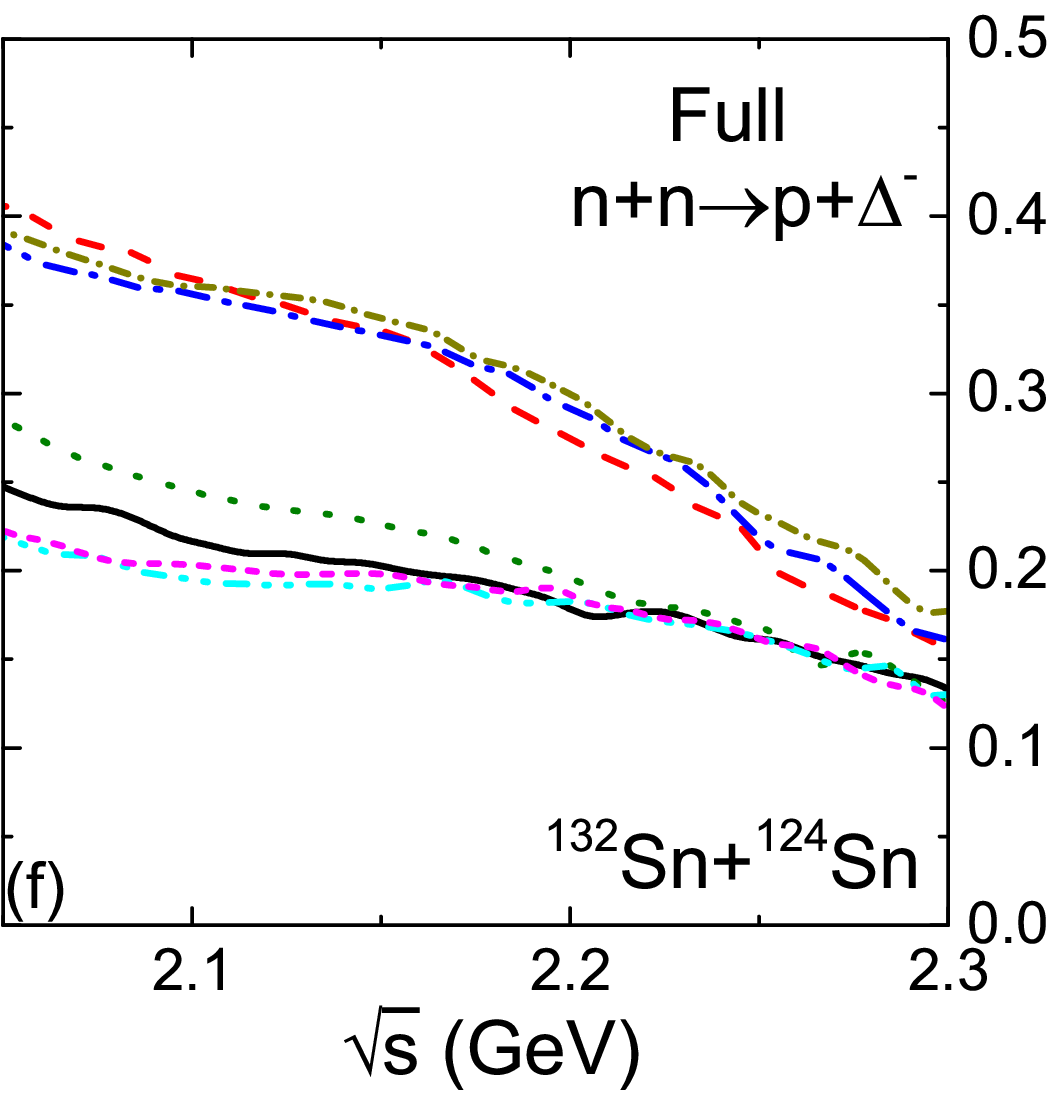}\\
\includegraphics[scale=0.22]{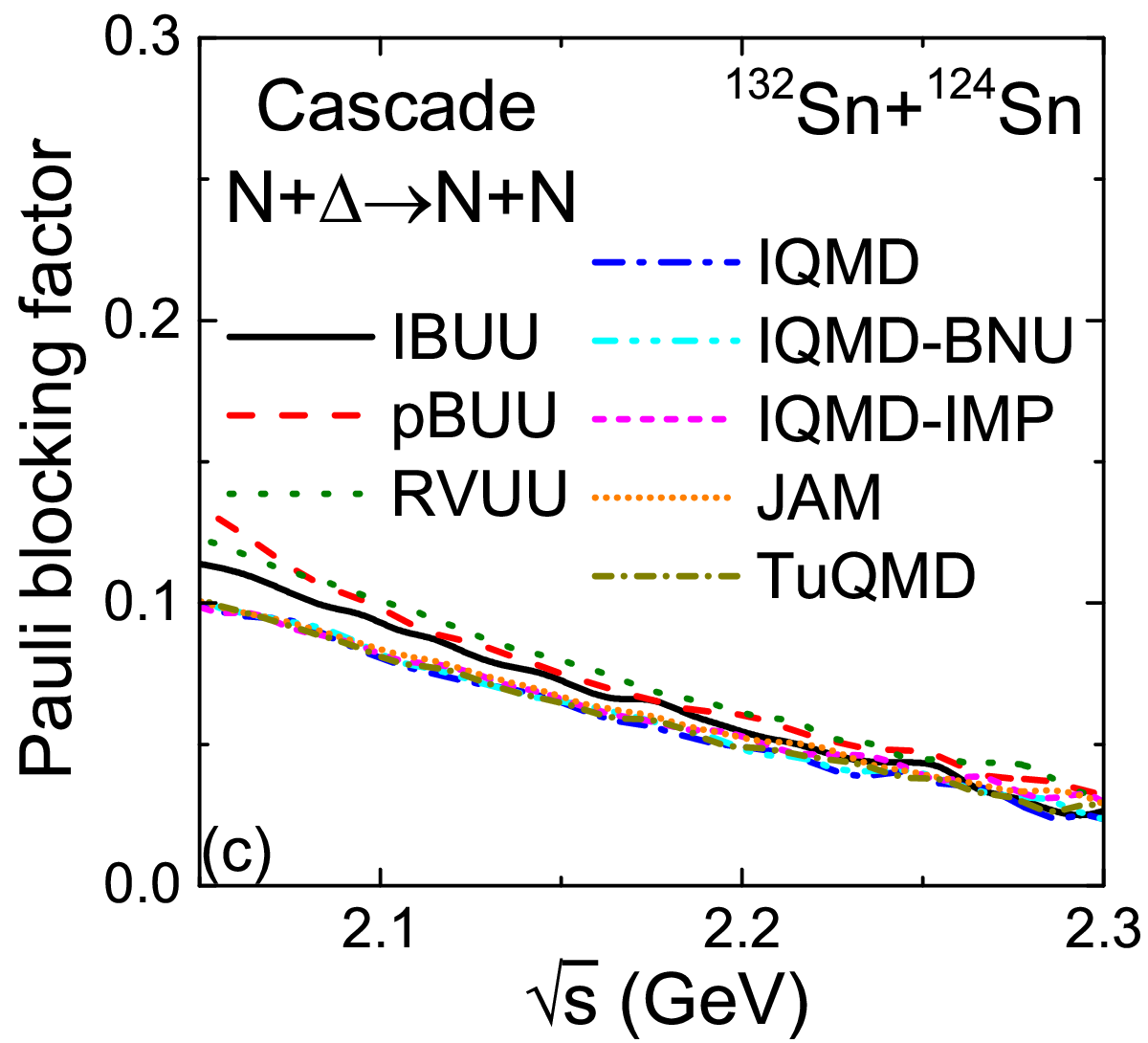}
\includegraphics[scale=0.22]{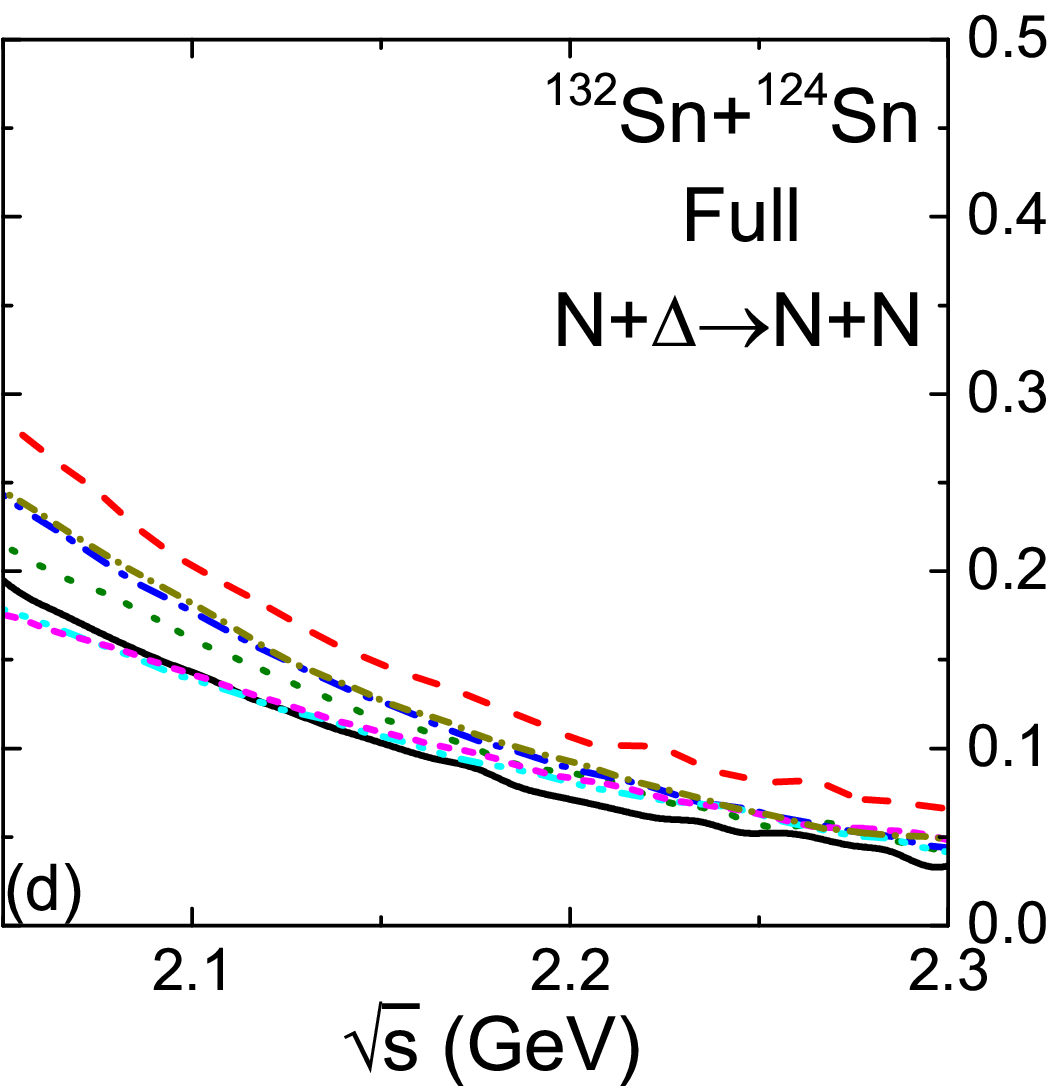}
\includegraphics[scale=0.22]{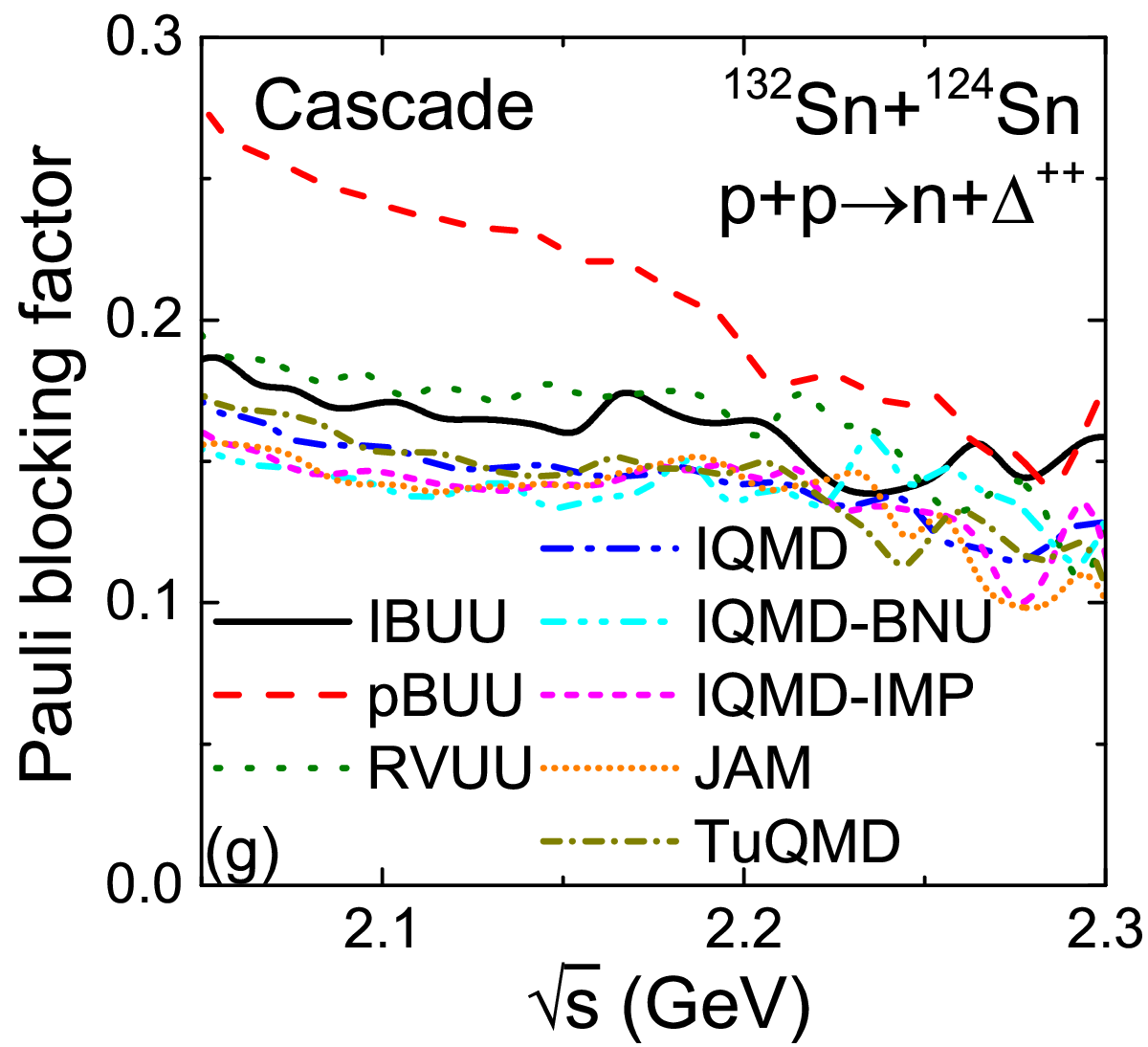}
\includegraphics[scale=0.22]{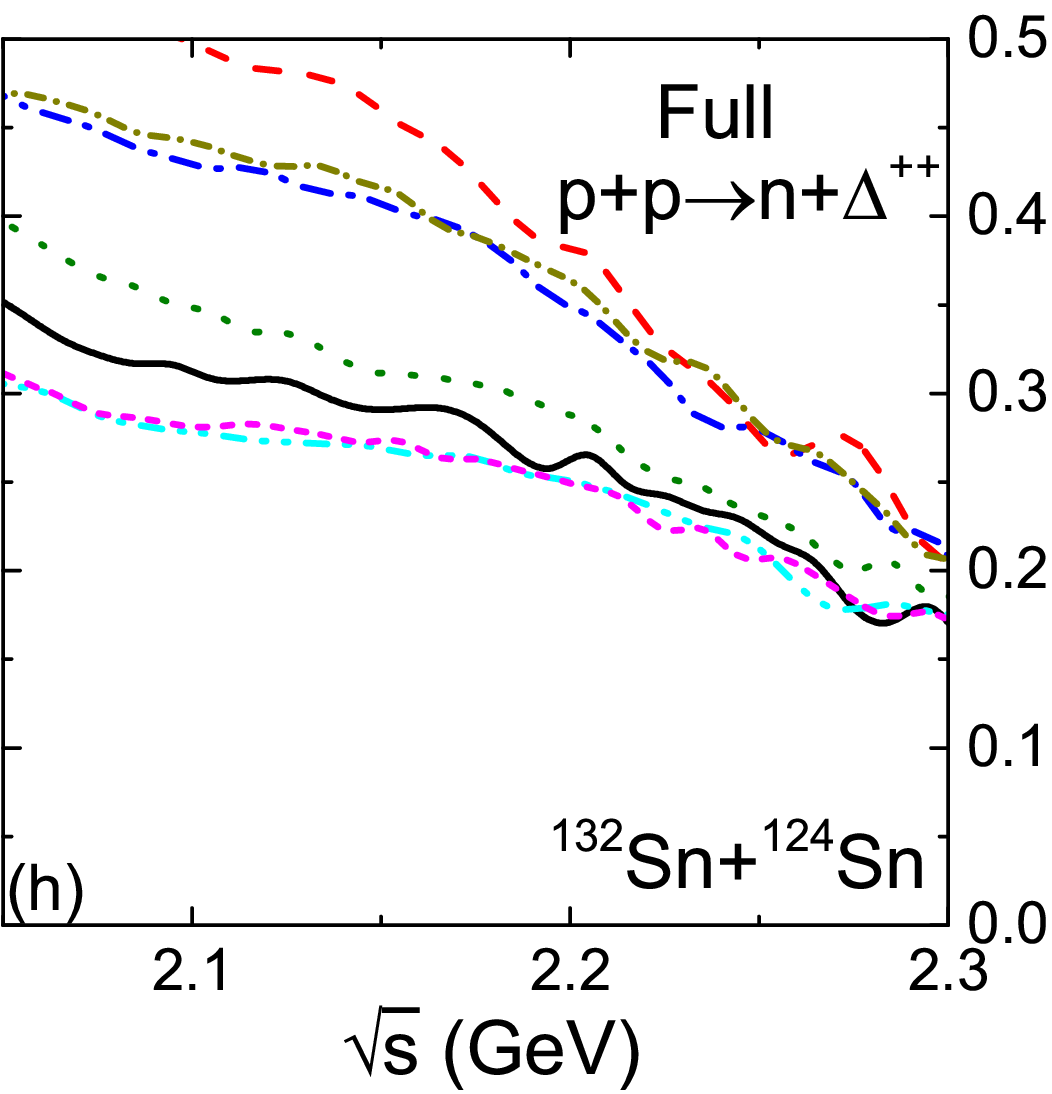}
\caption{(Color online) Left: Pauli blocking factor for $NN \rightarrow N\Delta$ (upper) and $N\Delta \rightarrow NN$ (lower) reactions as a function of C.M. energy in the Cascade [(a), (c)] and Full [(b), (d)] mode; Right: Pauli blocking factor for $nn \rightarrow p\Delta^-$ (upper) and $pp \rightarrow n\Delta^{++}$ (lower) reactions as a function of C.M. energy in the Cascade [(e), (g)] and Full [(f), (h)] mode.
} \label{NNND_rate}
\end{figure}

In Fig.~\ref{NNND_rate} we compare Pauli blocking factors, i.e., the ratio of blocked to attempted collision processes, as a function of C.M. energy from Cascade- and Full-mode calculations. The scales are different between the two modes, since the blocking is generally weaker in the Cascade mode. The range of C.M. energies has been slightly reduced compared to the previous figures to avoid plotting large fluctuations where the rates are very low. The left window [panels (a)-(d)] shows the Pauli blocking factors for the $NN \rightarrow N\Delta$ reaction and its inverse reaction averaged over the isospin states. Generally, the factor is larger at lower C.M. energies, when/where the phase space is more densely populated, especially around the maximum compression stage. In the $NN \rightarrow N\Delta$ reaction, systematically larger Pauli blocking factors in BUU models than in QMD models are observed in the Cascade mode, as a result of the smaller fluctuations in the occupation probability from BUU models than QMD models, similar to what we have already observed in the box-Cascade study~\cite{Zha18}. With the local phase-space distribution fitted with a superpositions of two Fermi-Dirac distributions, pBUU shows the highest Pauli blocking factor in the Cascade mode, while IQMD and TuQMD with the surface correction to the Pauli blocking give very high Pauli blocking factors in the Full-mode calculation, though the surface correction in these two codes is not active in the Cascade mode. RVUU using 1000 TPs suffers less from the fluctuation effect in the calculation of the occupation probability, and thus has a stronger Pauli blocking than IBUU using 100 TPs, and the reason has also been discussed in Ref.~\cite{Zha18}. On the other hand, the overall Pauli blocking factor in the $N\Delta \rightarrow NN$ reaction is smaller, and the agreement among the codes is better. This is understandable since nucleons in the final state of $N\Delta \rightarrow NN$ are more energetic compared to those in the final state of $NN \rightarrow N\Delta$, leading to a phase space that is not as dense and thus has a weaker Pauli blocking effect. Since the net $\Delta$ production rate is exactly equal to the pion production rate, the total pion multiplicity is mostly sensitive to the Pauli blocking factor in the $NN \rightarrow N\Delta$ reaction and thus depends strongly on how the Pauli blocking is implemented in a code for this reaction channel.

We further investigate the effect of the isospin-dependent Pauli blocking in the $NN\leftrightarrow N\Delta$ reaction. The right panels of Fig.~\ref{NNND_rate} compare the Pauli blocking factors in the two typical production channels of $\Delta^-$ and $\Delta^{++}$, i.e., $nn \rightarrow p\Delta^-$ and $pp \rightarrow n\Delta^{++}$ in both Cascade and Full modes. The blocking factors for the isospin-dependent reactions generally follow the behavior of the isospin-average blocking factors, shown in the left window of the figure. However, there are interesting isospin-dependent effects. A stronger Pauli blocking is observed in the $pp \rightarrow n\Delta^{++}$ reaction than in the $nn \rightarrow p\Delta^-$ reaction, especially at lower C.M. energies. This is due to the larger neutron occupation probability in the neutron-rich system, especially in the high-density phase where $\Delta$ resonances are mostly produced. Since the Pauli blocking factors generally agree among the codes in the inverse reaction, i.e., $N\Delta \rightarrow NN$, the discrepancies of the calculated Pauli blockings in the $pp \rightarrow n\Delta^{++}$ and the $nn \rightarrow p\Delta^-$ reactions are responsible for those in the final $\pi^-/\pi^+$ yield ratio, as $\Delta^-$ ($\Delta^{++}$) eventually decays into $\pi^-$ ($\pi^+$).

\begin{figure}[ht]
\includegraphics[scale=0.25]{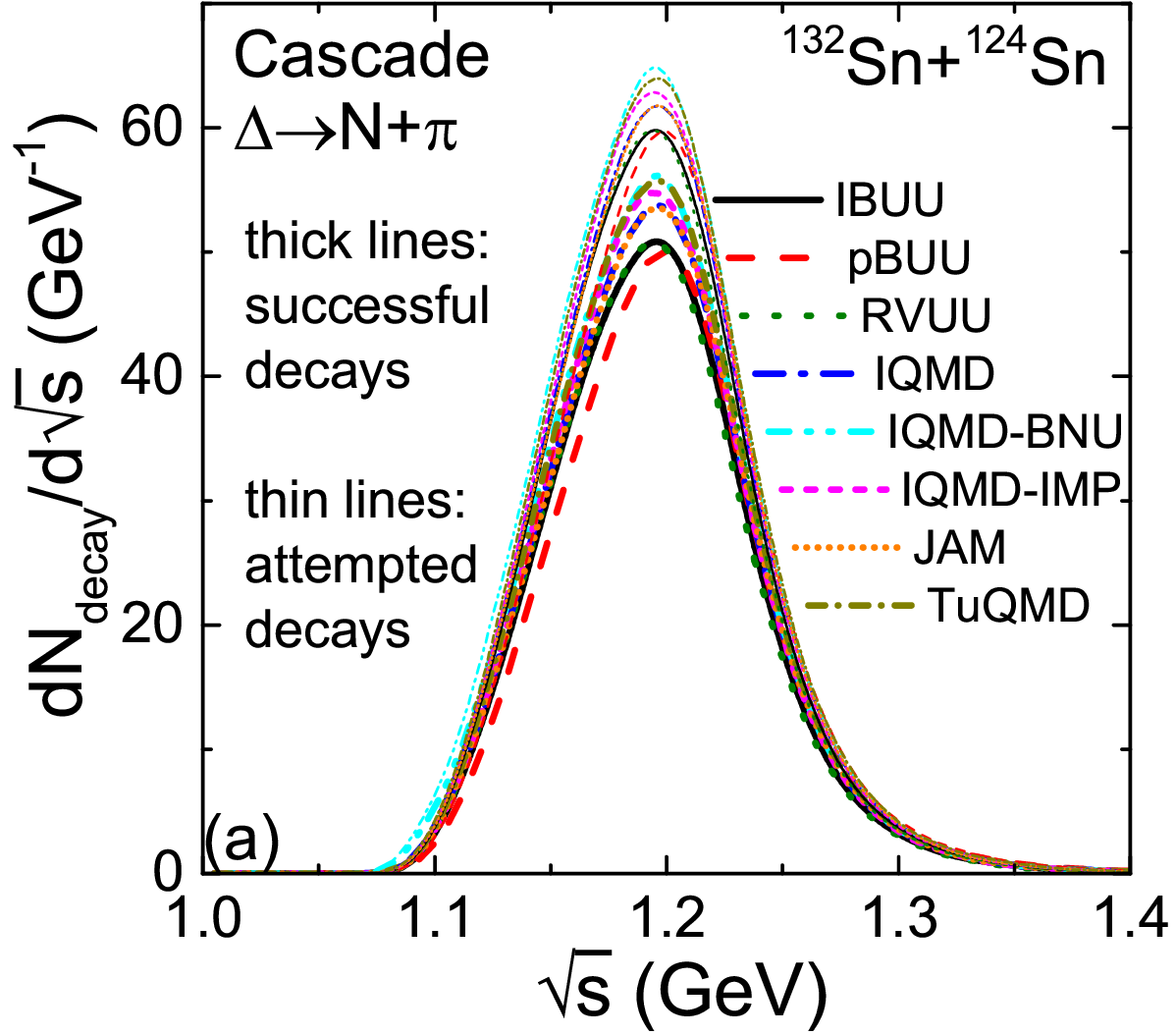}
\includegraphics[scale=0.25]{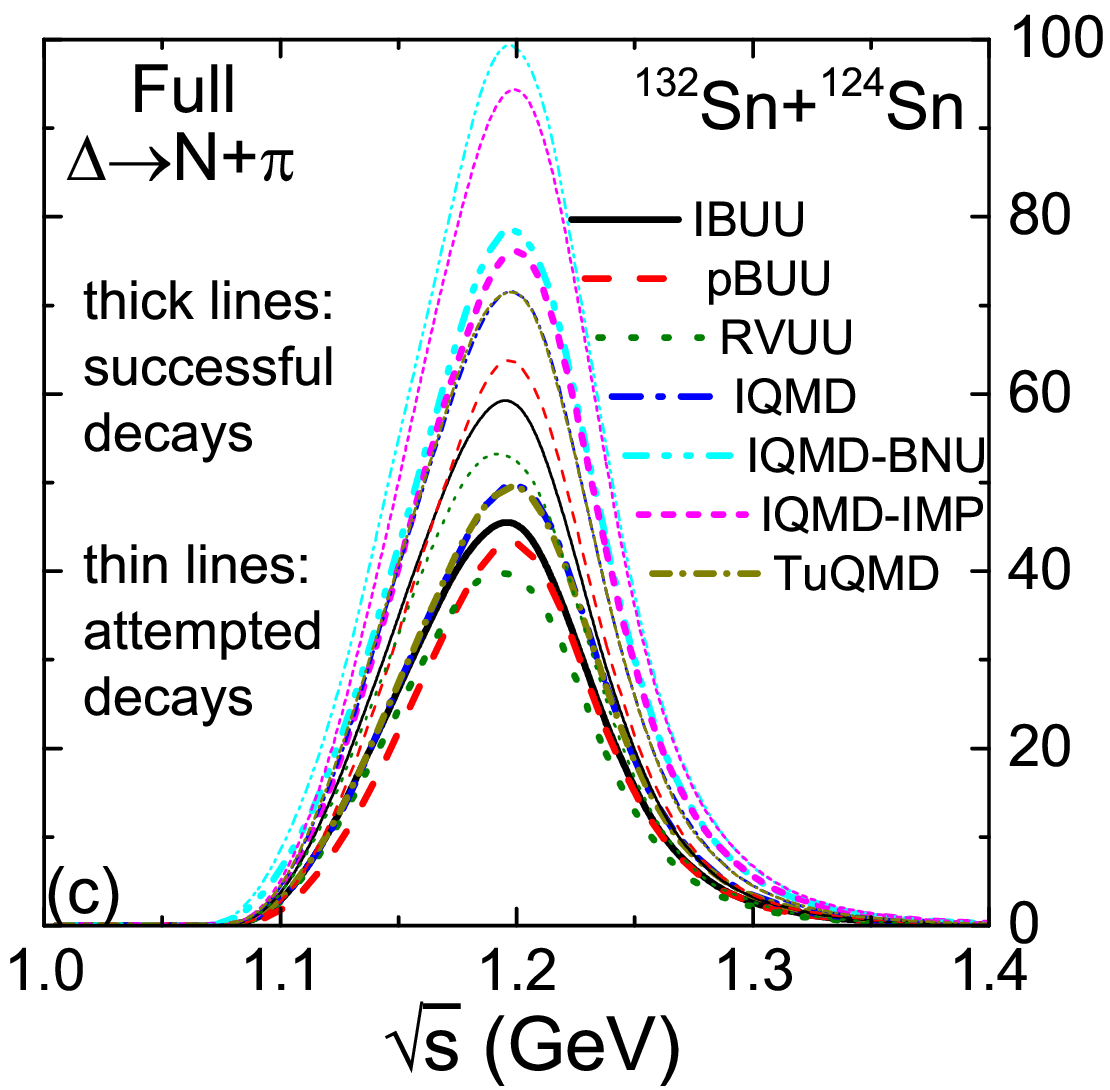}\\
\includegraphics[scale=0.25]{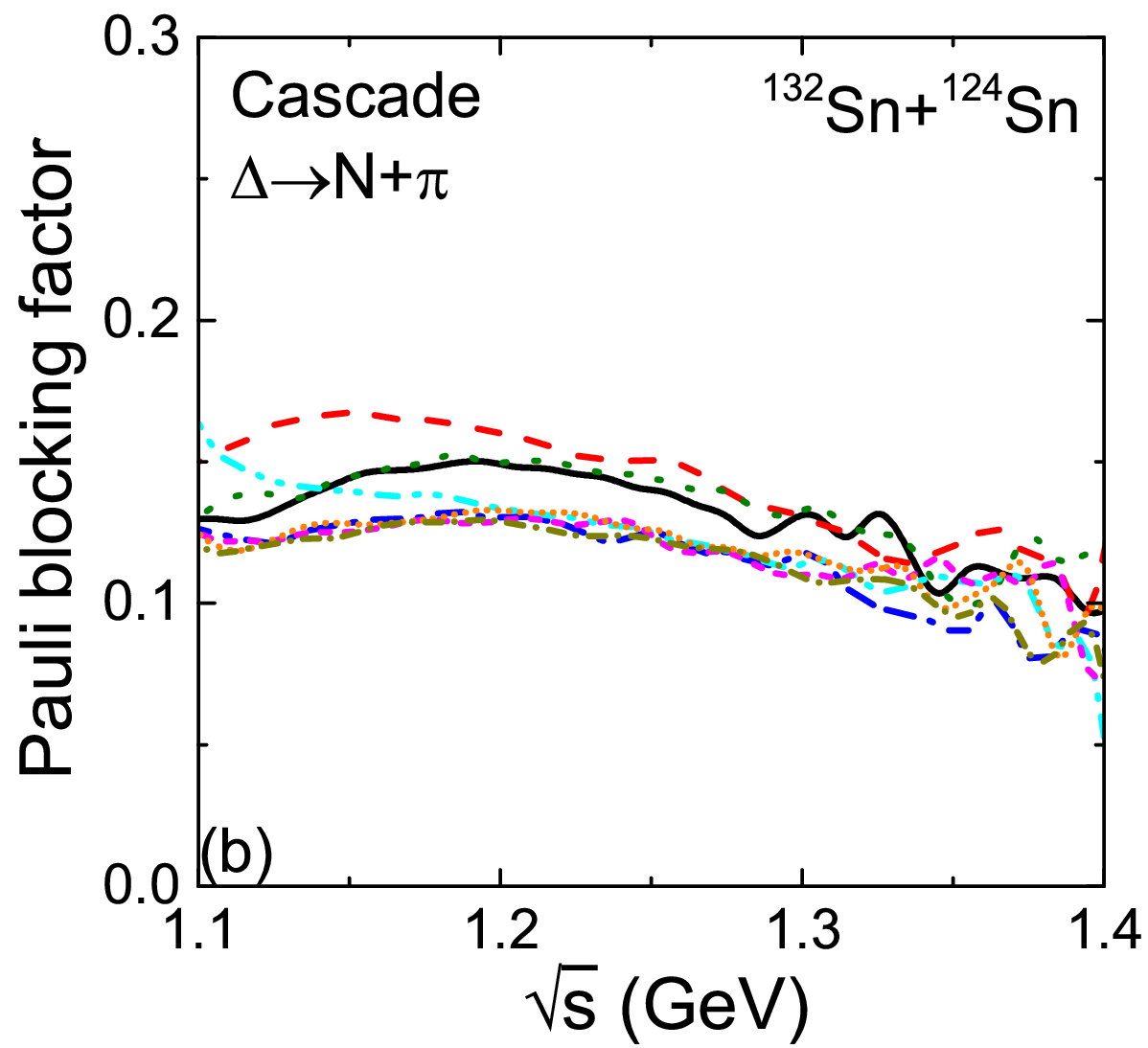}
\includegraphics[scale=0.25]{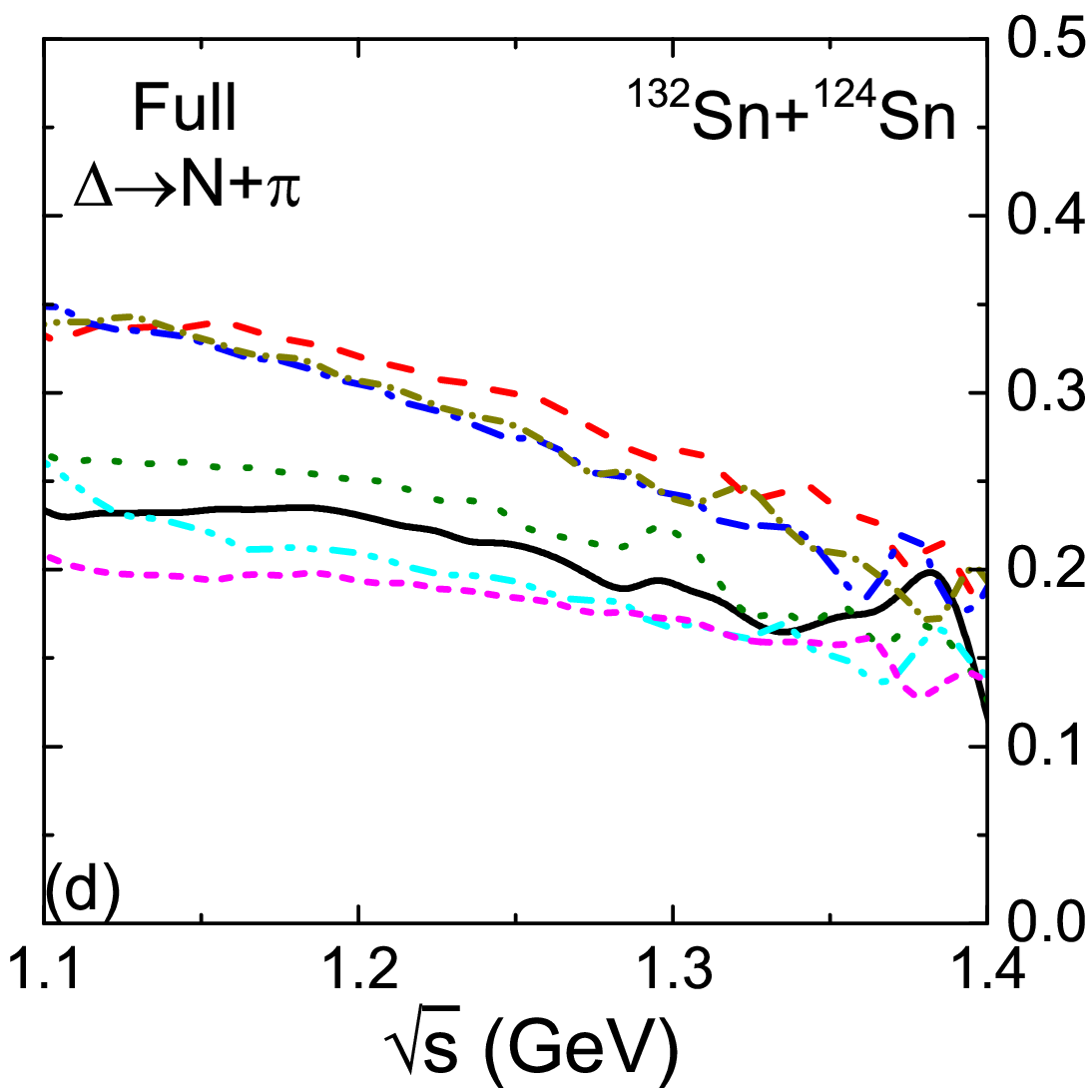}
\caption{(Color online) Upper: Successful and attempted rates of $\Delta \rightarrow N\pi$ decays as a function of C.M. energy in the Cascade (left) and Full (right) mode; Lower: Pauli blocking factor for $\Delta \rightarrow N\pi$ decays as a function of C.M. energy in the Cascade (left) and Full (right) mode.
} \label{DNPI}
\end{figure}

In Fig.~\ref{DNPI} we provide information about the $\Delta \rightarrow N\pi$ decay for the Cascade mode on the left and the Full mode on the right panels. The decay rates are shown in the upper panels, where again thin lines represent the attempted and thick lines the successful decay rates as a function of the C.M. energy, and the corresponding Pauli blocking factors are given in the lower panels. Here we only show the C.M. energy dependence of the rates and blocking factors as in Fig.~\ref{NNND_rate}. The blocking factors numerically fluctuate strongly at threshold and are therefore only shown from slightly above threshold. The decay rates as a function of C.M. energy are determined by the $\Delta$ mass distributions at the time of their decay. The peak is somewhat below the pole mass of the spectral function [Eq.~(\ref{BW})], since the $\Delta$ mass is sampled according to Eq.~(\ref{pd}) within $m_N^{}+m_\pi^{} < m_\Delta^{} < \sqrt{s} - m_N^{}$, and the limit of $\sqrt{s}$ shifts the $\Delta$ mass distribution to the lower side. The time dependence of the $\Delta$ decay (not shown here) peaks somewhat after the time of the maximum $\Delta$ production rate as shown in Fig.~\ref{NNND} because of the finite lifetime of $\Delta$ resonances. This delay effect is stronger in the Full-mode calculation, since the lifetime is effectively increased because of the stronger Pauli blocking, as is seen in the lower panels. The higher decay rates in most QMD models are due to the larger multiplicity of $\Delta$ resonances produced in the $NN\rightarrow N\Delta$ reactions, as a result of the higher density reached in QMD models compared to BUU models. Generally, the behavior of the decay rates is similar to that of the $\Delta$ production rates in Fig.~\ref{NNND} (f). However, the detailed relation depends on the strategy and collision sequence used in treating inelastic collisions and $\Delta$ decays in the collision term, as was analyzed in detail in a box calculation in Ref.~\cite{Ono19} for the case without Pauli blocking. In the Cascade-mode calculation [panel (a)], the QMD and BUU models are close together, and the particularly low rate for pBUU as in Fig.~\ref{NNND} (b) is not seen. In the Full-mode calculation, the behavior shows larger differences among the codes, similar to the $\Delta$ production rates for the same scenario in Fig.~\ref{NNND}. The blocking factors are generally higher for the $\Delta$ decay than for the $\Delta$ destruction reaction, i.e., $N\Delta \rightarrow NN$, since the pion takes part of the energy and the nucleon thus has a lower energy and is more blocked. The Pauli blocking effect in the $\Delta$ decay is small in the Cascade-mode calculation because of the lower density reached, but significant in the Full-mode calculation, where it is seen that about $15 - 30\%$ of the $\Delta$ decays in the maximum compression stage can be Pauli blocked. Here, pBUU and the QMD codes with surface correction in the Pauli blocking again have the highest blocking factors.

Considering the isospin-dependence of the Pauli blocking, neutrons in the $\Delta^- \rightarrow n\pi^-$ decays are more blocked than protons in $\Delta^{++} \rightarrow p \pi^+$. Therefore, the lifetime of $\Delta^-$ is increased compared to $\Delta^+$ in neutron-rich matter. In the present study, we have not considered a calculation mode with Pauli blockings only in the $N\Delta \leftrightarrow NN$ reaction and without Pauli blocking in the $\Delta \rightarrow N\pi$ channel. A test study of this case using IBUU shows that the Pauli blocking effect in the $\Delta \rightarrow N\pi$ reaction on the total pion yield is comparable to that in the $N\Delta \leftrightarrow NN$ channel, while the isospin-dependent Pauli blocking effect on the $\pi^-/\pi^+$ yield ratio mostly comes from the $N\Delta \leftrightarrow NN$ reaction rather than the $\Delta \rightarrow N\pi$ channel. For more details on the Pauli blocking effect of the $\Delta$ decay on the $\pi^-/\pi^+$ yield ratio, we refer the reader to Ref.~\cite{Ike20}.

Taking another look at Figs.~\ref{NNND}, \ref{NNND_rate}, and \ref{DNPI}, one sees that the collision rates are generally higher in the Full mode compared with the Cascade mode, due to the higher density reached in the Full-mode calculation. However, the discrepancies among codes are also larger in the Full-mode calculation compared with those from the Cascade-mode calculation. We can explain these in most cases as due to different calculations of the mean-field potential and different strategies in implementing the Pauli blocking. However, as a consequence we have to expect similar differences in the pion observables, to be shown in the next subsection.

\subsection{Pion multiplicity}

\begin{figure}[ht]
\includegraphics[scale=0.37]{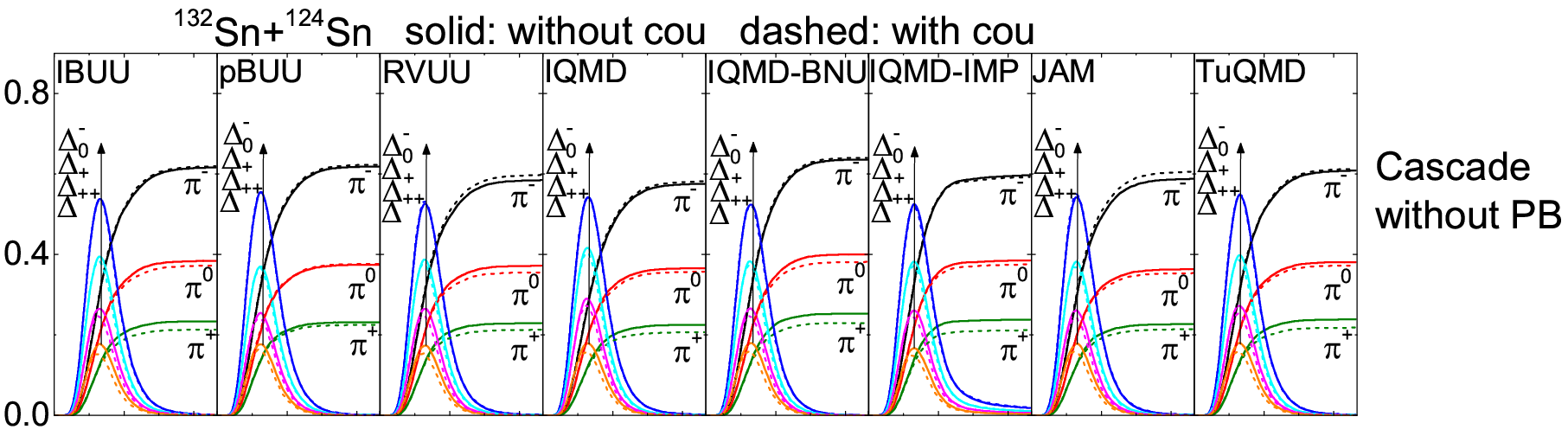}\\
\hspace{-2.5mm}\includegraphics[scale=0.37]{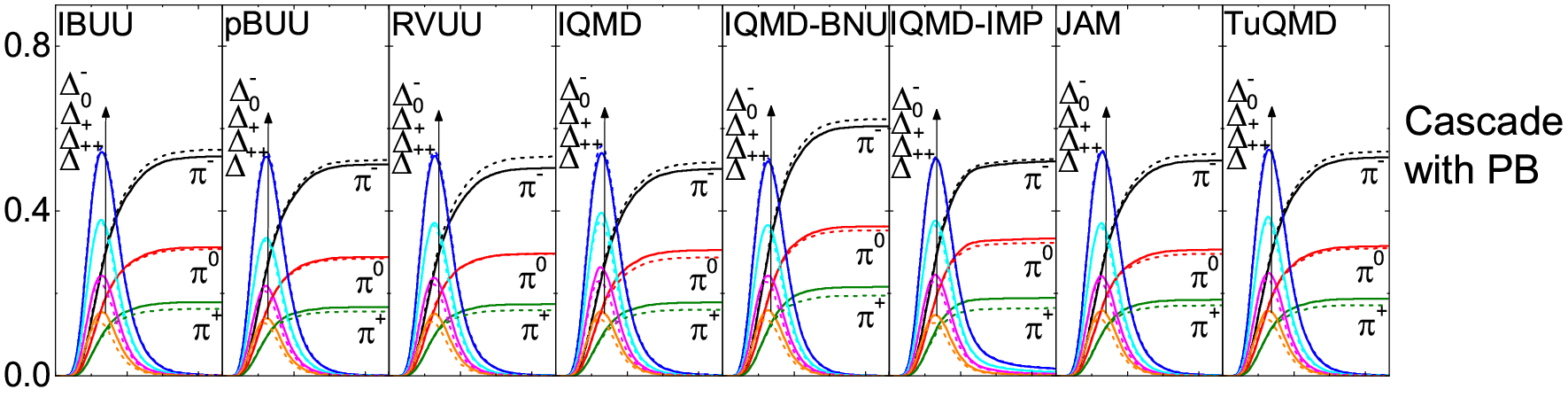}\\
\includegraphics[scale=0.37]{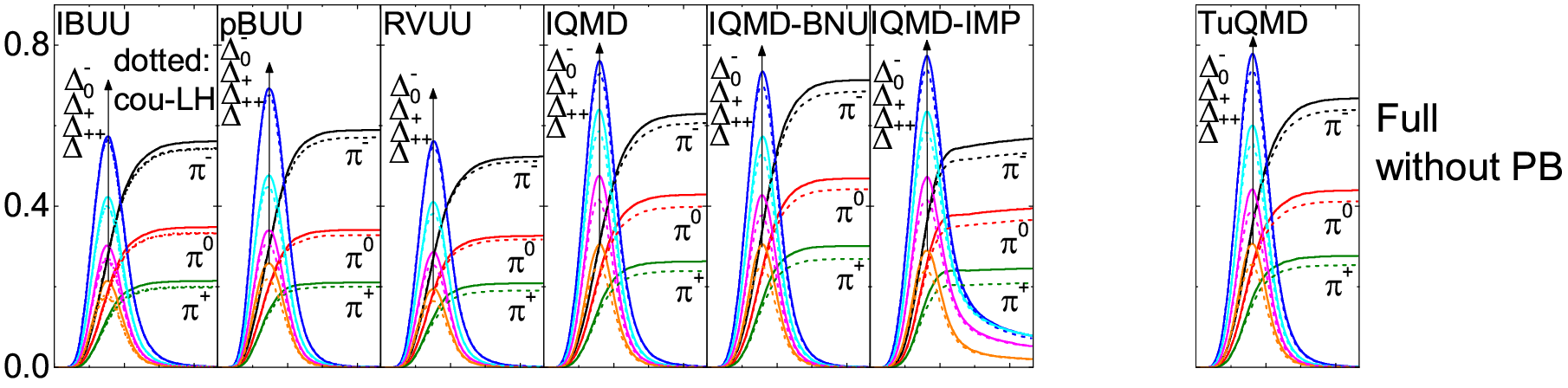}\\
\hspace{-4mm}\includegraphics[scale=0.37]{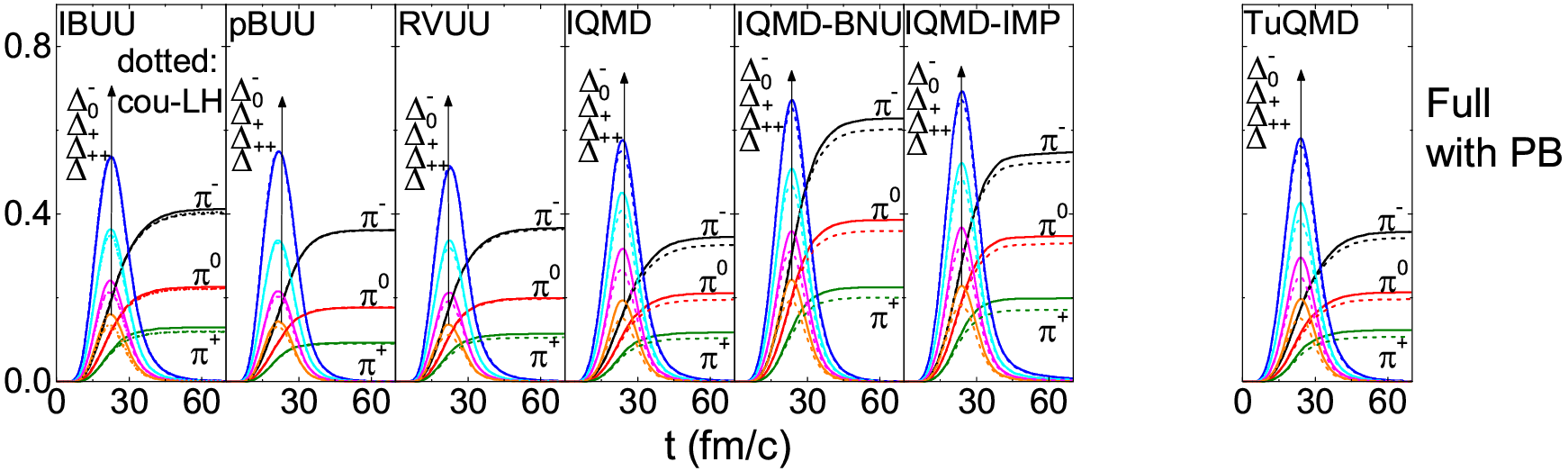}
\caption{(Color online) Time evolution of the multiplicities of pion-like particles in different scenarios (see legend and text).
} \label{PID}
\end{figure}

In this section we compare the pion multiplicities and study the role of the Coulomb force on pion observables. Figure~\ref{PID} shows the time evolution of the multiplicities of pion-like particles, i.e., the different charge states of pions and $\Delta$ resonances, from different codes for all simulation modes. The results for Cascade calculations without and with Pauli blocking are shown in the two top rows, and correspondingly for the Full calculations in the two bottom rows. Calculations without the Coulomb force are shown as solid lines, and those with Coulomb as dashed curves. Generally, for all codes and for all modes, the multiplicities of $\Delta$ resonances reach a maximum around the maximum compression stage of heavy-ion collisions, and those of pions continue to increase due to successive $\Delta$ decays as the compressed nuclear matter expands. There are almost no $\Delta$ resonances at $t=70$ fm/c when the calculation ends, except for IQMD-IMP. The multiplicity of different charge states decreases from negative to positive values of their charges for both $\Delta$ resonances and pions, since the system is neutron-rich. Considerable differences are seen among the codes, even in the simplest case of the Cascade mode without Pauli blocking. This mode is comparable to but still different from the box calculation of Ref.~\cite{Ono19}, where the codes agreed much better. This is understandable, since the abundance of pion-like particles does not change after reaching chemical equilibrium in the box calculation~\cite{Ono19}, while it undergoes further evolution in heavy-ion collisions in the expansion stage. The effect of turning on the Coulomb force is qualitatively similar for each code. Due to its repulsive nature in the globally positively charged system, the Coulomb force reduces the density and also the total yield of pions. The Coulomb force also pushes protons outward and thus the high-density region becomes more neutron-rich. The change of the multiplicity of $\pi^-$ due to the Coulomb force is a competition between two effects, and the balance is different in the Cascade and the Full calculation, respectively: the Coulomb force increases the $\pi^-$ multiplicity in the Cascade calculation mostly due to the more neutron-rich high-density region, and it reduces the $\pi^-$ multiplicity in the Full calculation mostly due to the overall lower density, compared to the case without the Coulomb force.

\begin{figure}[ht]
\includegraphics[scale=0.3]{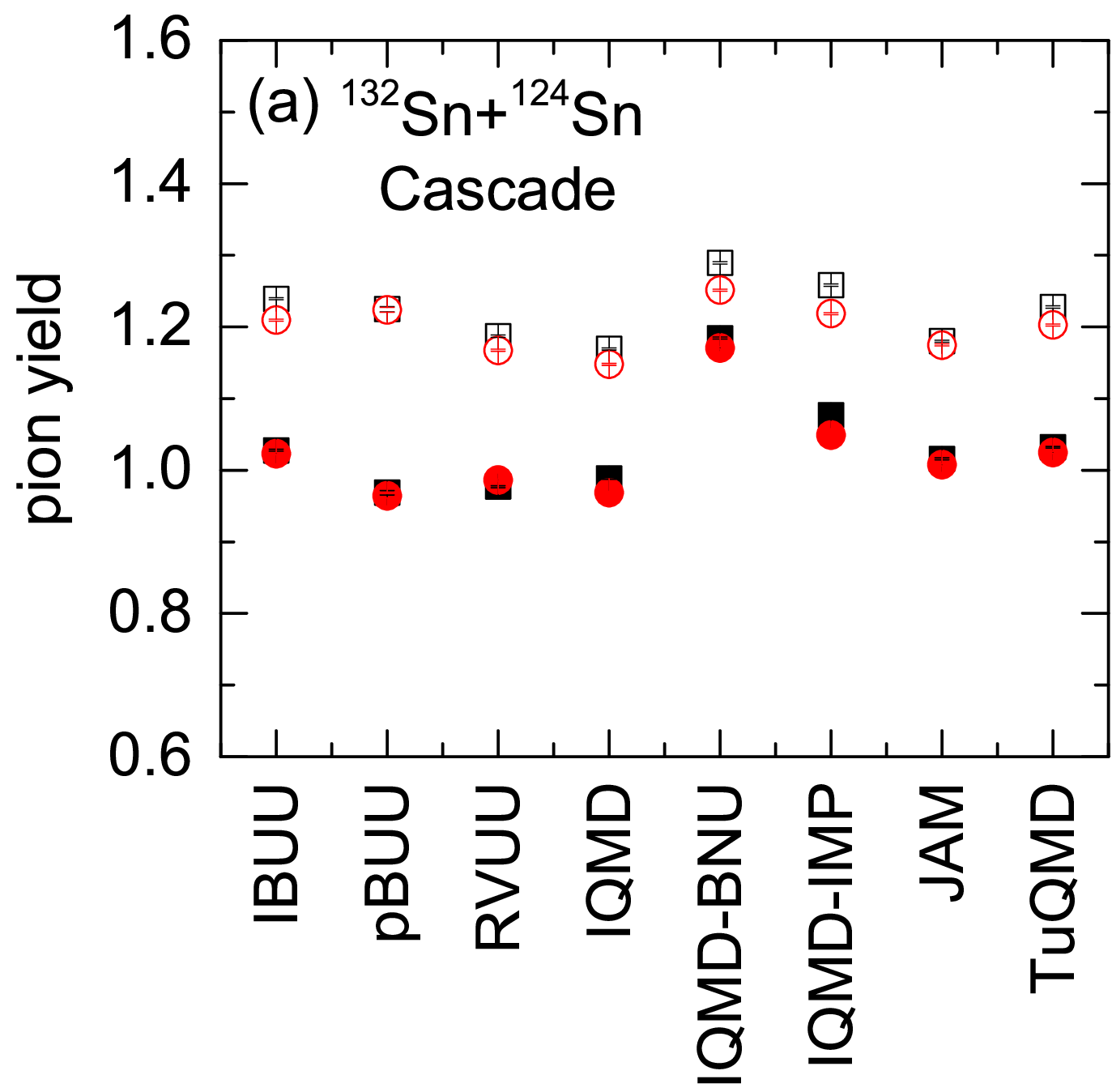}~~\includegraphics[scale=0.3]{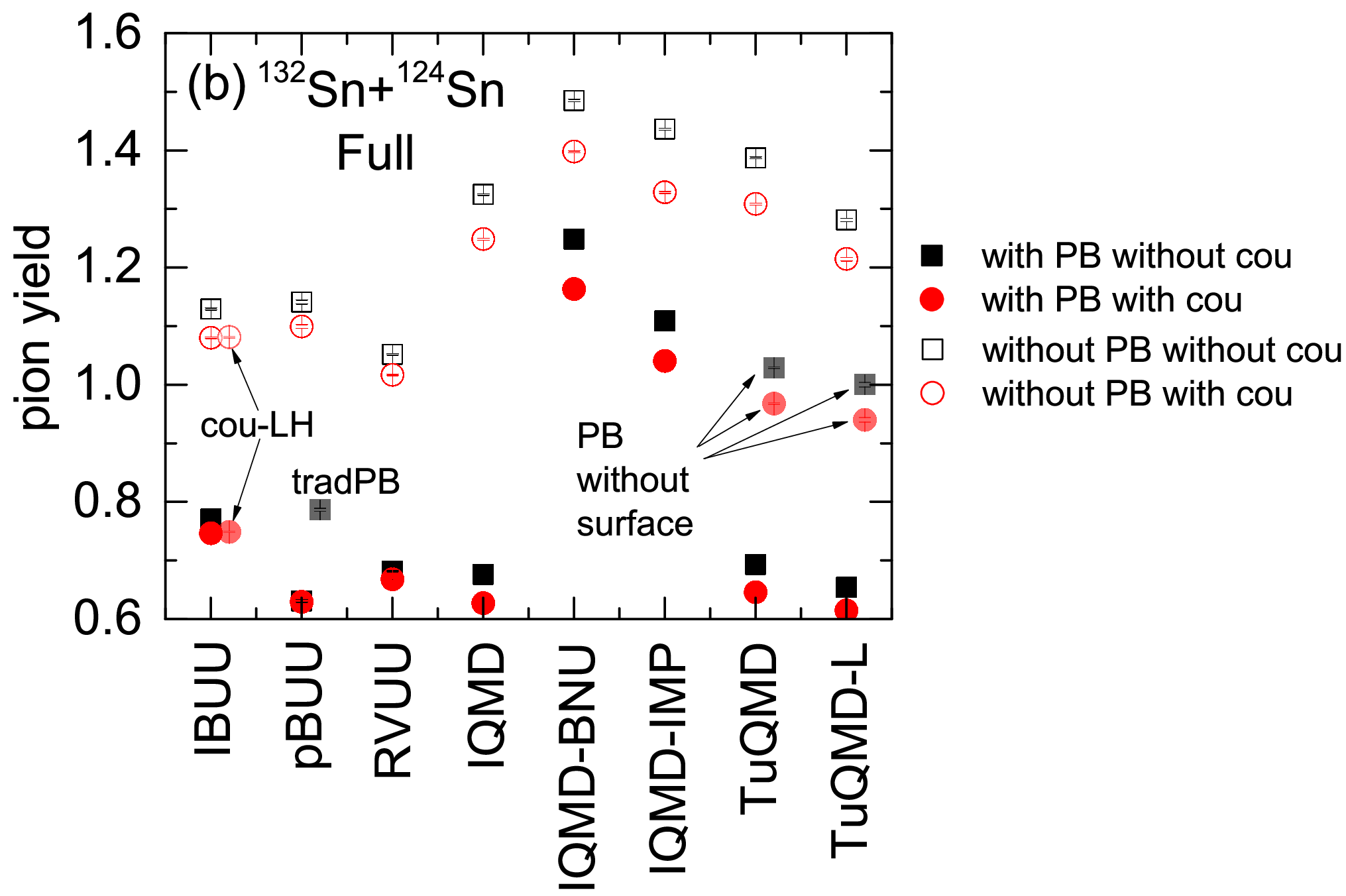}
\caption{(Color online) Final pion multiplicities in different scenarios (see legend and text).
} \label{pimul}
\end{figure}

For a more quantitative view, the total pion multiplicities are compared in Fig.~\ref{pimul}, by summing over all charge states and by enforcing the decay of all remaining $\Delta$ resonances at the final time of the calculation, in the left panel for Cascade and in the right panel for Full calculations. The symbol types are the same in both panels: open and closed symbols are used for calculations without and with Pauli blocking, respectively, and black squares and red circles show the results of the calculations without and with the Coulomb force, respectively. The right panel shows symbols for some variants in some of the codes, which are explained below. We first discuss the Coulomb effects for total pion multiplicities, i.e., compare the black and red symbols of the same type. Generally, these effects are small in the Cascade calculations, and are a bit larger in the Full calculations, since the mean-field potential leads to an overall higher density in the simulations. Systematic differences due to the different ways
of calculating the Coulomb potential, as explained in Sec. II C, are not very evident. In the Full calculations, the Coulomb effects are smaller for IBUU and RVUU, which use a cut-off distance in the calculation of the Coulomb force between point charged particles, and larger for most QMD models, which avoid singularities by a smearing of the particle charges from the finite width of the wave packets. The Coulomb effect is particularly small for pBUU which solves the Poisson equation. A lattice Hamiltonian method in IBUU (coul-LH) does not lead to a big difference relative to the standard cut-off method used in this code.

We now discuss the differences in the total pion yields from different codes and the effect of the Pauli blocking on the pion multiplicities, i.e., difference between the closed and open symbols of the same type in Fig.~\ref{pimul}. For the Cascade calculation, the total pion yields are reasonably similar among the codes, since the densities without the mean-field potential are also similar. IBUU, pBUU, IQMD-BNU, and TuQMD give slightly larger pion multiplicities at the final stage, which is consistent with the results in Ref.~\cite{Ono19} where these codes also predicted larger pion multiplicities especially with a finite time step. This has been understood as due to the way the sequence of $NN\leftrightarrow N\Delta$ and $\Delta\leftrightarrow N\pi$ reactions is treated in each time step by different codes. The high pion yield from IQMD-IMP can be related to the already high total yield of pions and $\Delta$ resonances in the box-pion study~\cite{Ono19}. With Pauli blocking for nucleons (full symbols), the final pion multiplicity is suppressed, since the Pauli blocking suppresses more $NN\rightarrow N\Delta$ than $N\Delta\rightarrow NN$. Therefore, one expects that the pion multiplicity should decrease with the increasing Pauli blocking factor, and vice versa, i.e., the pion multiplicities should reflect the behavior of the blocking factors, which were shown in Fig.~\ref{NNND_rate}. While this is generally true, especially for pBUU in which the pion multiplicity is more strongly suppressed by the Pauli blocking, this similarity is weakened by the discrepancies in the attempted collision rate shown in Fig.~\ref{NNND}, which did not similarly exist in Ref.~\cite{Ono19}. The slightly larger pion multiplicities due to the sequence effect in IBUU, IQMD-BNU, IQMD-IMP, and TuQMD remain also with Pauli blocking.

With the inclusion of the mean-field potential, QMD models generally lead to a higher density during the maximum compression stage compared with BUU models. As a result, a larger multiplicity of $\Delta$ resonances around the maximum compression stage in heavy-ion collisions (Fig.~\ref{PID}) and also a larger total multiplicity of pions at the final stage are produced in QMD models [Fig.~\ref{pimul} (b)]. In Fig.~\ref{pimul} (b) we also show the results for TuQMD-L with its more accurate calculation of the non-linear term leading to a lower density in Fig.~\ref{den}. Consistent with this, it gives a lower pion multiplicity than the conventional TuQMD, though not as low as the one for the BUU models. Besides this large separation between the BUU models and the QMD models for the Full modes, we see in Fig.~\ref{pimul} that the relative differences among the codes in the Cascade-mode calculation of the pion multiplicity remain in the Full-mode calculation, particularly with the inclusion of Pauli blocking. pBUU shows a strong suppression due to its more effective Pauli blocking. We show the result of a separate calculation, labelled as ``tradPB", where the traditional method for implementing the Pauli blocking by calculating the final-state occupation probabilities is used. The Pauli blocking from this method is less effective, leading to an increased pion multiplicity, with the magnitude similar to that in IBUU and RVUU. In IQMD and TuQMD, the surface correction to the Pauli blocking discussed in relation to Figs.~\ref{NNND} and \ref{NNND_rate} leads to a strong suppression of the pion multiplicity, such that the pion yields of these two codes are comparable to those from BUU models. It appears that the surface correction in QMD models compensates the coarser representation of the phase space relative to BUU models. Here we show the results of separate calculations for TuQMD without a surface correction labelled ``PB without surface", and these calculations lead to a considerably increased pion multiplicity, with the magnitude comparable to that in other QMD models.

\begin{figure}[ht]
\includegraphics[scale=0.45]{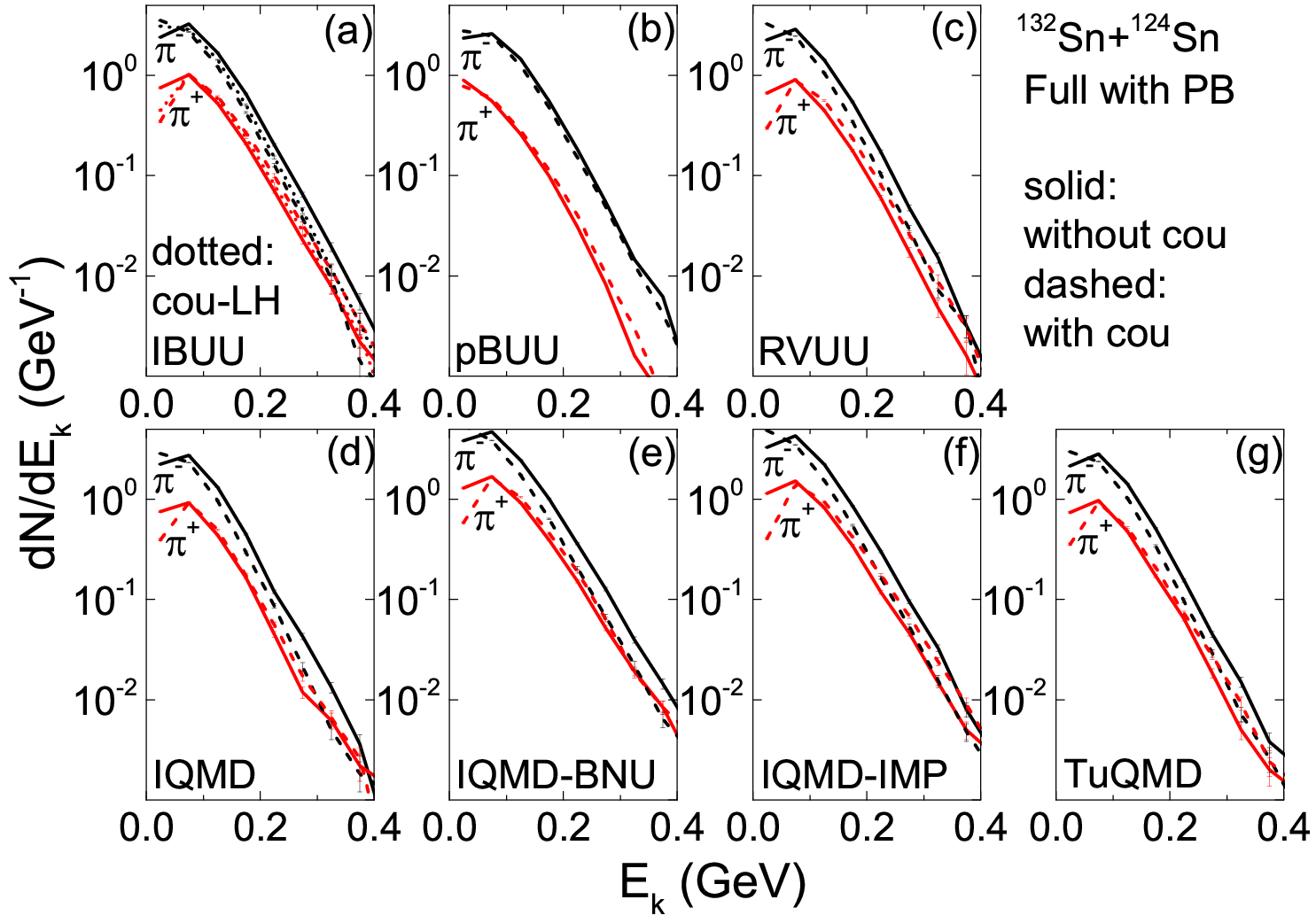}
\caption{(Color online) Pion kinetic spectra in the Full modes with Pauli blocking.
} \label{pike}
\end{figure}

In Fig.~\ref{pike}, we display the pion kinetic spectra from the Full-mode calculation with Pauli blocking, separately for $\pi^-$ and $\pi^+$. The spectra have a plateau at low energies and an exponential tail at higher energies. Thus the total pion multiplicities are dominated by low-energy pions. For the neutron-rich collision system, the $\pi^-$ are naturally more abundant. It is interesting to see that the Coulomb potential has a strong effect on the kinetic spectra of $\pi^-$ and $\pi^+$. Due to the attractive (repulsive) potential from the Coulomb force on $\pi^-$ ($\pi^+$), they are pulled into (pushed out of) the low-energy region, and their kinetic energy spectrum is softened (stiffened). In pBUU the separation between the $\pi^-$ and $\pi^+$ energy spectra is larger than that in other codes, due to the s-wave isovector potentials for pions. For IBUU we show separately the result from the lattice version of the Coulomb potential (cou-LH, dotted lines), which gives similar total pion yields as the default cut-off method in Fig.~\ref{pimul}, but is seen here to reduce the Coulomb effect on the kinetic energy spectra of different pion charge states.

\subsection{$\pi^-/\pi^+$ yield ratio}

\begin{figure}[ht]
\includegraphics[scale=0.3]{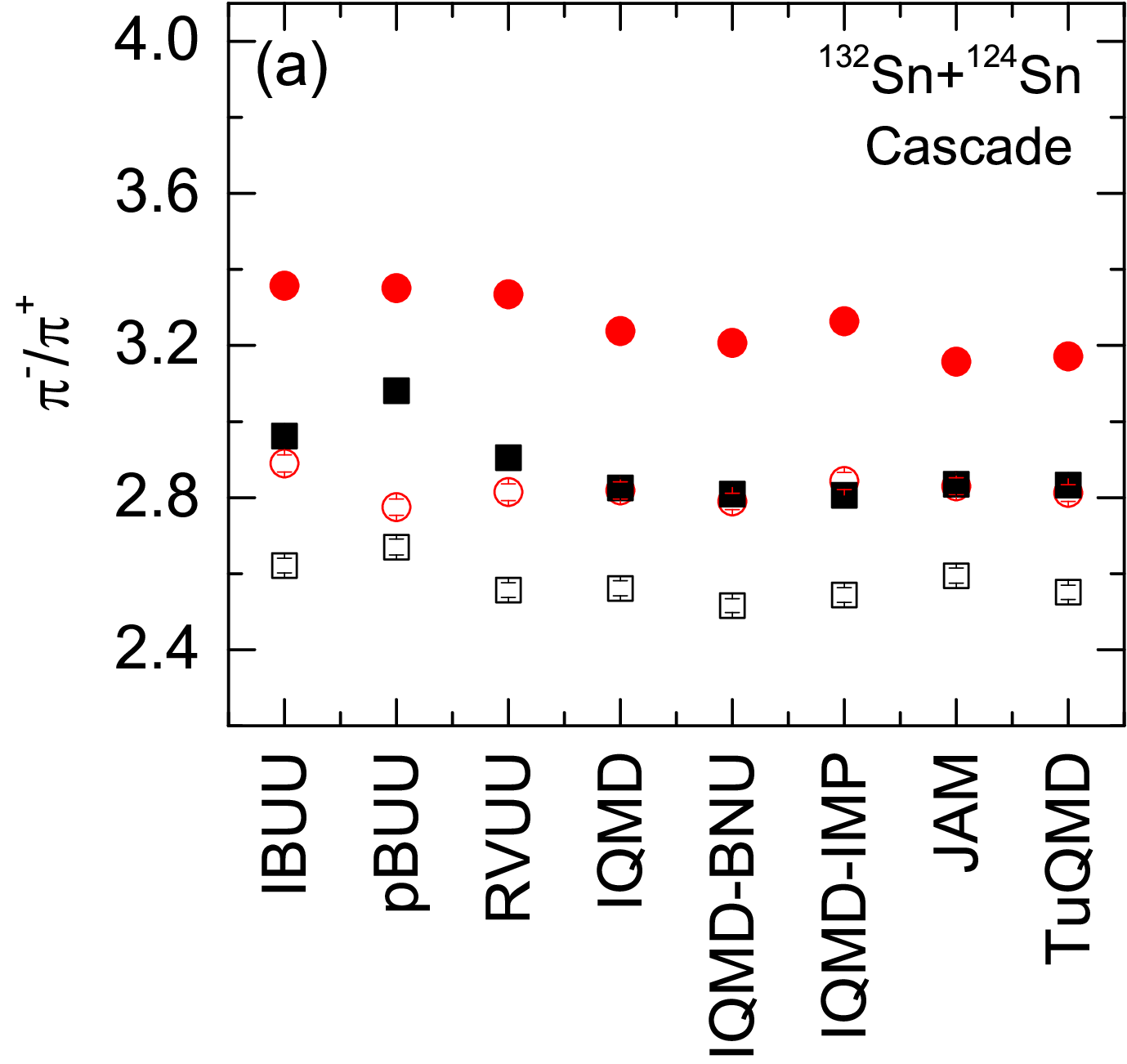}~~\includegraphics[scale=0.3]{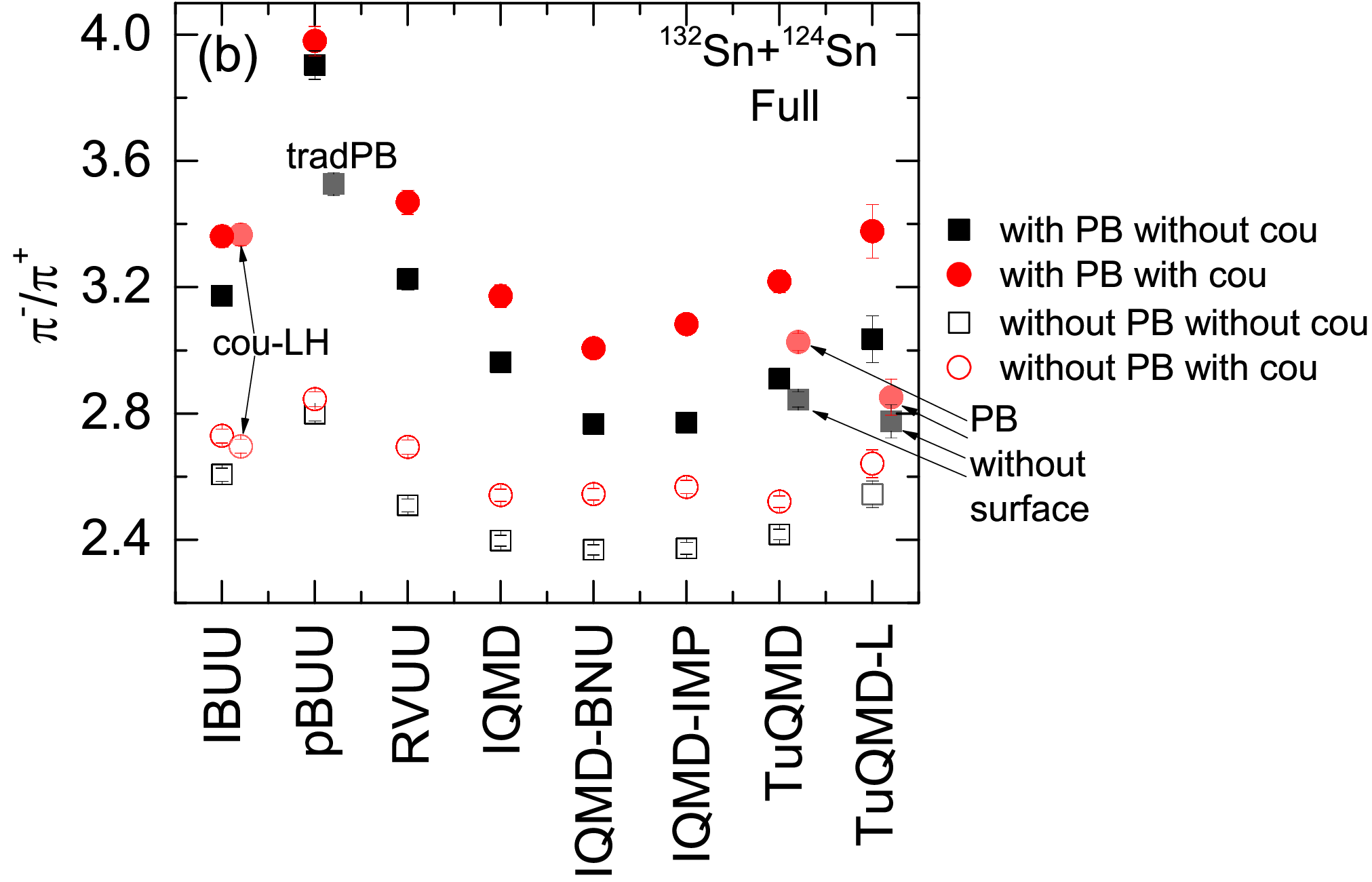}
\caption{(Color online) $\pi^-/\pi^+$ yield ratios in different scenarios (see legend and text).
} \label{ratio}
\end{figure}

The final $\pi^-/\pi^+$ yield ratios from different simulation modes for all participant codes are compared in Fig.~\ref{ratio}, where the symbols have the same meaning as in Fig.~\ref{pimul} for the total pion multiplicities. As in Fig.~\ref{pimul}, all remaining $\Delta$ resonances are forced to decay at the final time. We first discuss the ratios in the Cascade mode in the left panel. The small differences in the $\pi^-/\pi^+$ yield ratio in the Cascade mode without Pauli blocking and Coulomb potential (open black symbols) are consistent with those seen in the corresponding box calculation in Ref.~\cite{Ono19}. Including the Coulomb potential increases the ratios strongly, since, as discussed above, the Coulomb force pushes out the protons and makes the high-density region more neutron-rich. This is also seen in Fig.~\ref{pike} from the behavior of the low-energy pions, which dominate the total pion yields. Regarding IBUU and RVUU, it is seen that a larger cut-off distance ($r_c$) in IBUU and a small cut-off distance together with mixing events in RVUU (see Table \ref{T2}) lead to similar Coulomb effects. For IBUU, we note that the $\pi^-/\pi^+$ yield ratio changes from $3.09 \pm 0.02$ to $2.84 \pm 0.02$ when $r_c$ changes from 0.1 to 2 fm, compared to $2.89 \pm 0.02$ from the default $r_c=1$ fm. The Coulomb effect is smaller in pBUU using a Poisson approach, as was also seen in Fig.~\ref{pimul}. Except for JAM, all QMD models incorporate the Coulomb potential in the standard way (see Table \ref{T2}), and they give similar $\pi^-/\pi^+$ yield ratios. With the use of a width parameter of $\Delta x = 1.41$ fm, they have similar Coulomb effects as IBUU and RVUU with their default cut-off parameters. With isospin-dependent Pauli blocking included for nucleons (solid symbols), the $\pi^-/\pi^+$ yield ratios are significantly increased, consistent with the discussions regarding Fig.~\ref{NNND_rate} (e)-(h), that the $pp\rightarrow n\Delta^{++}$ reaction, which leads to $\pi^+$ production, is more blocked than the $nn\rightarrow p\Delta^-$ reaction, which leads to $\pi^-$ production. We note that generally one expects an anti-correlation between the pion yield and the charged pion yield ratio, since with larger pion yields, differences between the charge states become less important. BUU models with a stronger Pauli blocking generally give a larger $\pi^-/\pi^+$ yield ratio compared with QMD models, and this is particularly true for pBUU with a more effective Pauli blocking. With Pauli blocking included, turning on the Coulomb potential leads to a similar increase of the $\pi^-/\pi^+$ yield ratio. As a whole, the pion ratios are rather consistent among all models in the Cascade mode, with the exception of pBUU which has a rather small Coulomb effect with and without Pauli blocking.

Differences in the $\pi^-/\pi^+$ yield ratio among models become larger after the inclusion of mean-field potentials, as seen in panel (b) of Fig.~\ref{ratio}. A systematically larger $\pi^-/\pi^+$ yield ratio is seen for BUU models compared with QMD models, already in the Full-nopb mode and more pronounced with Pauli blocking included. This is related to the lower total pion multiplicity in BUU models compared with QMD models, as a result of lower densities reached in BUU models. For QMD models, the $\pi^-/\pi^+$ yield ratios are similar among codes that use the traditional method for the mean-field calculation. We show again a lattice calculation with TuQMD as in Fig.~\ref{pimul}, which gives a larger $\pi^-/\pi^+$ yield ratio compared to other QMD models, related to the smaller total pion yield as shown in Fig.~\ref{pimul}\footnote{The TuQMD-L results show a larger error bar of about $6\%$ compared to others, since these were obtained from a time-consuming calculation with a considerably smaller number of events.}. Generally, the $\pi^-/\pi^+$ yield ratios are increased after incorporating the Coulomb potential. With the inclusion of the isospin-dependent Pauli blocking, the $\pi^-/\pi^+$ yield ratios are further increased, especially in pBUU among BUU models and in IQMD and TuQMD among QMD models. The higher $\pi^-/\pi^+$ yield ratios for TuQMD and IQMD are due to the use of the surface correction to the Pauli blocking, which reduces the pion yields as was seen in Fig.~\ref{pimul} and thus increases the $\pi^-/\pi^+$ yield ratios, making them comparable to those from the IBUU and RVUU codes. We also show the results of TuQMD calculations without the surface correction (PB without surface) which have larger pion multiplicities (Fig.~\ref{pimul}) and consequently smaller $\pi^-/\pi^+$ yield ratios. Compared to the traditional TuQMD, the reduction of the $\pi^-/\pi^+$ yield ratio without the surface correction for the Pauli blocking is particularly strong in the TuQMD-L calculation, probably due to the isospin-dependent effect from the surface correction as well as the different neutron and proton densities from TuQMD and TuQMD-L (see Fig.~\ref{den} (d) as an illustration). It is remarkable to see the good agreement between TuQMD-L calculations with the surface correction on the Pauli blocking and the results of IBUU and RVUU (full red circles in the figure). The large charged pion yield ratios for pBUU is related to the fact that in this code pions and $\Delta$ resonances are subject also to the symmetry energy in contrast to all other codes, unlike the setup in the homework specifications. Thus, the results of this code are not really comparable to the other codes. We further show the ``tradPB'' scenario from pBUU using a traditional Pauli blocking approach, which then has a weaker Pauli blocking and thus gives a smaller $\pi^-/\pi^+$ yield ratio, comparable to that in IBUU and RVUU. The stronger Pauli blocking in RVUU relative to IBUU as seen in Fig.~\ref{NNND_rate} leads to a larger increase in the $\pi^-/\pi^+$ yield ratio after incorporating the Pauli blocking. Incorporating the Coulomb potential increases the $\pi^-/\pi^+$ yield ratios from all codes and in all scenarios, as was already seen in the Cascade mode. The Pauli blocking effect on the $\pi^-/\pi^+$ yield ratio is seen to be stronger (especially for TuQMD) after incorporating the Coulomb potential. This is understandable, since the Coulomb potential makes the high-density phase more neutron-rich, and thus enhances the isospin dependence of the Pauli blocking.

\hspace{1cm}
\begin{figure}[ht]
\includegraphics[scale=0.4]{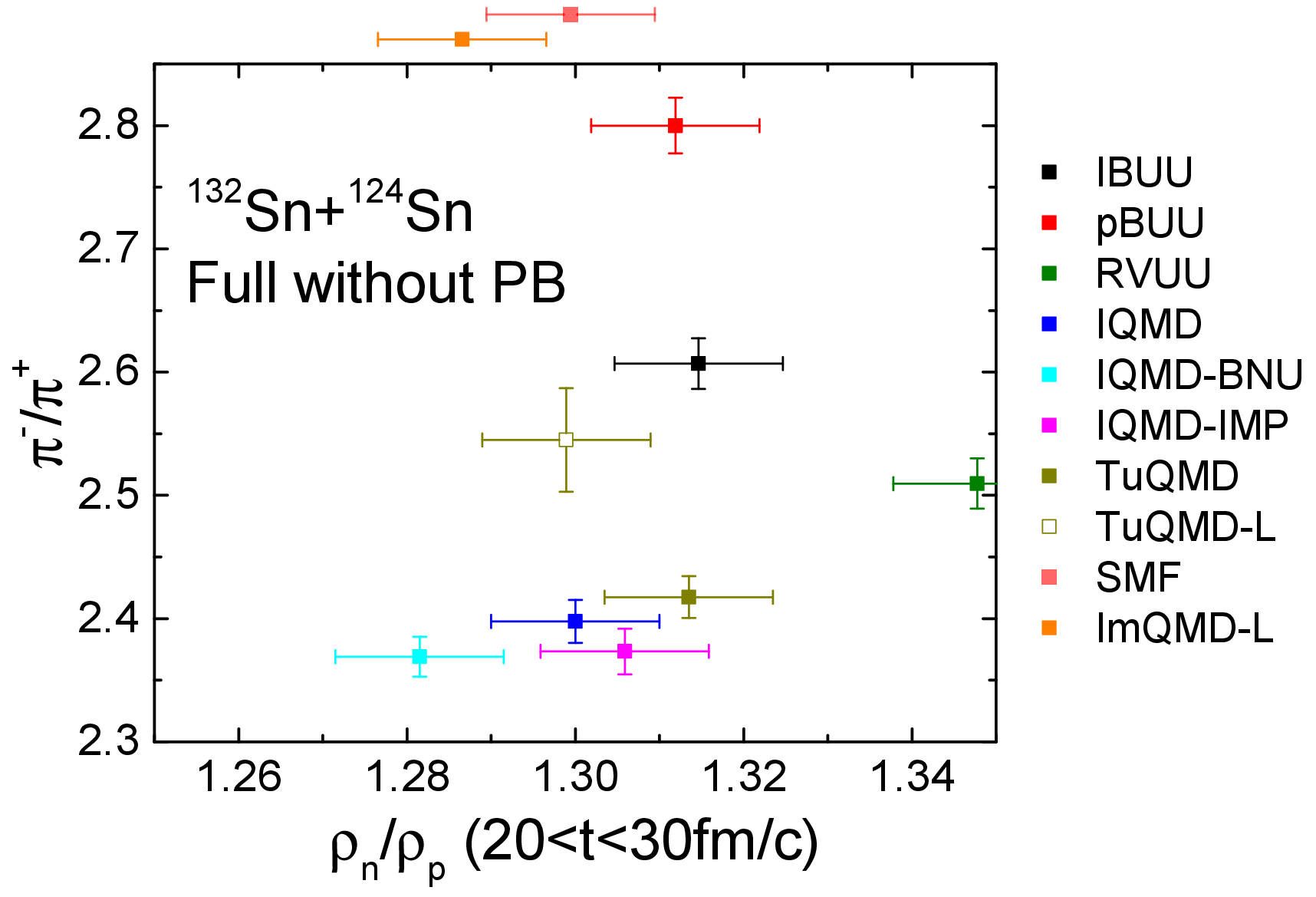}
\caption{(Color online) Correlation between the $\pi^-/\pi^+$ yield ratio and the average asymmetry ratio $\rho_n/\rho_p$ for centroid densities in the central region of $^{132}$Sn+$^{124}$Sn collisions in the time period of $20<t<30$ fm/c in the Full-nopb mode. For SMF and ImQMD-L, which do not calculate pions, the asymmetry ranges are indicated by symbols outside the frame.
} \label{corr}
\end{figure}

The assumption of chemical equilibrium would lead to the relation $\pi^-/\pi^+ \sim (\rho_n/\rho_p)^2$~\cite{Li02a,Li02b}. Besides a lack of strong evidence for chemical equilibrium, there are also other arguments that such relation is not valid for various reasons~\cite{Ike16}. For example, the different treatments of the Pauli blocking may have minor effects on the evolutions of the density and isospin asymmetry, but will affect significantly the $\pi^-/\pi^+$ yield ratio, as already observed in Fig.~\ref{ratio}. To investigate this more quantitatively, we display in Fig.~\ref{corr} the correlation between the $\pi^-/\pi^+$ yield ratio and the central asymmetry ratio $\rho_n/\rho_p$ for centroid densities around the maximum compression stage ($20 < t < 30$ fm/c, see Fig.~\ref{den}) for the Full-nopb mode. For the SMF and ImQMD models, which cannot calculate pions, we only give the asymmetry values indicated by symbols outside the frame. We assume a conservative error bar of $\pm 0.01$ for the central $\rho_n/\rho_p$ to account for the statistical error with 1000 events and the uncertainties of the time and volume for choosing $\rho_n/\rho_p$. Qualitatively, the results from QMD models are close together while BUU models give strongly different results. As was already seen in Fig.~\ref{den}, all codes including SMF and ImQMD-L lead to similar central $\rho_n/\rho_p$, while, as discussed there, RVUU shows a much larger asymmetry. However, the $\pi^-/\pi^+$ yield ratios differ considerably, as was already discussed in Fig.~\ref{ratio}. The different $\pi^-/\pi^+$ yield ratios for IBUU, pBUU, and RVUU are already present in the Cascade mode but the differences are larger in the Full mode. TuQMD-L leads to a comparable $\pi^-/\pi^+$ yield ratio to IBUU and RVUU. Generally it can be seen, that a clear correlation between the asymmetry in the central region of the reaction and the $\pi^-/\pi^+$ yield ratio is not evident, showing that the pion ratio is not only a reflection of the nucleon asymmetry, but also influenced by other aspects of the simulations as discussed above, which include the method of implementing the Pauli blocking, the sequence effect in treating inelastic collisions and decay, and the basic differences between BUU and QMD models.

\begin{figure}[ht]
\includegraphics[scale=0.3]{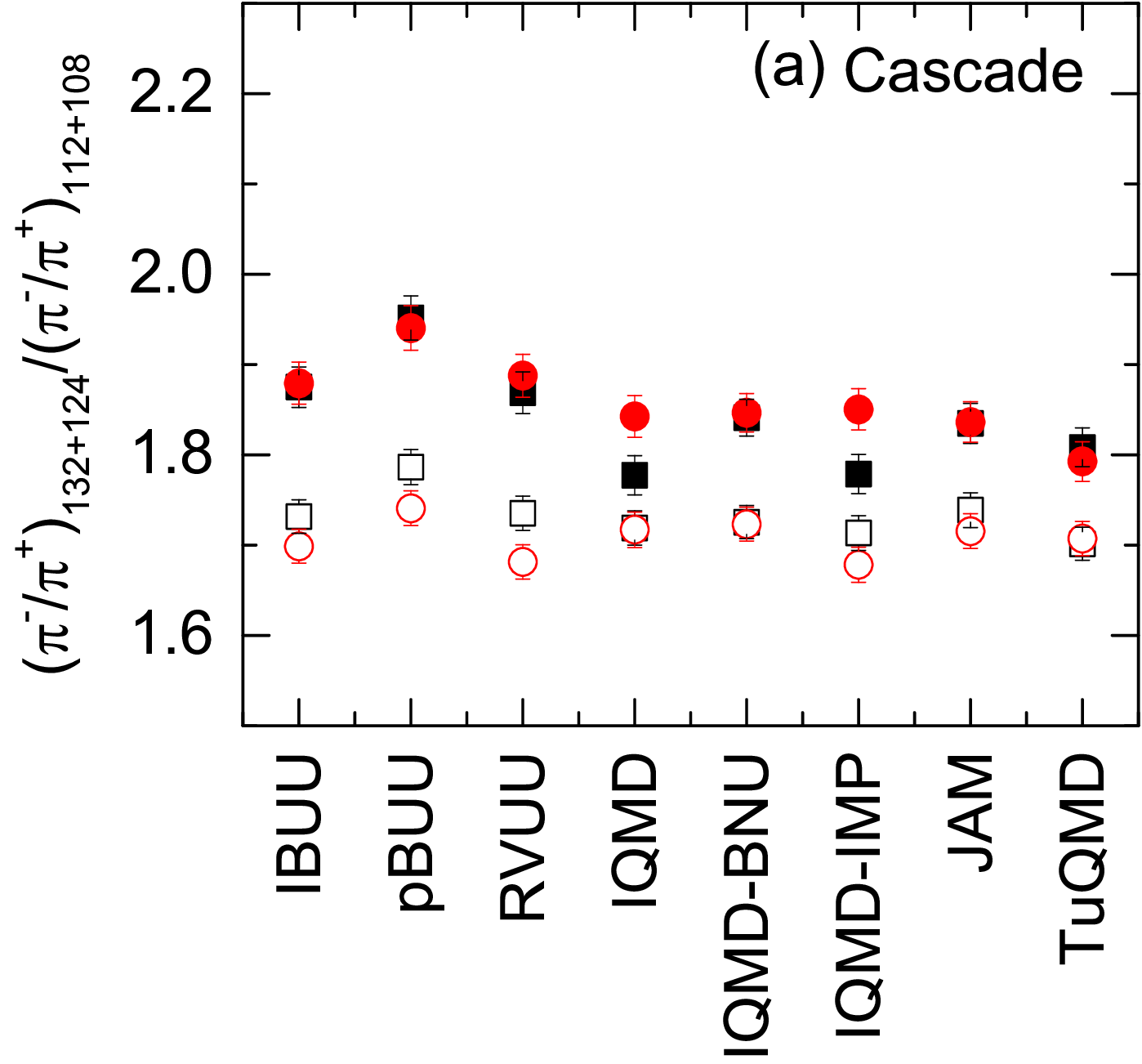}~~\includegraphics[scale=0.3]{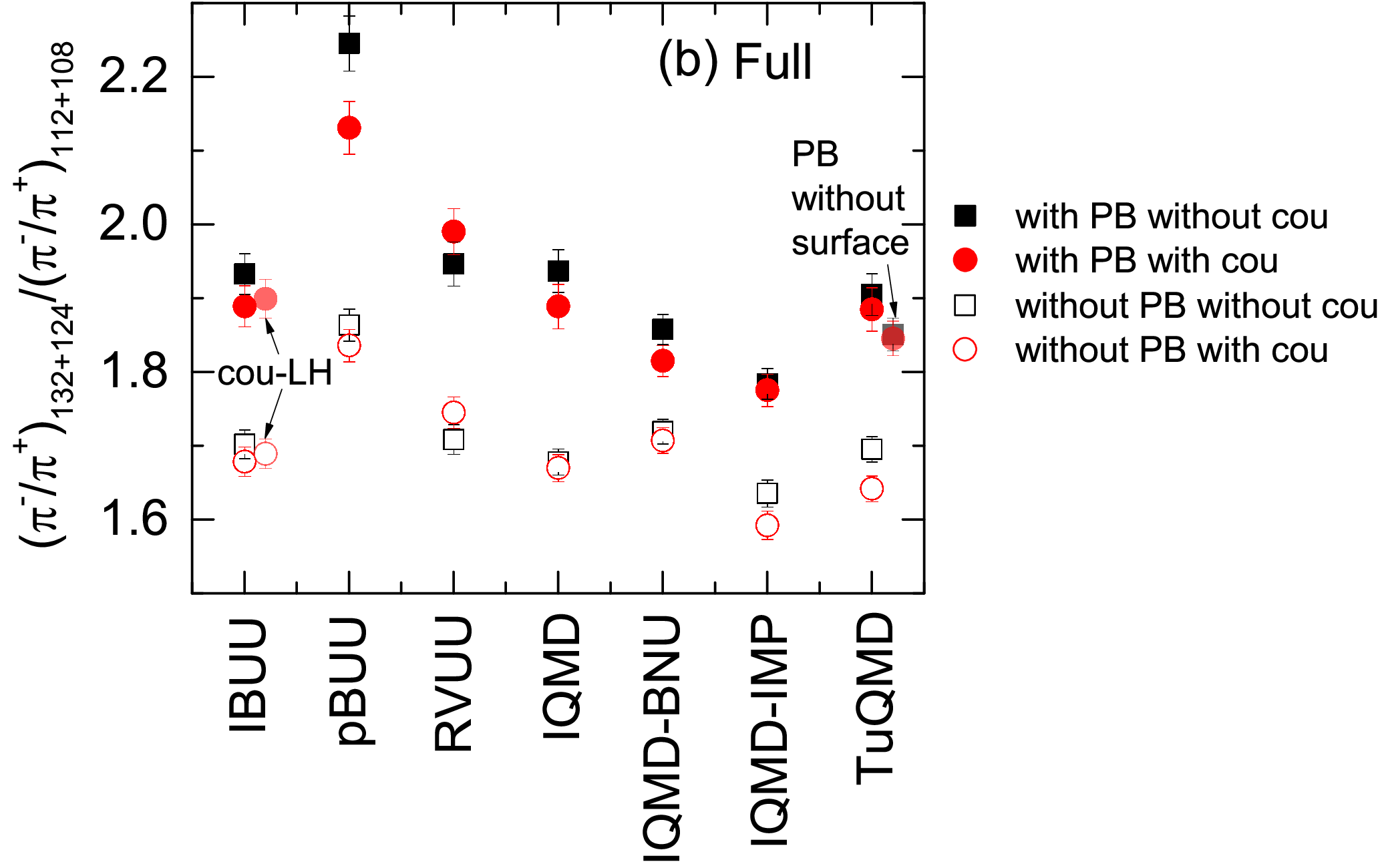}
\caption{(Color online) Double $\pi^-/\pi^+$ yield ratios in different scenarios (see legend and text). } \label{doubleratio}
\end{figure}

In Fig.~\ref{doubleratio} we show the ratios of the $\pi^-/\pi^+$ yield ratios of the neutron-rich system $^{132}$Sn+$^{124}$Sn over the neutron-deficient system $^{112}$Sn+$^{108}$Sn, the so-called double yield ratios, in a similar form as in Fig.~\ref{ratio}. These double ratios are often preferred in experiments, e.g., the double $n/p$ yield ratio, for which the different efficiencies for proton and neutron detection cancel approximately. Comparing Fig.~\ref{doubleratio} with Fig.~\ref{ratio} they are seen to be very similar, except that the Coulomb effect is strongly reduced, particularly in the Cascade mode. The approximate cancellation of the Coulomb effect is due to the identical charge in the two systems, and the differences in the procedure to calculate the Coulomb force also become less important. However, the cancellation is not exact, especially in the Full-mode calculation, where the charge densities evolve differently due to the different mean-field potentials in the two systems. Since the asymmetry effect, which leads to unequal multiplicities of $\pi^-$ and $\pi^+$, for the neutron-deficient system are considerably smaller, the double ratio is dominated by the asymmetry effect of the neutron-rich system, which explains why the behavior in Figs.~\ref{ratio} and \ref{doubleratio}, disregarding the Coulomb effect, are very similar. Generally, taking the double yield ratio may weaken the asymmetry effect, compared to the single yield ratio in the neutron-rich system, making the double yield ratio a less sensitive isovector probe.

\begin{figure}[ht]
\includegraphics[scale=0.3]{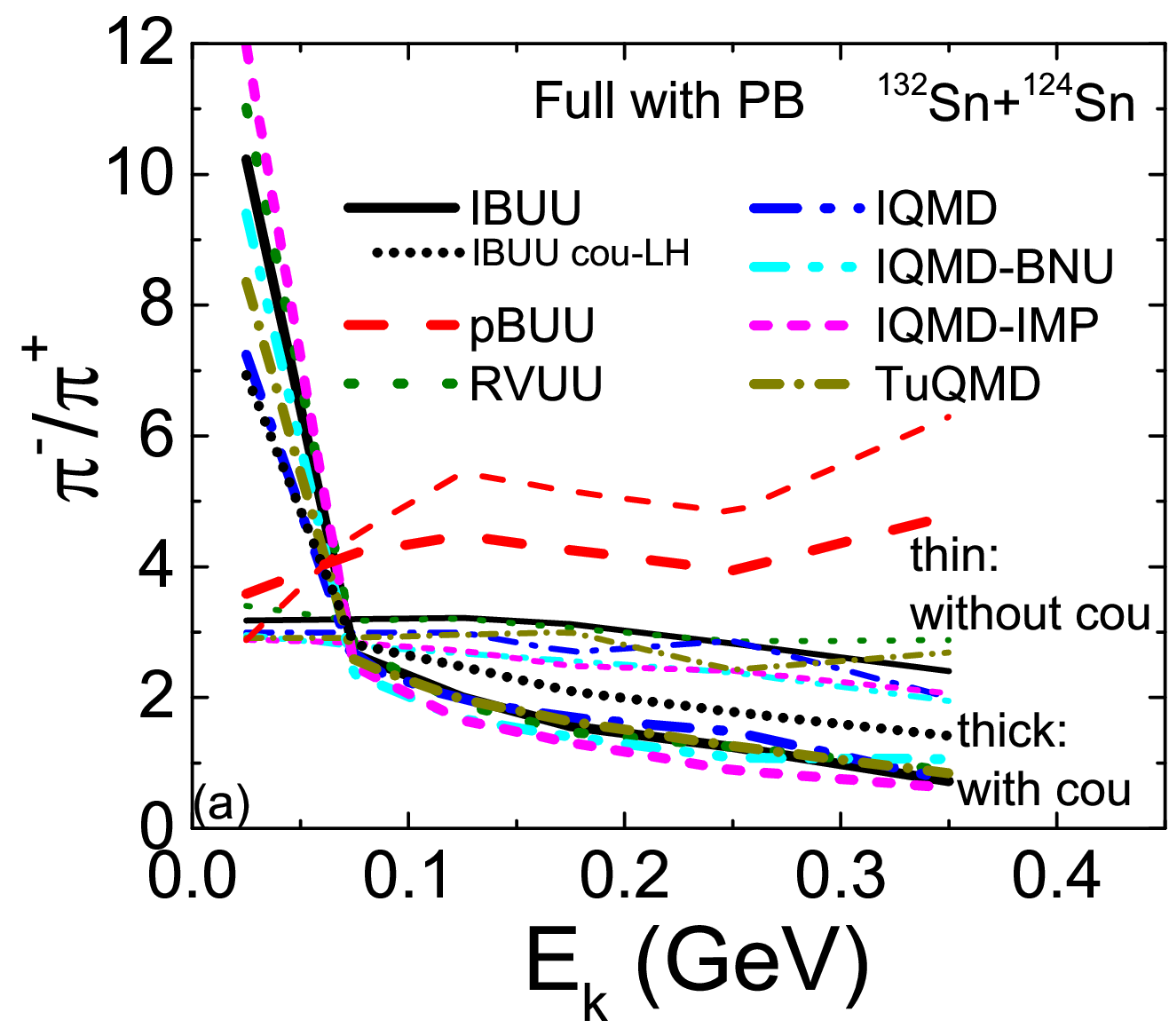}\includegraphics[scale=0.3]{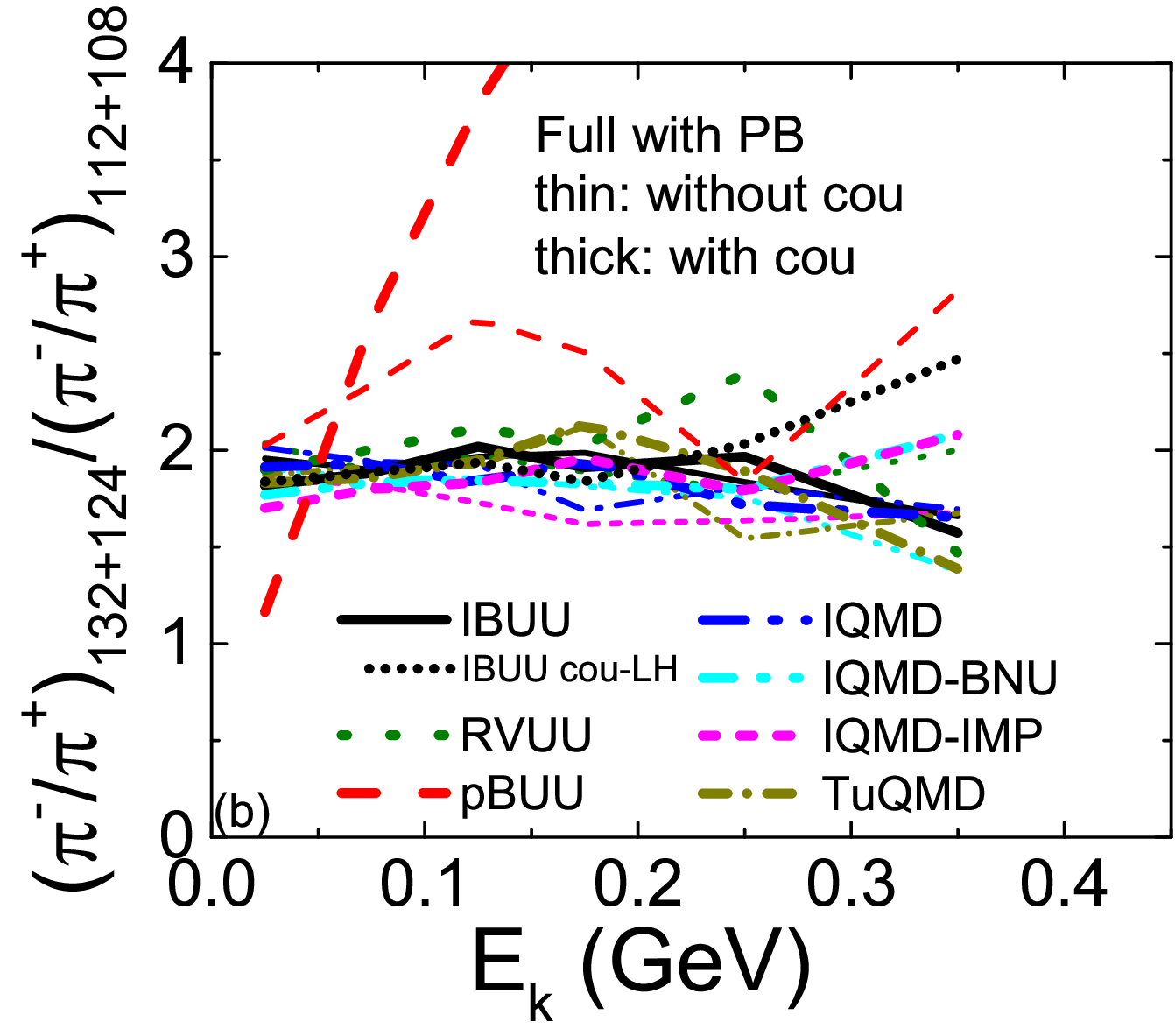}
\caption{(Color online) Left: $\pi^-/\pi^+$ kinetic spectra in the Full modes with Pauli blocking; Right: Double $\pi^-/\pi^+$ kinetic spectra ratio in the Full modes with Pauli blocking. } \label{pikeratio}
\end{figure}

We return to a more detailed discussion of the $\pi^-/\pi^+$ yield ratios. Taking the ratio of the $\pi^-$ and $\pi^+$ kinetic energy spectra in Fig.~\ref{pike} gives the kinetic energy dependence of the $\pi^-/\pi^+$ yield ratio shown in Fig.~\ref{pikeratio} (a). Let us first disregard the unusual behavior of pBUU. The Coulomb potential is seen to affect dramatically the kinetic energy dependence of the $\pi^-/\pi^+$ yield ratio. Without the Coulomb force (thin lines) all the codes collapse rather well into a single behavior, which is rather flat as a function of pion energy. With the Coulomb force, one sees opposing effects for the low- and high-energy pions: a strong enhancement of the ratio for the low-energy pions and similarly a common reduction for pions of energies above about 80 MeV. This was already seen in Fig.~\ref{pike}, and is due to the effect that the Coulomb force attracts $\pi^-$ and shifts them to lower energies, while it repels $\pi^+$ and shifts them to higher energies, in addition to the larger integrated $\pi^-/\pi^+$ yield ratio with Coulomb. The incorporation of the isovector potentials for pion-like particles in pBUU increases significantly the $\pi^-/\pi^+$ yield ratio at high kinetic energies, as already seen in Fig.~\ref{pike}. In the present study, the Coulomb potential for protons and charged pions are turned on and off simultaneously. While it is known that the integrated $\pi^-/\pi^+$ yield ratio is affected by the Coulomb force on protons, Ref.~\cite{Ike16} shows that activating or deactivating the Coulomb force only for charged pions changes the $\pi^-/\pi^+$ kinetic energy spectrum but not the integrated $\pi^-/\pi^+$ yield ratio. Because of the sufficiently small statistical errors for $E_k<300$ MeV, discrepancies in the results among the codes can still be seen, indicating that the results depend on the fine details of the code. Similar to the integrated ratio, taking the double ratio of the $\pi^-/\pi^+$ kinetic spectra in $^{132}$Sn+$^{124}$Sn to $^{112}$Sn+$^{108}$Sn systems largely cancels the Coulomb potential effect, as shown in Fig.~\ref{pikeratio} (b). The double ratio is seen to be nearly independent of the pion kinetic energy for $E_k<300$ MeV. Taking the double ratio generally reduces the difference from different implementation methods as shown in Fig.~\ref{pikeratio} (b), see, e.g., for IBUU the difference between results from the default cut-off method and the cou-LH method. Again, the pBUU code shows a rather different behavior for the differential double ratio, due to the incorporation of the isovector potentials for both $\Delta$ resonances and pions, compared to other codes.

\begin{figure}[ht]
\includegraphics[scale=0.3]{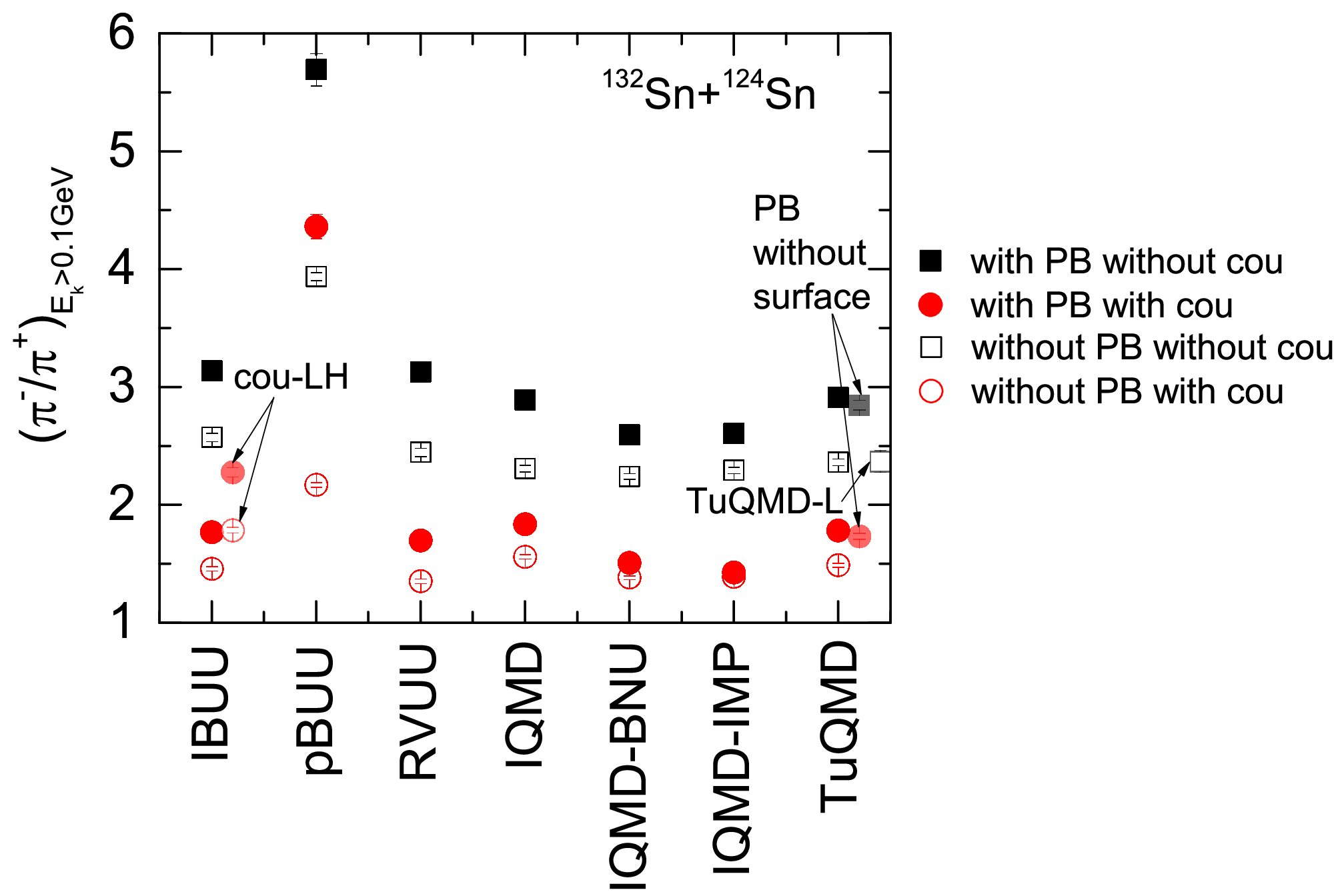}
\caption{(Color online) $\pi^-/\pi^+$ yield ratios for energetic pions ($E_k>0.1$ GeV) from Full-mode calculations.
} \label{ratioh}
\end{figure}

It has been argued that the $\pi^-/\pi^+$ yield ratios for energetic pions are more sensitive probes for extracting information on the symmetry energy at suprasaturation densities, and this was used, e.g., in Refs.~\cite{Li15,Tsa17,Spirit21}. Figure~\ref{ratioh} compares the $\pi^-/\pi^+$ yield ratios for energetic pions with kinetic energy $E_k>0.1$ GeV from Full-mode calculations. Except for pBUU, results without the Coulomb potential are similar to the total $\pi^-/\pi^+$ yield ratios, as can be seen from the weak kinetic energy dependence of the $\pi^-/\pi^+$ yield ratio from Fig.~\ref{pikeratio}. 
After incorporating the Coulomb potential, the $\pi^-/\pi^+$ yield ratio for energetic pions is generally reduced, opposite to the effect of the Coulomb potential on the total $\pi^-/\pi^+$ yield ratio, and this is understandable from the left panel of Fig.~\ref{pikeratio}. While the behaviors of different models are mostly similar to those observed in Fig.~\ref{ratio} (b) for the total $\pi^-/\pi^+$ yield ratio, one sees that the Coulomb potential effect is strongly enhanced. It is thus important to implement the Coulomb potential properly in order to extract accurate information on the symmetry energy from the $\pi^-/\pi^+$ yield ratios for energetic pions. We note the difference in vertical scales in this figure and in Fig.~\ref{ratio}. Since the pion ratios for the energetic pions are absolutely smaller than those for the total yields, the relative differences among the codes are actually even larger for this selection.

\section{Discussion}
\label{discussion}

In this study we compared different transport models under controlled conditions for intermediate-energy heavy-ion collisions to study the production of pions of different charge states. This comparison can be viewed as a particular stage of the TMEP activities: In 2016 we compared different transport models in similar heavy-ion collisions as the present one~\cite{Xu16}, and found considerable discrepancies, which were not easy to disentangle. Then, in different box calculations we investigated separately and under very transparent conditions the different ingredients of transport simulations: the mean-field propagation~\cite{Mar21}, elastic collisions as well as the Pauli blocking in the collision term~\cite{Zha18}, and inelastic collisions~\cite{Ono19}. While we did not see complete convergence of the codes in these studies, we could identify the reasons behind the differences due to different strategies of the simulations, and could eliminate some of the problems. In the present comparison we come back to investigate the results in full heavy-ion collisions. On the other hand, the present study can also be related to a recent TMEP comparison~\cite{Spirit20} of predictions of pion observables for the recent S$\pi$RIT experiment. In this comparison the physics inputs were not prescribed, and the predictions were requested before the release of the experimental data. It was found that the predictions of the codes differed from each other and from the experiment in such a way that a constraint on the symmetry energy could not be determined from pions. Thus the present study with controlled conditions for the same experimental system is also motivated by the attempt to better understand these differences, neglecting in this comparison some more realistic but also more complicated physics inputs, e.g., the momentum-dependent mean-field potential, pion in-medium effects, etc., which will certainly contribute to the discrepancies in Ref.~\cite{Spirit20}. The discussions in the previous sections have shown for the present comparison that the results of different codes still show considerable differences among each other. In the above, we discussed in considerable detail the different results and the possible explanations for the differences, in the light of what was understood from the box calculations. In this section we will critically discuss these differences from a broader perspective and try to derive some lessons from them.

We discussed first the nucleonic evolution, as it manifests itself in non-observables, like the evolution of the densities or the time and energy dependence of the collision rates, and in observables, such as the rapidity distribution and the sideward flow. If the evolution of the nucleon phase space already differs, we cannot expect pion observables to converge. The opposite does not necessarily hold, i.e., even if the bulk evolution were similar, the pion observables may still differ to a larger degree, especially for the pion charge states, since they may depend sensitively on subtle differences in the proton-neutron evolution. One has to keep in mind that in the open system of a heavy-ion collision, small differences in the evolution may propagate, amplify, and lead to large effects in final observables. This is in contrast to box calculations, where the system is effectively closed, and some quantities such as the number and isospin densities stay constant on the average. In this respect, the dependence of the results on the initialization of the colliding system may also become important as was seen in the first heavy-ion comparison~\cite{Xu16}. Here we tried to minimize this effects and provided a common initialization of the colliding nuclei. 

In the present comparison, as in all the previous ones, participating models are of the Boltzmann-Vlasov (BUU) or of the quantum-molecular (QMD) type. We discussed previously that the two types of models differ basically in the treatment of fluctuations, such that one cannot expect complete convergence of their results. It was found in box calculations that fluctuations have several effects in transport simulations. They influence the mean-field propagation, since they lead, on average, to stronger density gradients and thus to stronger forces~\cite{Mar21}, and stronger dissipations as well. They also affect the final-state blocking in the collision term, since the fluctuating phase-space density effectively reduces Pauli blocking~\cite{Zha18}, and thus increases the successful collision rate, enhances the dissipation, and affects particle production. However, in the present comparison the bombarding energy is considerably higher than the Fermi energy and thus the Pauli blocking, at least in primary elastic collisions, is not dominating for nucleonic observables. Thus, e.g., in the Cascade calculations of Fig.~\ref{dNdypx_Cascade}, there is no large difference seen between QMD and BUU models, and similarly in Fig.~\ref{NNND_rate} the blocking factors are similar and rather small. There are systematic differences seen in mean-field effects between QMD and BUU, which are also due to the different amounts of fluctuations, but mainly in an indirect way.

It was seen in Fig.~\ref{den} that the densities reached in QMD calculations are systematically higher than those in BUU models, similarly the inelastic NN collision rates are systematically higher, and this propagated into differences in pion production in the two approaches. This was traced back to the treatment of the non-linear term in the energy-density functional, and the fact that $\overline {\rho^\gamma}$ is not equal to $\overline{\rho}^\gamma$ for $\gamma>1$. The difference increases with the amount of fluctuations of the density. BUU calculations using point TPs do not have this problem, and for finite-size TPs this is treated consistently in the lattice-Hamiltonian method. On the other hand, the approximation is used in standard QMD models, which weakens the repulsive force and leads to stronger compression as seen in Fig.~\ref{den} (a) and (c). It was also shown that for QMD calculations without making this approximation (called lattice-QMD calculations here) the agreement with BUU calculations was better. Thus this difference can be avoided at a considerably larger numerical cost. Different calculations of the non-linear term in a stiff symmetry energy may also lead to different neutron-proton asymmetries, and can be investigated in further studies.

We have seen differences among BUU codes which we have traced to the use of different shapes/sizes of TPs, from point TPs (RVUU) to finite-size TPs of different shapes  (IBUU, pBUU, and SMF). With finite-size TPs, the lattice-Hamiltonian method is the proper formulation, which takes into account the density variations due to the shape of TPs. It was seen, e.g., in Fig.~\ref{dNdypx_Full} (b), that this does have some influence on the transverse flow. The width of the wave packets in QMD models, even without the effects for a non-linear term, averages out the forces according to Eq.~(\ref{qmddp}), so smaller wave packets generally lead to larger flows as seen in Fig.~\ref{dNdypx_Full} (d).

Different strategies are used for the collision probabilities and the Pauli blocking, as discussed in Sec.~\ref{pb}. In contrast to the usual geometric method, pBUU and SMF use a stochastic method to determine collision probabilities. As mentioned above, while the effect of the Pauli blocking on nucleonic bulk observables is not large, the Pauli blocking has considerable effects on pion observables. To reduce the effect of fluctuations in the calculation of the phase-space occupation, pBUU uses a local statistic method for the Pauli blocking, which leads to a more effective blocking of collisions, as is seen in Fig.~\ref{NNND_rate} (a) and (b). In box calculations this led to a good agreement with analytical results~\cite{Zha18}. TuQMD and IQMD employ a surface correction to the Pauli blocking in order to compensate for the non-uniform occupation of the final-state phase-space distribution, which leads to a much stronger blocking relative to the usual procedures [see also Fig.~\ref{NNND_rate} (b)]. The validity of different strategies to implement the Pauli blocking in the collision term has to be investigated further.

Now we turn to the influence of these differences in the nucleonic evolution on pion observables. A dominating pattern is the dependence of the pion multiplicities on the density reached in the nucleonic evolution. This explained the large differences between BUU and QMD models in inelastic collision rates in Fig.~\ref{NNND} [(e)-(h)], in total pion multiplicities in Fig.~\ref{NNND} (b), and in charged pion yield ratios in Fig.~\ref{ratio} (b). In order to increase the sensitivity to the symmetry energy, one commonly investigates differences in the production of pions of different charge states, and, in particular, ratios of charged pion observables to partially cancel some effects of different evolutions and of experimental uncertainties. In these ratios, small effects which are hardly seen in the bulk evolution can become much more visible. Generally, larger total pion multiplicities for similar asymmetries are expected to lead to smaller charged pion yield ratios, as is indeed observed in these comparisons. But there are more subtle effects, which are due to the increased sensitivity to the isospin-dependent Pauli blocking, the effect of using different strategies in the sequence of simulating inelastic collisions and $\Delta$ decays, and in higher-order correlations in inelastic processes~\cite{Ono19}. All these had to be considered to understand the different results in Figs.~\ref{pimul} and \ref{ratio}.

In this comparison we took particular care to investigate the effects of the Coulomb force on $\pi^-/\pi^+$ yield ratios. The Coulomb force is treated by various methods and approximations in different codes as discussed in Sec. II C. Of course, the Coulomb force has an influence on the bulk evolution of the system, i.e., it reduces slightly the maximum density in heavy-ion collisions and makes the compressed region more neutron-rich~\cite{Sto22}, and thus reduces slightly the overall pion yield and increases considerably the total $\pi^-/\pi^+$ yield ratio. The effect on the spectral ratios, however, is more involved, since the Coulomb force attracts $\pi^-$ and repels $\pi^+$. Therefore, the yield ratios increase for low-energy pions and decreases for higher-energy pions. As was discussed in connection with Fig.~\ref{NNND_rate}, these effects depend sensitively on the way the Pauli blocking acts in reactions in which $\Delta$ resonances or pions of different charges are produced. This leads to some differences among the various methods to implement the Coulomb force on the pion yield ratios, as seen in Fig.~\ref{ratio} (b). While Coulomb effects are not mainly responsible for the differences in the results in Fig.~\ref{ratio}, as seen by the largely similar behavior between the results from codes with (red symbols) and without (black symbols) Coulomb, they add to the discrepancies. To take double ratios between two different collision systems with the same charge numbers reduces strongly, but does not completely cancel, the Coulomb effect, as seen in Fig.~\ref{doubleratio} for the yield ratio and in Fig.~\ref{pikeratio} (b) for the spectra. In particular, it does not reduce the differences among the codes due to other effects, as seen in Fig.~\ref{doubleratio}.

Pions have been investigated in heavy-ion collisions, since they are considered to be sensitive probes to the symmetry energy~\cite{Li02a,Li02b}. The primary effect of the symmetry energy is the difference in the evolution of neutrons and protons, and pions should reflect these differences. In transport simulations, this expectation can be studied more quantitatively by considering the correlation between the nucleonic isospin asymmetry effect of the dense region and the final $\pi^-/\pi^+$ yield ratio. This correlation is shown in Fig.~\ref{corr} for the case without Pauli blocking or Coulomb force. The correlation among different codes is overall not very strong, while, of course, it may be present for a single code using different symmetry energies. One could argue from Fig.~\ref{corr} that for the QMD models there is a weak correlation, but this could be due to the fact that these codes are rather closely related, since they are derived from a common original code. The BUU models show large differences, which indicates that they also differ considerably in the different strategies adopted in the simulation. This demonstrates that the charged pion yield ratios are not just a reflection of the neutron-proton asymmetries, but are also strongly affected by the chosen strategies in the simulation.

Summarizing these discussions, we find that we do not achieve a complete convergence of the results, not even for the nucleonic evolution. The pionic observables also show considerable differences, not only because of the different bulk evolutions, but also because some of these effects are magnified in pion observables. We discuss this for the charged pion yield ratio in Fig.~\ref{ratio}, which is a very important observable, in particular for the determination of the symmetry energy. As discussed in Sec.~\ref{homework}, the results for the pBUU code, which differ most from the other codes, are really not comparable, since the code has some simulation features, which are rather different from the other codes, and, most importantly, uses ingredients which are different from the homework specifications. E.g., pions and $\Delta$ resonances feel the strong symmetry force in pBUU, while in the other codes they are treated as free particles only subject to the Coulomb force. Thus it should be left out of a final evaluation, which, of course, does not mean that the results are not valid. Then the mean and standard deviation of the charged pion yield ratio of the codes is $3.24\pm 0.17$, which amounts to an uncertainty of about $5.2\%$ in the complete calculation with Pauli blocking and the Coulomb force. The uncertainty is reduced to about $3.2\%$ if the Pauli blocking is artificially turned off, showing that the treatment of the Pauli blocking is a sensitive issue in simulations. Qualitatively, these differences are not reduced in double-ratio type of observables, and are not much changed when considering only high-energy pions as in Fig.~\ref{ratioh}. These differences among the codes would have to be compared with the sensitivity to different forms of the symmetry energy which was not investigated here. But from the results of the (uncontrolled) comparison in Ref.~\cite{Spirit20}, we see that the sensitivity is of the same order. Turning the argument around, it would mean that from the same measured pion yield ratio different transport codes could derive symmetry energies that may differ by that amount.

However, we can evaluate the results presented in Fig.~\ref{ratio} in more detail. As stated above we should not consider the results of the pBUU code. We can then discuss the remaining differences with respect to two issues which are treated differently in the codes. One is the surface correction to the Pauli blocking, used in some QMD codes (IQMD and TuQMD) but not in the others. The other is the approximate calculation of the non-linear term in the mean-field potential used in all traditional QMD codes. We can estimate the first effect by comparing the TuQMD calculations shown in Fig.~\ref{ratio} without and with surface correction, which amounts to an increase of the charged pion yield ratio by about $6.3\%$. If we increase the results of IQMD-BNU and IQMD-IMP by this amount, then a first group of codes including IQMD, IQMD-BNU, IQMD-IMP, and TuQMD have a mean and deviation of $3.22\pm0.05$ in the $\pi^-/\pi^+$ yield ratio, which is an uncertainty of about $1.4\%$. Such a scaling from the effect of the surface correction appears reasonable, as may be seen when looking at the results of these conventional QMD codes without Pauli blocking (and with Coulomb), and they agree without any scaling at $2.54\pm0.02$, with a variation of less than $1\%$. A second group includes two BUU-type codes, i.e., IBUU and RVUU, and TuQMD-L, where the latter has been shown to give similar $\pi^-/\pi^+$ yield ratio as BUU codes. For these three codes we find an agreement of the ratio of $3.40\pm0.06$, with a similar uncertainty of about $1.7\%$, by neglecting the statistical uncertainty of the TuQMD-L result. The two groups converge among themselves rather well, and do not overlap within their variances. We find a difference of about 0.18 between the two groups which relative to the mean $\pi^-/\pi^+$ yield ratio amounts to about $5.6\%$. From a similar estimate for the case with no Pauli blocking, we find a difference between the two groups of 0.14, with again an uncertainty of about $5.5\%$ of the mean.

We can also attempt to understand better this difference between the two groups from an estimate of the effect from different calculations of the non-linear term on the $\pi^-/\pi^+$ yield ratio from the difference between the TuQMD and TuQMD-L results. For the case with Pauli blocking and Coulomb it amounts to 0.16 (neglecting the statistical error $\pm0.06$ of the TuQMD-L result) or $4.8\%$. This agrees rather well with the difference (0.18) between the average results of two groups determined above. If we scale the conventional QMD calculations by these $4.8\%$, then we find for all the 6 codes a mean and deviation of $3.39 \pm 0.06$, or an uncertainty of $1.6\%$. For the case of calculations without Pauli blocking and with Coulomb, the corresponding numbers are 0.14 for the difference in the two groups and 0.12 for the difference between TuQMD and TuQMD-L, which correspond to a similar accuracy taking into account the lower absolute value of the $\pi^-/\pi^+$ yield ratio for this case. As a result of these discussions we find that we can explain very well the remaining differences in the results in Fig.~\ref{ratio} (b). One may argue that an agreement below a level of $2\%$ among all models can be achieved, as long as the same (or similar) ingredients, i.e., the surface correction to the Pauli blocking and the accurate calculation of the non-linear term, are included in each code.

\section{Conclusion and Outlook}
\label{summary}

In the present study, we compared simulations for typical heavy-ion collisions studied in the recent S$\pi$RIT experiment. These collisions reach a maximum density of about twice saturation density, and thus could be useful to constrain the nuclear matter EOS, in particular the symmetry energy, in the range of densities between those accessible to nuclear structure and those connected to neutron stars as well as their mergers. We included the production of pions and $\Delta$ resonances, since pions are expected to be messengers of the high-density symmetry energy. For this comparison, we used a simple physics model for the mean-field potential and the elastic and inelastic cross sections. 
The purpose of the present comparison was to investigate the status of simulations for heavy-ion collisions after extensive studies on the separate ingredients of transport calculations had been made in box simulations. Relative to previous comparisons of different codes in low- and intermediate-energy heavy-ion collisions, here we unify the collision setups and the physics model as much as possible and follow the time evolution of the system in detail. Thus, we focus on the uncertainties resulting from different implementations of transport codes rather than the effect of different assumptions of the physics model.

It is not easy to give a quantitative estimate of the differences among the codes in the charged pion yield ratio, which is the main observable used for constraining the EOS in these types of experiments. However, we tried to do so by estimating carefully the effects which result from systematic differences in the codes. Leaving out the pBUU code which is rather different in many respects, while the agreement amounts to about $5\%$ for the complete calculation, we expect that it can be reasonably reduced to $1.6\%$, once the same or similar ingredients, i.e., an improved Pauli blocking and calculation of the non-linear term, are included in each code. This may bring the accuracy of transport simulations into a range which would be sufficient for a determination of the EOS, especially the density dependence of the symmetry energy. We think that this is a notable achievement of the present comparison. Of course, unlike in box simulations, in this comparison it is difficult to make a statement on which code is more realistic.

These differences result from different strategies used in the solution of the transport equations by simulations, which are not fixed by the equations themselves. These strategies were partially checked in the box calculations, where exact analytic or numerical solutions are sometimes available. While box calculations are useful in calibrating each component of transport simulation, the effects of these different strategies are usually enhanced in the open system of a heavy-ion collision, which has a non-equilibrium dynamical evolution. As a positive result of this comparison, we have identified those strategies which influence pion observables sensitively, and should therefore be well controlled. Clearly, pion observables are sensitive to the nucleonic evolution, like the densities and isospin asymmetries reached. Thus the mean-field potentials should be precisely implemented, and observables such as momentum distributions should be controlled. The collision term and the NN collision rates influence directly $\Delta$ and pion production. The method chosen for implementing the Pauli blocking in the collision term is found to have a large effect on pion observables, via not only the nucleon evolution, but also the blocking of NN inelastic collisions and $\Delta$ decay processes. The representation of the phase space and the coarse-graining procedures also have a large influence, since they determine the fluctuations in the system. In BUU with finite TP sizes, the lattice-Hamiltonian method should be used, but it also contains parameters that affect the fluctuations. The Coulomb potential, of course, has to be included, and the particular method of implementation does not seem to be the main source of discrepancies in the charged pion yield ratios compared to other effects.

There are, however, open problems in transport simulations, where the proper treatment still has to be clarified. A major question is the correct amount of fluctuations in transport calculations, which, as discussed, is different in QMD- and BUU-type approaches. In the QMD approach, the width parameter of the wave packet gauges the magnitude of fluctuations, and is only roughly constrained by considerations of nuclear structure~\cite{Dan17,Yan21} or properties of the nuclear interaction. In BUU, the more exact is the solution of the BUU equation (e.g., with more TPs), the weaker are the fluctuations, and they should be included via a physically motivated fluctuation term as in the Boltzmann-Langevin approach. However, codes that include these are rather few and have to make approximations. Actually, it would be more meaningful to compare QMD with Boltzmann-Langevin calculations. Furthermore the treatment of the Pauli blocking has a large influence. We have seen that equally justifiable treatments based on overlaps of wave packets or TPs, or local statistic methods, lead to considerable differences.

There are important physics issues of heavy-ion collisions and pion production, which are not considered in this comparison. One of them is the consequences of the momentum-dependence of the isoscalar and isovector mean-field potentials, which in inelastic collisions lead to threshold shifts due to energy conservation. This is presently being investigated in a separate box simulation~\cite{Dan22}. Light clusters are produced abundantly in heavy-ion collisions, and the treatment of their dynamic formation in transport simulations, in contrast to a-posteriori coalescence or evaporation approaches, is a question of intensive debate. It has been shown that cluster formation can sensitively affect results including pion observables (see, e.g., Refs.~\cite{Ono19r,Ike16}). Short-range correlations in nuclei, which lead to a high-momentum tail in the momentum distribution of nucleons, have been clearly identified in nuclear structure studies, and they may also have an impact in transport calculations at energies near relevant thresholds. However, how they should be implemented in transport calculations, whether by empirical high-momentum tails, by three-body collisions, or in off-shell transport calculations, is a matter of much debate.

Besides the sensitivity to simulation strategies, the sensitivity to the physics inputs is of major interest, though it was not included in the present comparison. The relevant physics inputs could be the density-dependence of the symmetry energy, and the momentum dependence of the mean-field potential, or the nucleon effective masses as well as the neutron-proton effective mass difference. The choice of the $\Delta$ potential is a major challenge, since there are no direct experimental constraints. It is of great interest to study how similar the sensitivities to these physics inputs are among different codes.

Finally, an important question is how to quantify the uncertainty of transport model analyses of a given experiment. This will be a project in the near future by TMEP, and is needed in the present multi-messenger era, when results from different physics systems on the same quantity are compared, as in this case the EOS from nuclear structure, heavy-ion collisions, and astrophysical observations. As seen from the present comparison, different codes do not agree for the same physics model, and therefore will deduce different physics ingredients from describing the same experiment. The present study also shows, that it may not be possible to reach sufficiently good convergence in heavy-ion collisions for all codes to constrain tightly a given physics quantity, since the differences lie in different simulation strategies. To take the mean of the predictions of different codes and the variance as its uncertainty as the result of an analysis is certainly not satisfactory.

In the absence of exact solutions of transport equations, comparisons with experimental data could perhaps provide a way to error quantification. 
What is needed, at minimum, is to compare models to measured observables that correspond to the degrees of freedom, which here are the nucleons, that dominate the reaction dynamics. Any model that attempts to study the EOS or other quantities of interest, using rare probes such as pions, should first be calibrated to describe the global features of the reaction. While often this will already provide some information about the EOS, the rare probes may lead to an increased accuracy or perhaps a different range of probed densities, asymmetries, or temperatures. The quality and robustness of the predictions of a transport analysis could then be judged by the degree to which a code succeeds in this. In an analysis of an experiment with several codes, the result and its uncertainty could be found in a multi-observable multi-model Bayesian analysis. Schemes for Bayesian model averaging have been proposed, which average predictions of several models with weights depending on the capability to reproduce given data ~\cite{Eve21,Cir22,Qiu24}. Such methods could be applied to the problem of constraining the nuclear matter EOS in heavy-ion collisions.

\begin{acknowledgments}

We thank Swagata Mallik, Tatsuhiko Ogawa, Dmytro Oliinychenko, and Taesoo Song for participating the homework calculations in the early stage, and Rui Wang for undertaking calculations in relation to this work. We further thank William G. Lynch, Chun Yuen Tsang, and Yong-Jia Wang for helpful discussions. J. Xu acknowledges the support by the Strategic Priority Research Program of the Chinese Academy of Sciences under Grant No. XDB34030000, the National Natural Science Foundation of China under Grant Nos. 12375125 and 11922514, and the Fundamental Research Funds for the Central Universities. H. Wolter acknowledges the support by the Deutsche Forschungsgemeinschaft (DFG, German Research Foundation) under Germany's Excellence Strategy - EXC-2094 - 390783311, ORIGINS. M. D. Cozma acknowledges the support by the Romanian Ministery of Research, Innovation and Digitization through contract PN 23 21 01 01 / 2023. P. Danielewicz acknowledges the support from the US Department of Energy under Grant No. DESC0019209. C. M. Ko acknowledges the support by the US Department of Energy under Award No. DE-SC0015266. A. Ono acknowledges the support from Japan Society for the Promotion of Science KAKENHI Grant Nos. 21K03528JP and 17K05432JP. M. B. Tsang acknowledges the support by the US National Science Foundation Grant No. PHY-2209145. Y. X. Zhang acknowledges the support in part by National Science Foundation of China Nos. 11875323, 11961141003, 11475262, and 11365004, National Key Basic Research Development Program of China under Grant No. 2018YFA0404404, and the Continuous Basic Scientific Research Project (No. WDJC-2019-13, BJ20002501). N. Ikeno acknowledges the support from Japan Society for the Promotion of Science KAKENHI Grant Number 19K14709JP. Z. Zhang acknowledges the support by the National Natural Science Foundation of China under Grant No. 11905302. L. W. Chen acknowledges the support by the National Natural Science Foundation of China under Grant Nos. 12235010 and 11625521, and National SKA Program of China No. 2020SKA0120300. B. A. Li acknowledges the support by the U.S. Department of Energy, Office of Science, under Award Numbers DESC0013702 and DESC0009971. F. S. Zhang acknowledges the support by the National Natural Science Foundation of China under Grant Nos. 12135004, 11635003, and 11961141004.
\end{acknowledgments}

\end{document}